\documentclass[12pt]{article}

\usepackage[left=2.51cm,right=2.51cm,top=2.54cm,bottom=2.54cm]{geometry}
\usepackage[english=british]{csquotes}
\usepackage{amsfonts}
\usepackage{graphicx}
\usepackage{amsmath,amssymb, amsfonts}
\usepackage{IEEEtrantools}
\usepackage{latexsym}
\usepackage{mathrsfs}
\usepackage{color}
\usepackage{slashed}
\usepackage{cancel}
\usepackage{soul}
\usepackage{ulem} %
\usepackage{setspace} %
\usepackage{comment}
\usepackage{booktabs}
\usepackage{multirow}
\usepackage{tabularray}
\usepackage{tikz-feynman}
\tikzfeynmanset{compat=1.0.0}
\usepackage{cite}
\usepackage[small]{caption}
\usepackage{epsfig}
\usepackage{slashed}
\usepackage{epstopdf}
\usepackage{latexsym}
\usepackage{graphicx}
\usepackage{float}
\usepackage{subcaption}
\usepackage{rotating}
\usepackage{subcaption}
\usepackage{bbm}
\usepackage{bm}
\usepackage{bbold}
\usepackage{xfrac}
\usepackage{enumitem}
\usepackage{duckuments}

\usepackage{empheq}
\usepackage[colorlinks,linkcolor=black,citecolor=blue,urlcolor=blue,linktocpage]{hyperref}

\usepackage[colorinlistoftodos,prependcaption,textsize=tiny]{todonotes}

\usepackage[percent]{overpic}
\usepackage{tikz}               
\usetikzlibrary{tikzmark,calc,arrows,arrows.meta,shapes,decorations.pathreplacing,decorations.markings,intersections,angles,quotes,patterns,patterns.meta,shapes.misc,fadings,shapes.geometric,shadows,hobby}
\usepackage{subcaption} 
\usepackage{pgfplots}
\pgfplotsset{compat=1.11}
\usepgfplotslibrary{fillbetween}
\pgfdeclarelayer{bg}
\pgfsetlayers{bg,main}
\tikzset{every picture/.style={remember picture}}
\tikzset{
  end dots/.style={
    -,
    postaction={decorate},
    decoration={
      markings,
      mark=at position 1 with {
        \fill (2pt,0) circle[radius=0.6pt]
              (5pt,0) circle[radius=0.6pt]
              (8pt,0) circle[radius=0.6pt];
      }
    }
  }
}

\definecolor{mygreen}{HTML}{04af70}
\definecolor{mygreen-2}{HTML}{43af04}
\definecolor{myred}{HTML}{af0443}
\definecolor{myblue}{HTML}{7004af}
\definecolor{myorange}{HTML}{af7004}
\definecolor{mypink}{HTML}{ff1493}

\definecolor{turquoise}{RGB}{0, 247, 230}
\definecolor{goldenyellow}{RGB}{255, 218, 66}
\definecolor{fuchsia}{RGB}{255, 0, 172}
\definecolor{petrolgreen}{RGB}{66, 144, 145}

\definecolor{blond}{RGB}{251,231,161}
\definecolor{skincolor}{RGB}{224,177,132}
\definecolor{idea}{RGB}{254,231,2} 	        

\numberwithin{equation}{section}
\begin{document}

\thispagestyle{empty}

\begin{center}

	\vspace*{-.6cm}

	\begin{center}

		\vspace*{1.1cm}

		{\hspace{-0.3cm}\centering \Large\textbf{$\mathcal{N}=4$ Super Yang-Mills Flux Tube in AdS \\
        \hspace{-0.3cm}}}

	\end{center}

	\vspace{0.8cm}
	{\hspace{-0.8cm}\bf Victor Gorbenko, Orr Sela, Brian R. Zielazienski}

	\vspace{1.cm}

	{ Laboratory for Theoretical Fundamental Physics, Institute of Physics,
Ecole Polytechnique F\'{e}d\'{e}rale de Lausanne, Switzerland}

	\vspace{.3cm}

    \today

\end{center}

\vspace{1cm}

\centerline{\bf Abstract}
\vspace{2 mm}
\begin{quote}
We study a supersymmetric Wilson line on the boundary of $\mathcal{N}=4$ super Yang-Mills theory in AdS$_4$, or equivalently on a half-space, as a model for exploring the connection between holographic strings and confining flux tubes. The boundary breaks four-dimensional conformal symmetry and is expected to remove integrability, bringing this system closer to a confining string while preserving enough supersymmetry to constrain its dynamics. Working in the planar limit, we analyze the defect spectrum at weak and strong coupling and compute the superconformal index, thereby relating the perturbative gauge-theory description to the worldsheet excitations of a long holographic string. These complementary perspectives allow us to propose a map between operators in the two regimes, under which the worldsheet axion, in particular, becomes a two-particle state at strong coupling. The resulting picture supports a smooth interpolation from an exponentially growing gauge-theory spectrum to the Cardy-like spectrum of a string worldsheet, providing a non-integrable supersymmetric model for investigating the spectral reorganization expected in confining gauge theories.\end{quote}

\newpage

\tableofcontents


\section{Introduction}
\label{sec:introduction}
Color flux tubes are central objects in non-abelian gauge theories, and in confining theories they have finite width and tension and are responsible for the linear potential between colored sources. In a conformal theory, by contrast, flux lines remain spread out, and flux tubes are not well localized in space. If the conformal theory is also holographic, these flux tubes admit a dual description in terms of strings in the bulk. These strings, however, appear quite different from confining strings, in particular because their spectrum universally contains a massless graviton. Nevertheless, there is at least one setting in which confining strings and holographic strings have much in common. This setup is achieved by placing a confining theory in a rigid AdS space, in which case its flux tube corresponds to a conformal defect in a lower-dimensional nonlocal CFT on the AdS boundary. Varying the AdS radius from small to large then gives perturbative descriptions of this object in terms of gauge-theory and worldsheet degrees of freedom, respectively \cite{Barak:2026ym1,Barak:2026ym2}. In this sense, the setup is analogous to that of a supersymmetric Wilson line in $\mathcal{N}=4$ SYM \cite{Maldacena:1998wilson,Rey:1998wilson,Drukker:1999minimal,Giombi:2017cqn}, which has been extensively studied using integrability \cite{Grabner:2020nis,Cavaglia:2021bnz,Cavaglia:2022qpg,Julius:2026rosetta}. In the planar limit, bulk gravitational dynamics decouple from this long string, so the presence of the graviton in the short-string spectrum is not as relevant. Moreover, at strong coupling, the worldsheet degrees of freedom in both cases are Goldstone bosons or fermions associated with the symmetries spontaneously broken by the string.

Despite these similarities, important differences remain between the two objects. Unlike a pure Yang-Mills defect, the $\mathcal{N}=4$ defect is integrable, which leads to certain features in its spectrum. In particular, levels with the same regular quantum numbers can cross as the coupling, and hence the holographic dual curvature radius in string units, is varied. On the other hand, for a flux tube in a generic non-integrable confining theory such as pure Yang-Mills, levels with the same quantum numbers are expected to repel. Another difference is the extra dimension in the holographic description, which appears on the worldsheet as an additional Goldstone boson associated with conformal transformations. For these reasons it is interesting to consider an intermediate setup which shares some features of both theories. We therefore study a flux tube in $\mathcal{N}=4$ SYM theory placed on half-space, or equivalently on AdS$_4$, with appropriate boundary conditions. In the case of pure Yang-Mills it is important to use Neumann boundary conditions for the gauge fields in order to avoid a deconfinement phase transition at sufficiently small AdS radius \cite{Aharony:2013confinement,DiPietro:2024confinement}. By analogy, we will also impose Neumann-like boundary conditions on the gauge fields of $\mathcal{N}=4$ at weak coupling. Such boundary conditions were studied in \cite{Gaiotto:2009sbc,Gaiotto:2009sdua}, and their generalizations include introducing a 3d CFT whose global $SU(N)$ symmetry is gauged by the boundary values of the bulk gauge fields. In the microscopic brane construction, this 3d theory is associated with NS5 branes on which the D3 branes terminate. In particular, a single NS5 brane corresponds to pure Neumann boundary conditions and a trivial 3d theory, whereas multiple NS5 branes give rise to a nontrivial boundary theory. Note that introducing some number of D5 branes would instead fix Dirichlet boundary conditions for some of the gauge fields. In addition, since we work in the planar limit we cannot make any straightforward use of S-duality. 

The holographic spacetimes dual to strongly-coupled $\mathcal{N}=4$ SYM with these boundary conditions have also been studied before, see \cite{DHoker:2007ehb1,DHoker:2007ehb2,Aharony:2011nhs}. In these geometries, which we review below, the bulk spacetime consists of an AdS$_4$ factor, two $S^2$ factors and a two-dimensional noncompact manifold $\Sigma$, with the radii of the AdS$_4$ and sphere factors depending nontrivially on the coordinates of $\Sigma$. In particular, the geometry contains an NS5 throat with a singularity associated with the NS5 branes on which the D3 branes terminate. Since such backgrounds are believed to be generically non-integrable, we do not expect levels with the same quantum numbers to cross as the parameters of the model are varied.  

We next introduce a straight supersymmetric Wilson line on the boundary of the AdS$_4$ spacetime in which SYM is defined. Its holographic dual is a fundamental string whose worldsheet spans an AdS$_2$ inside the bulk AdS$_4$ factor \cite{Coccia:2022map,He:2024djr}, an embedding that already resembles that of a confining flux tube in AdS$_4$. A important question at this step is then the position of the string on $\Sigma$, which, as we will see, depends strongly on the number of NS5 branes. When this number scales with $N$, the string lies very far from the NS5's so that the spacetime is weakly curved. This regime is in a sense more similar to the critical long string in AdS$_5\times$S$_5$. On the other hand, when the number of NS5 branes becomes smaller, the string moves closer to the throat and the curvature of some of the transverse directions increases. To retain perturbative control of the worldsheet theory, it turns out that we still need a large number of branes, although this number does not have to scale with $N$. It appears that a small number of NS5 branes produces a string with some properties resembling more closely those of the Yang-Mills flux tube. In particular, there are some indications that for a single brane some of the transverse dimensions disappear. While we will not discuss this regime in detail here, it certainly deserves further study.

Supersymmetry plays an important role in our analysis. The boundary conditions break half of the $\mathcal{N}=4$ supercharges, while the Wilson line further breaks half of the remaining ones. Using the preserved supercharges, we identify the superconformal multiplets on the defect at weak and strong coupling and compute the characters that count their states. We then compute the superconformal index, which provides coupling-independent constraints on the spectrum and connects the two descriptions. Together, these calculations allow us to conjecture how the defect CFT spectrum evolves as the coupling, or equivalently the AdS radius in string units, is varied. The holographic description gives us all the reasons to believe that the interpolation is smooth and that we thus have an example of a non-integrable conformal defect which interpolates between a gauge-theory spectrum with an exponentially growing density of states and a Cardy-like spectrum at large AdS radius in string units.  In this way, the non-integrable conformal defect provides a toy model and a proof of principle for the spectral reorganization expected in confining theories on an AdS background. 

Among other states, we discuss in detail the operator corresponding to the worldsheet axion, expected to be present on generic four-dimensional flux tubes \cite{Dubovsky:2013axion,Dubovsky:2025axion}. At weak coupling, this operator is represented by a single insertion of the field strength with appropriate transverse indices into the Wilson line. At strong coupling, however, this operator is not present in the worldsheet theory and we identify it with a two-particle state, as in $\mathcal{N}=4$ SYM without a boundary \cite{Julius:2026rosetta}. Thus, unlike in pure Yang-Mills, the axion "decays" into two more fundamental particles and does not appear as a narrow worldsheet resonance. 

The rest of the paper is organized as follows. Section 2 introduces the field-theory setup and its holographic description, while section 3 analyzes the Wilson-line spectrum at weak and strong coupling and computes its superconformal index. Using these results, section 4 proposes a map between operators in the two regimes, and section 5 presents our conclusions and future directions. Appendices A and B discuss the Wilson-line index without a boundary and the discrete symmetries of the line, respectively.

\section{$\mathbf{\mathcal{N}=4}$ SYM on half space with a Wilson line on the boundary}
\label{sec:SYM-on-AdS-and-half-space}

We begin in this section by introducing the BCFT setup and its holographic realization, providing the ingredients required for the weak- and strong-coupling analyses that follow. We first review $\mathcal{N}=4$ SYM on a half-space with NS5-like $1/2$-BPS boundary conditions, including both pure Neumann boundary condition for the gauge field and its generalization obtained by coupling the bulk theory to a three-dimensional SCFT on the boundary. We then describe the corresponding type-IIB supergravity solutions, emphasizing their asymptotic regions associated in the field-theory description with the bulk and boundary degrees of freedom. Finally, we introduce a $1/2$-BPS Wilson line supported on the boundary and identify its holographic dual as a fundamental string wrapping $AdS_2$. We determine the position of this string and analyze the dilaton and string-frame curvature there, thereby identifying the regime in which the supergravity description used in the subsequent strong-coupling analysis is reliable. 

\subsection{$\mathbf{\mathcal{N}=4}$ SYM on half space with NS5-like boundary conditions}\label{sec:NS5-boundary-conditions}

Our starting point is $\mathcal{N}=4$ SYM on $\mathbb{R}^{1,3}$ with gauge group $SU(N)$, which has the superconformal symmetry $PSU(2,2|4)$ and consists of the following matter content, 
\begin{itemize}
    \item Six scalars $\Phi_I$, with $I=1,\dots,6$, transforming in the $\boldsymbol{6}$ of the $\textrm{Spin}(6)_{R}\simeq SU(4)_{R}$ $R$-symmetry group;
    \item Four Weyl fermions $\Psi_{A\alpha}$ and their conjugates $\bar{\Psi}_{\dot{\alpha}}^{A}$,  with $A=1,\dots,4$, transforming in the $\mathbf{4}$ and $\mathbf{\bar{4}}$ of the $R$-symmetry group, respectively;
    \item A gauge field $A_\mu$, $\mu = 0,\dots,3$.
\end{itemize}
We then seek to introduce a maximally supersymmetric planar boundary at $x^3 = 0$ and consider the $4d$ theory on $x^3\geq 0$, which requires imposing suitable boundary conditions to the bulk fields at $x^3 = 0$. In \cite{Gaiotto:2009sbc} (see also \cite{Wang:2020td}) it was shown that 1/2-BPS boundary conditions can be obtained, preserving an $OSp\left(4|4\right)$ subgroup of the bulk $PSU(2,2|4)$ and corresponding to $3d$ $\mathcal{N}=4$ superconformal symmetry. Since our interest in this work would be on Neumann-like boundary conditions for the gauge field, corresponding to either pure Neumann or to coupling the gauge field on the boundary to a nontrivial 3d CFT, we will next sketch briefly how such 1/2-BPS boundary conditions are obtained for the bulk theory, starting from the pure Neumann case. 

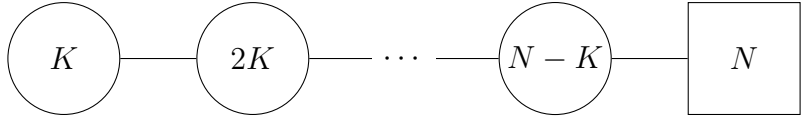
\begin{figure}
    \begin{center}
        \begin{tikzpicture}
            \node[draw,circle,inner sep=2pt,minimum width=1.48cm] (K) {$K$};
            \node[draw,circle,inner sep=2pt,right of=K,minimum width=1.48cm,xshift=1.5cm] (2K) {$2K$};
            \node[draw,circle,inner sep=2pt,right of=2K,xshift=3cm] (N-K) {$N-K$};
            \node[draw,inner sep=2pt,right of=N-K,minimum width=1.48cm,minimum height=1.48cm,xshift=1.5cm] (N) {$N$};
            \draw (K) -- (2K); 
            \draw (2K) -- (N-K) node[midway,fill=white,align=center]{$\cdots$};
            \draw (N-K) -- (N);
        \end{tikzpicture}
    \end{center}
    \caption{Quiver diagram of the 3d gauge theory flowing to the $T_{[K^{N/K}]}[SU(N)]$ SCFT that appears on the boundary in our setup, which in turn corresponds to $N$ D3-branes ending on a stack of $N_{\text{NS5}} =N/K$ NS5-branes such that $K$ D3-branes terminate on each NS5-brane. A circle labeled $n$ represents a $3d$ $\mathcal{N}=4$ $U(n)$ vector multiplet, and a line connecting two such nodes denoted $n$ and $m$ corresponds to a bifundamental hypermultiplet transforming under $U(n)\times U(m)$. The square at the right stands for an $SU(N)$ global symmetry, which in the BCFT setup is gauged using the boundary restriction of the $SU(N)$ bulk gauge field. Meanwhile, the line connecting this square to the rightmost circle corresponds to a bifundamental hypermultiplet, analogously to the other lines.}
    \label{fig:quiver}
\end{figure}

First, in order to render the $OSp\left(4|4\right)$ symmetry of the boundary transparent, let us look at the 4d $\mathcal{N}=4$ SYM theory as a 3d $\mathcal{N}=4$ one, i.e. decompose the 4d $\mathcal{N}=4$ vector multiplet into multiplets (or representations) of the 3d $\mathcal{N}=4$ subgroup. This decomposition, in turn, results in a 3d $\mathcal{N}=4$ vector multiplet and a hypermultiplet obtained in the following way. Defining 
\begin{equation}
\label{XY}
    \vec{X}:=(\Phi_1,\Phi_2,\Phi_3)\in (\mathbf{3}_{SO(3)_X},\mathbf{1}_{SO(3)_Y})\quad ,\quad \vec{Y}:=(\Phi_4,\Phi_5,\Phi_6)\in(\mathbf{1}_{SO(3)_X},\mathbf{3}_{SO(3)_Y})
\end{equation}
with $SO(3)_X\times SO(3)_Y$ a subgroup of the $\textrm{Spin}(6)$ $R$-symmetry group, $\vec{X}$ combines with $A_3$, the gauge-field component normal to the boundary, to form the four real scalars of a 3d $\mathcal{N}=4$ hypermultiplet. Then, $\vec{Y}$ corresponds to the three real scalars of a vector multiplet in the three-dimensional sense, with $A_\mu$, $\mu=0,1,2$, the associated gauge field. Schematically, we have the following decomposition:
\begin{equation}\label{eq:3d/4d-multiplets}
    4d\ \mathcal{N}=4:\ \text{vector}=\{\Phi,\Psi, A_\mu\}\quad \rightarrow\quad 3d\ \mathcal{N}=4:\
    \left\{
    \begin{IEEEeqnarraybox}[\IEEEeqnarraystrutmode][c]{l}
        \text{hyper}=\{(\vec{X},A_3),\Psi_-\}\\
        \text{vector}=\{\vec{Y},\Psi_+,A_\mu\}
    \end{IEEEeqnarraybox}
    \right.\ , 
\end{equation}
where the fermions $\Psi_\pm$ on the right are appropriately obtained from $\Psi$. Using these notations, the 1/2-BPS pure Neumann boundary condition preserving $OSp\left(4|4\right)$ is then given by imposing Neumann boundary conditions for the 3d $\mathcal{N}=4$ vector multiplet and Dirichlet boundary conditions for the hypermultiplet,\footnote{In fact, imposing Neumann boundary conditions on the vector multiplet enforces Dirichlet boundary conditions for the hypermultiplet. This can be seen either by studying the full $3d$ $\mathcal{N}=4$ multiplet structure, including auxiliary fields, or by solving the BPS equations for preserving $OSp(4|4)$, as in \cite{Gaiotto:2009sbc}.} that is 
\begin{equation}\label{eq:pure-neumann-boundary-conditions}
    x^3 = 0:\ F_{3\mu} = \vec{X} = A_3 = \Psi_- = \mathcal{D}_3\vec{Y} = 0\ .
\end{equation}
As discussed in \cite{Gaiotto:2009sbc}, this setup, which is given by 4d $\mathcal{N}=4$ SYM on half space with the 1/2-BPS pure Neumann boundary condition above, arises from $N$ semi-infinite parallel D3-branes ending on a single NS5-brane. Meanwhile, the more general brane construction involving $N$ D3-branes ending on a stack of $N_{\text{NS5}}=N/K$ NS5-branes, such that $K$ D3-branes terminate on each NS5-brane, corresponds to the more general boundary condition in which the 3d vector multiplet in (\ref{eq:3d/4d-multiplets}) couples to a $3d$ $\mathcal{N}=4$ SCFT on the boundary called $T_{[K^{N/K}]}[SU(N)]$ \cite{Gaiotto:2009sbc}. The latter has an $SU(N_{\text{NS5}})\times SU(N)$ global symmetry and the coupling to the bulk theory is done by gauging the $SU(N)$ part with the bulk gauge field (in a 3d $\mathcal{N}=4$ way). The $T_{[K^{N/K}]}[SU(N)]$ SCFT itself can be obtained as the IR fixed point of the gauge theory described by the quiver diagram in figure \ref{fig:quiver}, with the coupling to the bulk 4d $\mathcal{N}=4$ SYM theory in the BCFT setup done by the $SU(N)$ gauging we described. Note that in this BCFT setup the $4d$ gauge coupling remains exactly marginal. Moreover, at finite coupling, the Dirichlet boundary conditions imposed on the triplet $\vec{X}$ of the hypermultiplet are shifted to
\begin{equation}
\label{X_mu}
    \vec{X}|_{x^3 = 0} = -\vec{\mu}\ ,
\end{equation}
where $\vec{\mu}$ is the moment map operator for the $SU(N)$ flavor symmetry of the 3d SCFT, i.e. the operator lying at the bottom of the corresponding conserved-current multiplet.
Overall, the $PSU(2,2|4)$ superconformal symmetry of $4d$ $\mathcal{N}=4$ SYM is minimally broken by the planar boundary at $x^3 = 0$ down to an $OSp\left(4|4\right)$ subgroup, corresponding to $3d$ $\mathcal{N}=4$ superconformal symmetry. Among the possible boundary conditions preserving this symmetry, we focus on those corresponding to D3/NS5-brane constructions, where the $3d$ $\mathcal{N}=4$ vector multiplet satisfies the Neumann boundary condition for a single NS5-brane while being coupled to the 3d SCFT $T_{[K^{N/K}]}[SU(N)]$ for the setup with multiple NS5-branes. Given their origin, we refer to both of these boundary conditions as NS5-like, in accordance with \cite{Gaiotto:2009sbc}, and next turn to discuss their holographic description.  

\subsection[The AdS$_4\times S^2\times S^2\times\Sigma$ holographic dual]{The AdS$\mathbf{_4\times S^2\times S^2\times\Sigma}$ holographic dual}\label{sec:holographic-dual}

The $OSp(4|4)$ symmetry preserved by the 1/2-BPS 3d boundary contains the bosonic subgroup
\begin{equation}
    SO(2,3)\times SO(3)\times SO(3)\, ,
\end{equation}
suggesting that the holographic dual should involve AdS$_4\times S^2\times S^2$, which has the same isometry group. Indeed, it was shown in \cite{DHoker:2007ehb1,DHoker:2007ehb2,Aharony:2011nhs} that $OSp(4|4)$-symmetric solutions of type-IIB supergravity with all the desired properties take the form
\begin{equation}\label{eq:supergravity-solution}
    \mathrm{d}s^2 = f_4^2\mathrm{d}s_{\text{AdS}_4}^2 + f_1^2\mathrm{d}s_{S_1^2}^2 + f_2^2\mathrm{d}s_{S_2^2}^2 + f_\Sigma^2|\mathrm{d}z|^2\, ,
\end{equation}
where $z = re^{i\theta} = x + iy$ parametrizes the first quadrant of the complex plane
\begin{equation}
    \Sigma := \{x + iy\ |\ x\geq 0,\ y\geq 0\}\, .
\end{equation}
In Einstein-frame,\footnote{Recall that the Einstein-frame metric $G_{\mu\nu}^{(E)}$ is related to the string-frame one $G_{\mu\nu}$ by $G_{\mu\nu}^{(E)}=e^{-\frac{\Phi}{2}}G_{\mu\nu}$, with $\Phi$ the dilaton.} the metric functions in (\ref{eq:supergravity-solution}) are given by 
\begin{equation}\label{eq:metric-coefficient-functions}
    f_4^8 = 16\frac{\mathcal{N}_1\mathcal{N}_2}{\mathcal{W}^2}\,,\quad f_1^8 = 16h_1^8\frac{\mathcal{N}_2\mathcal{W}^2}{\mathcal{N}_1^3}\,,\quad f_2^8 = 16h_2^8\frac{\mathcal{N}_1\mathcal{W}^2}{\mathcal{N}_2^3}\,,\quad f_\Sigma^8 = 16^2\frac{\mathcal{N}_1\mathcal{N}_2\mathcal{W}^2}{h_1^4h_2^4}
\end{equation}
with
\begin{equation}
\label{WNi}
    \mathcal{W} = \partial_z h_1\partial_{\bar{z}} h_2 + \partial_{\bar{z}} h_1 \partial_z h_2\, ,\quad \mathcal{N}_i = 2h_1h_2|\partial_z h_i|^2 - h_i^2\mathcal{W}\,,
\end{equation}
such that the solution is completely determined in terms of the two functions $h_1$ and $h_2$, which in turn are harmonic functions on $\Sigma$. Similarly,  the rest of the supergravity fields can be expressed in terms of $h_1$ and $h_2$, with the dilaton given by 
\begin{equation}\label{eq:dilaton}
e^{2\Phi} = \frac{\mathcal{N}_2}{\mathcal{N}_1}
\end{equation}
and the $B$-field by 
\begin{equation}\label{eq:B_field}
B=b_{1}\sin\theta_{1}d\theta_{1}\wedge d\phi_{1}\,,\quad b_{1}=2h_{2}^{D}+2ih_{1}^{2}h_{2}\left(\partial_{z}h_{1}\partial_{\bar{z}}h_{2}-\partial_{\bar{z}}h_{1}\partial_{z}h_{2}\right)/\mathcal{N}_{1}\,.
\end{equation}
Here, $\theta_{1}$ and $\phi_{1}$ are the coordinates on $S^2_1$,\footnote{$\,$Note that the $B$-field takes the form $B=b_1\omega$ with $\omega = \sin\theta_1d\theta_1\wedge d\phi_1$ the volume form of $S^2_1$. The corresponding NS-NS field strength is then given by $H_{3}=dB=db_{1}\wedge\omega$.} and $h_{2}^{D}$ is the harmonic dual of $h_2$ obtained from  
\begin{equation}
    \begin{IEEEeqnarraybox}[\IEEEeqnarraystrutmode][c]{lcl}
        h_1(z,\bar{z})\ &=&\ -i\left(\mathcal{A}_1(z) - \bar{\mathcal{A}}_1(\bar{z})\right) \\[4pt]
        h_2(z,\bar{z})\ &=&\ \mathcal{A}_2(z) + \bar{\mathcal{A}}_2(\bar{z})
    \end{IEEEeqnarraybox}
    \quad\implies\quad 
    \begin{IEEEeqnarraybox}[\IEEEeqnarraystrutmode][c]{lcl}
        h_1^D(z,\bar{z})\ &=&\ \mathcal{A}_1(z) + \bar{\mathcal{A}}_1(\bar{z}) \\[4pt]
        h_2^D(z, \bar{z})\ &=&\  i\left(\mathcal{A}_2(z) - \bar{\mathcal{A}}_2(\bar{z})\right)
    \end{IEEEeqnarraybox}
\end{equation}
where $\mathcal{A}_1(z)$ and $\mathcal{A}_2(z)$ are holomorphic. We will not need below the expressions for the other supergravity fields, and will therefore avoid writing them explicitly (see \cite{VanRaamsdonk:2021hlc,Coccia:2022map} for recent discussions that include the expressions for all the fields). Nevertheless, note that, in addition to the NSNS fields described above, there is an RR 3-form flux $F_3=\mathrm{d}C_2$, with $C_2$ proportional to the volume form on $S_2^2$, as well as a self-dual RR 5-form flux $F_5$. Our background has no RR 0-form potential.

In the general case where D5-branes are also involved in the brane construction (in addition to the NS5 ones we consider), $h_1$ and $h_2$ are given by\footnote{We here follow the conventions of \cite{VanRaamsdonk:2021hlc}, except that the coordinates $x$ and $y$ on $\Sigma$ in our case correspond to dividing those of \cite{VanRaamsdonk:2021hlc} by $\sqrt{N}$. Correspondingly, we have the relations  $f_{\Sigma}^{2}=4N\rho^{2}$, $\mathcal{W}=NW$ and $\mathcal{N}_{i}=NN_{i}$ to the quantities $\rho$, $W$ and $N_{i}$ used in \cite{VanRaamsdonk:2021hlc}.} 
\begin{equation}
\label{h1_h_2_gen}
    \begin{aligned}
   h_{1} &= \frac{\pi}{2}L_{AdS}^{2}\sqrt{\frac{N}{\lambda}}x+\frac{L_{AdS}^{2}}{8\sqrt{\pi N}}\sum_{A}N_{\textrm{D5}}^{\left(A\right)}\log\left(\frac{\left(x+\mathcal{L}_{A}\right)^{2}+y^{2}}{\left(x-\mathcal{L}_{A}\right)^{2}+y^{2}}\right),\\  
    h_2 &=\frac{1}{8}L_{AdS}^{2}\sqrt{\frac{\lambda}{N}}y+\frac{L_{AdS}^{2}}{8\sqrt{\pi N}}\sum_{B}N_{\textrm{NS5}}^{\left(B\right)}\log\left(\frac{x^{2}+\left(y+\mathcal{K}_{B}\right)^{2}}{x^{2}+\left(y-\mathcal{K}_{B}\right)^{2}}\right),
    \end{aligned}
\end{equation}
where $\lambda = g_{\text{YM}}^2 N$ is the 't Hooft coupling and $L_{AdS}$ the Einstein-frame AdS radius, such that $L_{AdS}^{4} = 4\pi Nl_s^4$ with $l_s = \sqrt{\alpha'}$ the string length. The locations of the poles of (the derivatives of) $h_1$ and $h_2$ on the $x$ and $y$ axes of $\Sigma$, denoted by $\mathcal{L}_A$ and $\mathcal{K}_B$, correspond to the locations of the D5 and NS5 branes in the associated brane construction, respectively, and satisfy the following relation (see \cite{VanRaamsdonk:2021hlc} for a recent discussion),
\begin{equation}\label{eq:relation-rank-poles}
    N = \sqrt{\frac{\lambda}{4\pi}}\sum\limits_A N_{\text{D5}}^{(A)}\,\mathcal{L}_A + N\sqrt{\frac{4\pi}{\lambda}}\sum\limits_B N_{\text{NS5}}^{(B)}\,\mathcal{K}_B\, .
\end{equation}
In the case we are interested in, we have $N$ D3-branes ending on a stack of $N_{\text{NS5}}=N/K$ NS5-branes (with $K$ possibly being equal to $N$ such that $N_{\text{NS5}}=1$), and in order to keep track of the effect of the $N$-scaling of $K$ on the result, we write 
\begin{equation}
K = \kappa N^\alpha\,,\quad 0\leq\alpha\leq 1\,,\quad \kappa = \mathcal{O}(N^0)\,.
\end{equation}
The special case $\alpha =\kappa = 1$, in particular, corresponds to imposing Neumann boundary conditions to the 3d vector multiplet in (\ref{eq:3d/4d-multiplets}), while the other cases correspond to coupling it to the 3d boundary SCFT. We then have $N_{\text{D5}}^{(A)} = 0$ and $N_{\text{NS5}}^{(B)} = \delta_{1B}N^{1-\alpha}/\kappa$, which after substituting in \eqref{h1_h_2_gen} (and using \eqref{eq:relation-rank-poles}) results in 
\begin{equation}
    h_{1}=\frac{\pi}{2}L_{AdS}^{2}\sqrt{\frac{N}{\lambda}}x\,,\;\,h_{2}=\frac{1}{8}L_{AdS}^{2}\sqrt{\frac{\lambda}{N}}y+\frac{L_{AdS}^{2}}{8\kappa\sqrt{\pi}}N^{1/2-\alpha}\log\left(\frac{x^{2}+\left(y+\sqrt{\frac{\lambda}{4\pi}}\kappa N^{-\left(1-\alpha\right)}\right)^{2}}{x^{2}+\left(y-\sqrt{\frac{\lambda}{4\pi}}\kappa N^{-\left(1-\alpha\right)}\right)^{2}}\right).   
    \label{h1_h2_gen}
\end{equation}
This completely specifies the holographic dual of the field-theory setup under consideration, and in order to get a better understanding of its structure, we will next examine the explicit form of the metric at interesting locations on the $\Sigma$ space (recall that the coordinate of $\Sigma$ is $z=re^{i\theta}=x+iy$). \\

\noindent\underline{\textit{The asymptotic region $r\rightarrow\infty$}:}\\

\noindent Here, we have 
\begin{equation}
\label{h1h2_r_inf}
    h_{1}=\frac{\pi}{2}L_{AdS}^{2}\sqrt{\frac{N}{\lambda}}\,r\cos\left(\theta\right)\,,\quad h_{2}=\frac{1}{8}L_{AdS}^{2}\sqrt{\frac{\lambda}{N}}\left(r+\frac{2}{\pi r}\right)\sin\left(\theta\right)+\mathcal{O}\left(r^{-3}\right)
\end{equation}
and therefore 
\begin{equation}
\label{ds2_large_r}
    ds^{2}\simeq L_{AdS}^{2}\left[\left(\frac{dr^{2}}{r^{2}}+\frac{\pi}{4}r^{2}ds_{AdS_{4}}^{2}\right)+\left(d\theta^{2}+\cos^{2}\left(\theta\right)ds_{S_{1}^{2}}^{2}+\sin^{2}\left(\theta\right)ds_{S_{2}^{2}}^{2}\right)\right].
\end{equation}
The first part in \eqref{ds2_large_r} is an $AdS_{5}$ metric written in $AdS_{4}$ slicing at large $r$, as can be seen by the coordinate change $r=e^{\chi}/\sqrt{\pi}$, 
\begin{equation}
ds_{AdS_{5}}^{2}=d\chi^{2}+\cosh^{2}\left(\chi\right)ds_{AdS_{4}}^{2}\simeq d\chi^{2}+\frac{1}{4}e^{2\chi}ds_{AdS_{4}}^{2}=\frac{dr^{2}}{r^{2}}+\frac{\pi}{4}r^{2}ds_{AdS_{4}}^{2}\,,
\end{equation}
while the second part is the metric of an $S^{5}$. Overall, we have
\begin{equation}
ds^{2}\simeq L_{AdS}^{2}\left(ds_{AdS_{5}}^{2}+ds_{S^{5}}^{2}\right)\end{equation}
and we see that the metric in the limit $r\rightarrow\infty$ is that of an $AdS_{5}\times S^{5}$ with a common radius $L_{AdS}$. This form, in turn, is obtained from one coordinate of $\Sigma$ (i.e. $r$) combining with the $AdS_{4}$ part to form $AdS_{5}$ and from the other coordinate of $\Sigma$ (i.e. $\theta$) combining with the two remaining spheres to form $S^{5}$. This asymptotic part of the geometry therefore corresponds to the 4d degrees of freedom in the field-theory dual, that is to the $\mathcal{N}=4$ SYM part. Let us also mention that the dilaton, given in \eqref{eq:dilaton}, takes at $r\rightarrow\infty$ the asymptotic value 
\begin{equation}
\underset{r\rightarrow\infty}{\lim}\,\frac{\mathcal{N}_2}{\mathcal{N}_1}=\frac{\lambda^{2}}{16\pi^{2}N^{2}}=g_{s}^{2}
\end{equation}
corresponding to the $\mathcal{N}=4$ SYM gauge coupling (through $g_{YM}^{2}=4\pi g_{s}$), and that one can also check that there is a 5-form flux $N$ on the $S^{5}$.\\

\noindent\underline{\textit{The limit $r\rightarrow0$}:}\\
    
\noindent Here, 
\begin{equation}
h_{2}=\frac{L_{AdS}^{2}N^{2(1-\alpha)}}{\lambda\kappa^{2}}\sqrt{\frac{\lambda}{N}}\,r\sin\theta\left(1+\frac{\lambda\kappa^{2}}{8}N^{-2(1-\alpha)}-\frac{4\pi}{3\lambda\kappa^{2}}N^{2(1-\alpha)}r^{2}(1+2\cos(2\theta))\right)+\mathcal{O}(r^{5})\ ,
\end{equation}
with $h_1$ as in \eqref{h1h2_r_inf}, and we find  
\begin{equation}
    \begin{aligned}
   ds^{2}\simeq&\,\frac{1}{16}\left(8+\lambda\kappa^{2}N^{-2(1-\alpha)}\right)L_{AdS}^{2}\mathrm{d}s_{AdS_{4}}^{2}+\\
   &+\frac{4\pi}{\lambda\kappa^{2}}N^{2(1-\alpha)}L_{AdS}^{2}\left[dr^{2}+r^{2}\left(d\theta^{2}+\cos^{2}\left(\theta\right)ds_{S_{1}^{2}}^{2}+\sin^{2}\left(\theta\right)ds_{S_{2}^{2}}^{2}\right)\right]\\  
    =&\,\frac{1}{16}\left(8+\lambda\kappa^{2}N^{-2(1-\alpha)}\right)L_{AdS}^{2}\mathrm{d}s_{AdS_{4}}^{2}+\frac{4\pi}{\lambda\kappa^{2}}N^{2(1-\alpha)}L_{AdS}^{2}ds_{\mathbb{R}^{6}}^{2}
    \end{aligned}
\end{equation}
such that the two coordinates of $\Sigma$ (i.e. $r$ and $\theta$) combine with the two spheres to form $\mathbb{R}^{6}$. In particular, the point $r=0$ is a regular interior point of the full 10d geometry. This is in fact true for all the boundary points of $\Sigma$, except for the point where (the derivative of) $h_{2}$ has a pole, such that even though all these points are on the boundary of $\Sigma$, they are not boundary points of the 10d geometry. \\

\noindent\underline{\textit{The near-pole limit}:}\\
    
\noindent Let us next examine the vicinity of the pole of the derivative of $h_{2}$, 
\begin{equation}\label{z_pole}
    z\rightarrow i\,\sqrt{\frac{\lambda}{4\pi}}\,\kappa\, N^{-\left(1-\alpha\right)}\,,
\end{equation}
which corresponds to the location of the stack of NS5-branes. Denoting 
\begin{equation}
z=i\,\sqrt{\frac{\lambda}{4\pi}}\,\kappa\,N^{-\left(1-\alpha\right)}+R\,e^{i\phi}\,,\quad-\pi/2\leq\phi\leq\pi/2   
\end{equation}
and considering the leading behavior at the limit $R\rightarrow0$, we obtain 
\begin{equation}
h_{1}=\frac{\pi}{2}L_{AdS}^{2}\sqrt{\frac{N}{\lambda}}\,R\cos\left(\phi\right)\,,\quad h_{2}=\frac{L_{AdS}^{2}}{4\kappa\sqrt{\pi}}\,N^{1/2-\alpha}\,|\log\left(R\right)|+\mathcal{O}\left(R^{0}\right)
\end{equation}
and therefore
\begin{equation}
ds^{2}\simeq L_{AdS}^{2}\left(\frac{\pi}{4\kappa^{2}\lambda}\right)^{\frac{1}{4}}N^{\frac{1}{2}\left(1-\alpha\right)}\sqrt{R}\,|\log\left(R\right)|^{\frac{3}{4}}\left[ds_{AdS_{4}}^{2}+ds_{S_{2}^{2}}^{2}+|\log\left(R\right)|^{-1}\left(\frac{dR^{2}}{R^{2}}+ds_{S^{3}}^{2}\right)\right]
\end{equation}
where the $ds_{S^{3}}^{2}$ part is obtained as $ds_{S^{3}}^{2}=d\phi^{2}+\cos^{2}\left(\phi\right)ds_{S_{1}^{2}}^{2}$. We hence get near the pole an $AdS_{4}\times S^{2}$ factor and an $S^{3}$ one. This corresponds to the 3d boundary (or 3d degrees of freedom) in the field-theory dual, and to the $N^{1-\alpha}/\kappa$ NS5-branes (on which the D3-branes end) wrapping the $AdS_{4}\times S^{2}$ part. Indeed, one can check that there is a corresponding NS5 flux on the $S^{3}$ and that the dilaton diverges at $R\rightarrow0$. \\

To summarize, we see that the 10d supergravity solution corresponding to the D3/NS5-brane system we consider is characterized by an $AdS_{5}\times S^5$ throat, dual to the 4d $\mathcal{N}=4$ SYM theory part, as well as an $AdS_{4}\times S^2$ throat associated with the NS5 branes. See figure \ref{fig:geometry} for a depiction of this geometry.

\subsection{1/2-BPS Wilson line on the boundary}\label{sec:wilson-line-on-the-boundary}

Now that the field-theory and supergravity descriptions have been established, we place on the 3d boundary a Wilson line of the 4d $\mathcal{N}=4$ SYM theory in the fundamental representation of its $SU(N)$ gauge group. We choose the Wilson line to preserve the maximal amount of supersymmetry, which results in half of the supersymmetry preserved by the boundary. In total, the 4d $\mathcal{N}=4$ supersymmetry of SYM undergoes the following breaking pattern, 
\begin{equation}
\label{SUSY_breaking}
   4d\ \mathcal{N}=4\ \xrightarrow{\text{boundary}}\ 3d\ \mathcal{N}=4\ \xrightarrow{\text{\textcolor{white}{aa}line\textcolor{white}{aa}}}\ 1d\ \mathcal{N}=4\,,
\end{equation}
thereby retaining 8 of the original 32 supercharges. Such a Wilson line incorporates in addition to the gauge field also one of the six scalars of $\mathcal{N}=4$ SYM, which we choose to be $\Phi_6$. Indeed, this scalar is in the group of scalars $\vec{Y}$ which is not fixed by the boundary conditions we consider, see Eqs. \eqref{XY}-\eqref{X_mu}. The resulting Wilson line, which takes the form 
\begin{equation}
\label{W}
W=\mathrm{tr}\,\mathcal{P}\exp\left(i\int\mathrm{d}t\hspace{1pt}(A_t + i\Phi_6)\right),
\end{equation}
then preserves 1d $\mathcal{N}=4$ superconformal symmetry and breaks the $SO(3)_Y$ symmetry associated with $\vec{Y}$ to $SO(2)$.

In the holographic description, the Wilson line \eqref{W} corresponds to a fundamental string wrapping an $AdS_{2}$ inside $AdS_{4}$ while being located at a point in $S_{1}^{2}\times S_{2}^{2}\times\Sigma$. As shown in \cite{Coccia:2022map}, the condition for the string to preserve supersymmetry as in \eqref{SUSY_breaking} fixes its position in $S_{1}^{2}\times S_{2}^{2}\times\Sigma$ to be at a point where $h_1 = \partial_z h_2 = 0$. This condition, in turn, implies that $f_1^2 = 0$ (see (\ref{eq:metric-coefficient-functions})) such that the sphere $S_1^2$ shrinks to zero size and its $SO(3)$ isometry is preserved. This is in contrast to the sphere $S_2^2$ which does not shrink and whose associated $SO(3)$ symmetry is broken to $SO(2)$ by the string. In the field theory side this corresponds to the boundary conditions setting $\Vec{X}|_{x^3 = 0}=0$ (or $\vec{X}|_{x^3=0}=-\vec{\mu}$ at finite coupling when there is a nontrivial theory on the boundary), preserving the $SO(3)_X$ symmetry under which $\vec{X}$ transforms as a triplet, and to the inclusion of $\Phi_6$ in the BPS Wilson line which as we discussed breaks $SO(3)_Y$ to $SO(2)$. 

\begin{figure}
    \centering
    \begin{tikzpicture}
        \node (cmap) at (0,0) {\includegraphics[trim= 0cm 0cm 2cm 0cm, clip, width=0.55\textwidth]{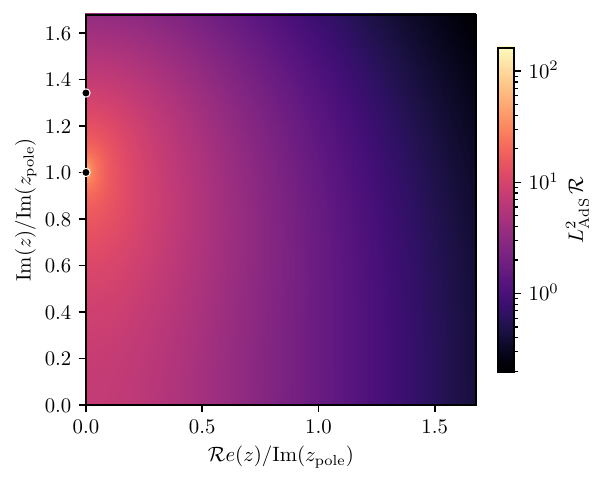}};
        \node[color=white] at (3.25,3.85) {AdS$_5\times S^5$};
        \node[left of=cmap,xshift=-1.4cm,yshift=15pt,rotate=90,color=white] {$S_1^2 \rightarrow 0$};
        \node[below of=cmap,xshift=0.675cm,yshift=-1.55cm,color=white] {$S_2^2\rightarrow 0$};
        \node[color=white] at (-2.05,2.75) {String};
        \node (throat) at (-6.945,1.2925) {\scalebox{0.5}{\definecolor{edge}{HTML}{286578}
\definecolor{mesh}{HTML}{6196A4}
\definecolor{pale}{HTML}{EAF4F5}

\begin{tikzpicture}[rotate=-90,line cap=round,line join=round]
    \def\H{6.4}
    \def\perspective{.28}

    \pgfmathdeclarefunction{radius}{1}{%
      \pgfmathparse{3.3*cos(90*#1/\H)/(1+.65*#1)}%
    }
    \pgfmathdeclarefunction{slope}{1}{%
      \pgfmathparse{
        -3.3*.65*cos(90*#1/\H)/(1+.65*#1)^2
        -3.3*pi*sin(90*#1/\H)/(2*\H*(1+.65*#1))
      }%
    }
    \pgfmathdeclarefunction{sx}{1}{%
      \pgfmathparse{radius(#1)*sqrt(1-(\perspective*slope(#1))^2)}%
    }
    \pgfmathdeclarefunction{sy}{1}{%
      \pgfmathparse{-#1+\perspective^2*radius(#1)*slope(#1)}%
    }
    \pgfmathsetmacro{\tangent}{asin(\perspective*slope(0))}
    \pgfmathsetmacro{\openingdepth}{\perspective*radius(0)}
    \coordinate (NS5) at (0,-\H);

    \path[shade,left color=edge!20!white,middle color=pale,
          right color=edge!30!white,shading angle=0]
      (-3.3,0)
      arc[start angle=180,end angle=0,
          x radius=3.3,y radius=\openingdepth]
      arc[start angle=0,end angle=\tangent,
          x radius=3.3,y radius=\openingdepth]
      -- plot[domain=0:\H,samples=100,variable=\u]
        ({sx(\u)},{sy(\u)})
      -- (NS5)
      -- plot[domain=\H:0,samples=100,variable=\u]
        ({-sx(\u)},{sy(\u)})
      arc[start angle={180-\tangent},end angle=180,
          x radius=3.3,y radius=\openingdepth] -- cycle;

    \foreach \angle in {20,40,...,160} {
      \draw[mesh!65,line width=.35pt,densely dashed]
        plot[domain=0:\H,samples=90,variable=\u]
          ({radius(\u)*cos(\angle)},
           {-\u+\perspective*radius(\u)*sin(\angle)});
    }

    \foreach \u in {.4,.9,1.5,2.2,3,3.9,4.9,5.8} {
      \pgfmathsetmacro{\a}{radius(\u)}
      \pgfmathsetmacro{\b}{\perspective*\a}
      \draw[mesh!75,line width=.4pt,densely dashed]
        (\a,-\u) arc[start angle=0,end angle=180,
                      x radius=\a,y radius=\b];
      \draw[mesh,line width=.45pt]
        (-\a,-\u) arc[start angle=180,end angle=360,
                       x radius=\a,y radius=\b];
    }

    \foreach \angle in {200,220,...,340} {
      \draw[mesh,line width=.45pt]
        plot[domain=0:\H,samples=90,variable=\u]
          ({radius(\u)*cos(\angle)},
           {-\u+\perspective*radius(\u)*sin(\angle)});
    }

    \foreach \sign in {-1,1} {
      \draw[edge,line width=.85pt]
        plot[domain=0:\H,samples=100,variable=\u]
          ({\sign*sx(\u)},{sy(\u)});
    }
    \draw[edge!75,line width=.65pt]
      (3.3,0) arc[start angle=0,end angle=180,
                  x radius=3.3,y radius=\openingdepth];
    \draw[edge,line width=.9pt]
      (-3.3,0) arc[start angle=180,end angle=360,
                   x radius=3.3,y radius=\openingdepth];

    \draw[black!75,line width=1.2pt,dash pattern=on 3pt off 2pt]
      (NS5) -- (0,-\openingdepth);
    \draw[black,line width=1.2pt,
          -{Latex[length=4.5mm,width=3mm]}]
      (0,-\openingdepth) -- (0,2.5);

    \pgfmathsetmacro{\a}{radius(3)}
    \pgfmathsetmacro{\b}{\perspective*\a}
    \draw[black!75,line width=1pt,densely dashed]
      (\a,-3) arc[start angle=0,end angle=180,
                  x radius=\a,y radius=\b];
    \draw[black,line width=1.2pt]
      (-\a,-3) arc[start angle=180,end angle=360,
                   x radius=\a,y radius=\b];

\end{tikzpicture}}};
        \node[left of=throat,xshift=-1.3cm,yshift=-0.4cm] {NS5-stack};
        \node[above of=throat,yshift=5pt,xshift=-1cm] (transverse-geometry) {AdS$_4\times S^2\times S^3$};
        \draw[-{Latex}] (transverse-geometry.south) -- ($(throat) - (16pt,-4pt)$);
        \node[right of=throat,xshift=1.15cm,yshift=-9pt] {$R$};
        \node[circle,fill=black,minimum width=0pt,inner sep=1.35pt,left of=throat,xshift=-1.2cm] {};
        \draw ($(throat)-(0.5cm,0)$) ellipse [x radius=3cm,y radius=2.25cm];
        \node[draw,circle] (circle) at (-2.945,1.2925) {};
        \draw (circle.north) -- ($(throat.north east)-(0.67cm,0.4cm)$);
        \draw[dash pattern=on 0.75cm off 1cm] (circle.south) -- ($(throat.south east)-(0.67cm,-0.4cm)$);
    \end{tikzpicture}
    \caption{A depiction of the $\Sigma$ part of the 10d geometry dual to the BCFT we consider, given by the quadrant $\Sigma=\{z\in\mathbb{C}\,|\, \text{arg}(z)\in\left[0,\pi/2\right]\,\}$ and over which the rest of the geometry is fibered. The positions of the string and the NS5 pole are shown for $\alpha=1$ and $v=\lambda\kappa^2 = 10$, with the coordinates normalized by the pole location on the $\mathrm{Im}(z)$ axis, $\mathrm{Im}(z_{\text{pole}})=\sqrt{v/(4\pi)}$. In the asymptotic region $|z|\rightarrow \infty$, the total geometry approaches AdS$_5\times S^5$ with an RR self-dual 5-form flux on the 5-sphere. On the boundaries of $\Sigma$, either one of the 2-spheres collapses to zero size, namely $S_1^2$ at $\mathcal{R}e(z)\rightarrow 0$ and $S_2^2$ at $\mathrm{Im}(z)\rightarrow 0$. In the vicinity of the NS5 pole, one finds a throat whose geometry at fixed radial coordinate $R = |z-z_{\text{pole}}|$ factors into an AdS$_4\times S^2$ part, which serves as the worldvolume of the NS5 branes, and an $S^3$ part on which an NSNS 3-form flux is supported.}
    \label{fig:geometry}
\end{figure}
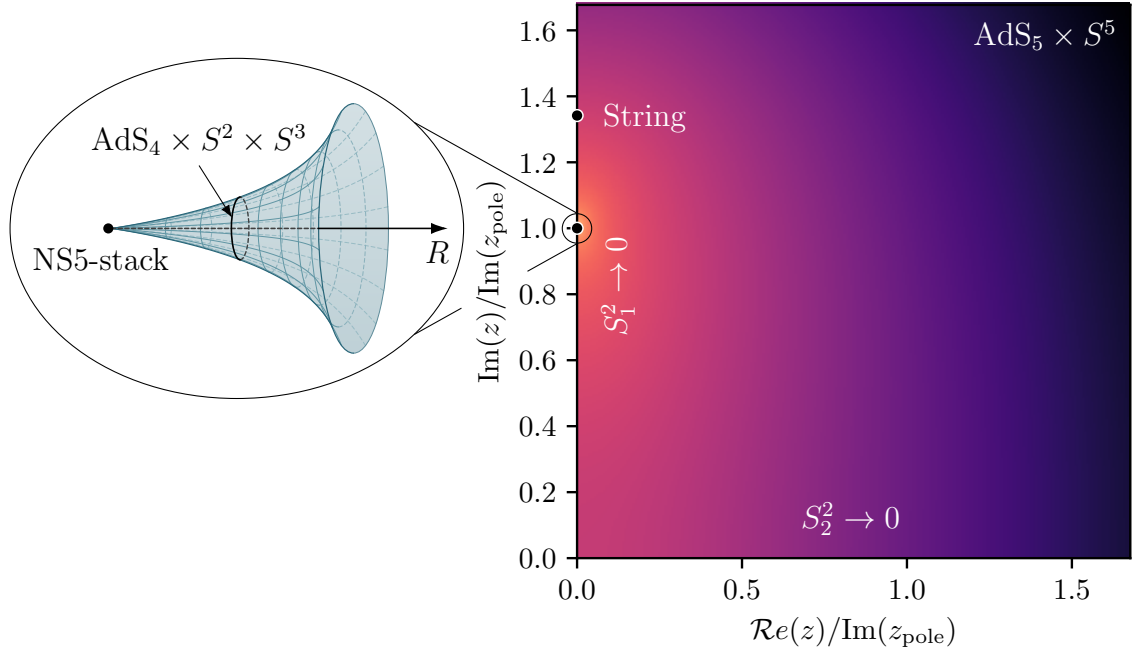

Using \eqref{h1_h2_gen}, the condition $h_1 = \partial_z h_2 = 0$ for the string to preserve half of the supersymmetry preserved by the boundary determines its location on $\Sigma$ to be 
\begin{equation}
\label{z_string}
    z_{\textrm{string}}=i\,\sqrt{\frac{2}{\pi}+\frac{\lambda\kappa^{2}}{4\pi}N^{-2(1-\alpha)}}\,.
\end{equation}
Since we are only interested in the leading large-$N$ limit, we see that $z_{\textrm{string}}$ has a different form for $\alpha=1$ versus $\alpha<1$, 
\begin{equation}\label{eq:string-position}
    z_{\text{string}} = 
    \left\{
    \begin{IEEEeqnarraybox}[\IEEEeqnarraystrutmode][c]{cl}
        i\,\sqrt{\frac{2}{\pi}}\quad &,\quad 0\leq\alpha < 1 \\[4pt]
        i\,\sqrt{\frac{8+\lambda\kappa^{2}}{4\pi}}\quad &,\quad \alpha = 1
    \end{IEEEeqnarraybox}
    \right.
\end{equation}
where the expressions are the leading large-$N$ ones (which we will use from now on). It is useful to compare this to the large-$N$ limit of the pole position in \eqref{z_pole}, given by 
\begin{equation}\label{eq:pole-position}
    z_{\text{pole}} = 
    \left\{
    \begin{IEEEeqnarraybox}[\IEEEeqnarraystrutmode][c]{cl}
        \hspace{1ex}0\text{\textcolor{white}{$\sqrt{\frac{2}{\pi}}$}} &,\quad 0\leq\alpha < 1 \\[4pt]
        i\,\sqrt{\frac{\lambda\kappa^{2}}{4\pi}}\quad &,\quad \alpha = 1
    \end{IEEEeqnarraybox}
    \right.\ .
\end{equation}
The locations in (\ref{eq:string-position}) and (\ref{eq:pole-position}) are roughly shown on figure \ref{fig:geometry}, alongside the representation of the 10d geometry. 

We therefore see that the string approaches the NS5 pole for $\alpha=1$ and $\lambda\kappa^{2}\gg1$, which might raise concerns for the validity of the supergravity approximation in studying the string for this parameter regime. In order to check this carefully, we should find whether the dilaton and the string-frame curvature are small at the location of the string, ensuring that string-loop corrections and $\alpha'$-corrections are suppressed. Expanding $h_2$ (see \eqref{h1_h2_gen}) around $z_{\text{string}}$ and computing the resulting metric and dilaton, we find that\footnote{Note that in this supergravity discussion we already assume that $N\rightarrow\infty$ and $\lambda\gg1$.} 
\begin{equation}
\label{R_dil_string_alpha}
0\leq\alpha<1:\quad\alpha'\mathcal{R}^{\left(\textrm{str}\right)}|_{z_{\text{string}}}=\frac{27}{\sqrt{8\lambda}}\ll1\,,\quad e^{2\Phi}|_{z_{\text{string}}}=\frac{\lambda^{2}}{4\pi^{2}N^{2}}\rightarrow0
\end{equation}
and 
\begin{equation}\label{R_dil_string_alpha1}
\begin{aligned}
\alpha=1:&\quad\alpha'\mathcal{R}^{\left(\textrm{str}\right)}|_{z_{\text{string}}}=\frac{6(v+6)^{2}}{\sqrt{\lambda}(v+8)^{3/2}}\,\,,\quad v=\lambda\kappa^{2}
\\
e^{2\Phi}|_{z_{\text{string}}}=\,&\frac{\lambda^{2}}{64\pi^{2}N^{2}}\left(v+8+8\sqrt{1+8/v}\,\textrm{arsinh}\sqrt{v/8}\right)\rightarrow0
\end{aligned}
\end{equation}
where $\mathcal{R}^{\left(\textrm{str}\right)}$ is the string-frame Ricci scalar (note also the distinction between $\alpha'=l_{s}^{2}$ and $\alpha$).\footnote{The (orthonormal) components of the Riemann tensor have the same parametric behavior, such that the use of the Ricci scalar here does not hide some parametrically larger curvature.} As a result, even though the dilaton is small at $z_{\text{string}}$ for any $\alpha$, the curvature is only small for $\alpha<1$ and for $\alpha=1$ with $\kappa\ll1$. Indeed, for $\alpha=1$ we get for both small $v$ (or $\kappa\ll\lambda^{-1/2}$) and $v=\mathcal{O}(1)$ (or $\kappa=\mathcal{O}(\lambda^{-1/2})$) that 
\begin{equation}
v\lesssim1:\quad\alpha'\mathcal{R}^{\left(\textrm{str}\right)}|_{z_{\text{string}}}=\mathcal{O}(\lambda^{-1/2})\ll1\,,
\end{equation}
while for large $v$ that 
\begin{equation}\label{eq:curvature-large-t}
v\gg1:\quad\alpha'\mathcal{R}^{\left(\textrm{str}\right)}|_{z_{\text{string}}}=6\kappa\left(1+\mathcal{O}(v^{-2})\right)
\end{equation}
which requires $\kappa\ll1$ for small curvature. In other words, as far as $\kappa\ll1$ we obtain for $\alpha=1$ a small curvature at the string location, regardless of the exact relation between $\kappa$ and $\lambda$.\footnote{Notice in particular that for $\lambda^{-1/2}\ll\kappa\ll1$ the curvature is small even though the string and the pole have the same location at the leading large-$\lambda\kappa^{2}$ limit.} Moreover, since $N_{\textrm{NS5}}=1/\kappa$ for $\alpha=1$, this condition is equivalent to $N_{\textrm{NS5}}\gg1$. Combining this with the fact that $N_{\textrm{NS5}}=N^{1-\alpha}/\kappa\gg1$ for $\alpha<1$, we then obtain the following simple result, 
\begin{equation}\label{sugra_criterion}
N_{\textrm{NS5}}\gg1\quad\Longleftrightarrow\quad\alpha'\mathcal{R}^{\left(\textrm{str}\right)}|_{z_{\text{string}}}\ll1\,.
\end{equation}
We will therefore mainly focus in what follows on the case $N_{\textrm{NS5}}\gg1$, corresponding to having a nontrivial 3d CFT on the boundary (in the field-theory description), while only commenting about the $N_{\textrm{NS5}}=\mathcal{O}(1)$ case. 

To illustrate the above discussion and demonstrate what the curvature and the dilaton look like across the $\Sigma$ space in a specific example, let us consider the regime $0\leq\alpha<1$. At the leading large-$N$ limit, the expression for $h_2$ in \eqref{h1_h2_gen} then simplifies as follows,\footnote{Note that \eqref{h2} is identical to the $r\rightarrow\infty$ form of $h_{2}$ in \eqref{h1h2_r_inf}.} 
\begin{equation}
\label{h2}
\alpha<1:\quad h_{2}\underset{N\rightarrow\infty}{\longrightarrow}\frac{1}{8}L_{AdS}^{2}\sqrt{\frac{\lambda}{N}}y\left(1+\frac{2}{\pi\left(x^{2}+y^{2}\right)}\right)=\frac{1}{8}L_{AdS}^{2}\sqrt{\frac{\lambda}{N}}\left(r+\frac{2}{\pi r}\right)\sin\left(\theta\right)
\end{equation}
and results in the following string-frame metric functions and dilaton (see \eqref{eq:supergravity-solution}), 
\begin{equation}
\label{fs}
    \textrm{String frame, \ensuremath{N\rightarrow\infty} :}\begin{array}{c}
f_{4}^{2}=\frac{L_{AdS}^{2}}{8\pi r}\sqrt{\frac{\lambda}{N}\left(2+\pi r^{2}\right)\left(8+2\pi r^{2}+\pi^{2}r^{4}-2\left(2-\pi r^{2}\right)\cos\left(2\theta\right)\right)}\,,\\
\\
f_{1}^{2}=L_{AdS}^{2}\sqrt{\frac{\lambda}{N}}\cos^{2}\left(\theta\right)\frac{\sqrt{\left(2+\pi r^{2}\right)\left(8+2\pi r^{2}+\pi^{2}r^{4}-2\left(2-\pi r^{2}\right)\cos\left(2\theta\right)\right)}}{8\pi r+2\pi^{2}r^{3}+4\pi r\cos\left(2\theta\right)}\,,\\
\\
f_{2}^{2}=L_{AdS}^{2}\sqrt{\frac{\lambda}{N}}\sin^{2}\left(\theta\right)\frac{\left(2+\pi r^{2}\right)^{3/2}}{2\pi r\sqrt{8+2\pi r^{2}+\pi^{2}r^{4}-2\left(2-\pi r^{2}\right)\cos\left(2\theta\right)}}\,,\\
\\
f_{\Sigma}^{2}=\frac{L_{AdS}^{2}}{2\pi r^{3}}\sqrt{\frac{\lambda\left(8+2\pi r^{2}+\pi^{2}r^{4}-2\left(2-\pi r^{2}\right)\cos\left(2\theta\right)\right)}{N\left(2+\pi r^{2}\right)}}\,,\\
\\
e^{2\Phi}=g_{s}^{2}\,\frac{\left(2+\pi r^{2}\right)\left(8+2\pi r^{2}+\pi^{2}r^{4}-2\left(2-\pi r^{2}\right)\cos\left(2\theta\right)\right)}{\pi^{2}r^{4}\left(4+\pi r^{2}+2\cos\left(2\theta\right)\right)}\,.
\end{array}
\end{equation}
While the dilaton diverges at the origin (where the pole lies), the curvature remains finite and very small (of order $\lambda^{-1/2}$) everywhere, taking the form $\alpha'\mathcal{R}^{\left(\textrm{str}\right)}=f\left(r,\theta\right)/\sqrt{\lambda}$ for some bounded function $f$. In particular, as already shown in Eq. \eqref{R_dil_string_alpha}, the curvature and the dilaton are small at the location of the string. 

We will next proceed in the following section to the analysis of the Wilson-line spectrum at the planar limit at weak and strong 't Hooft coupling, making use of the supergravity description for the strong-coupling regime when $N_{\textrm{NS5}}\gg1$.


\section{Perturbative descriptions at weak and strong 't Hooft coupling}
\label{sec:dictionaries}

The setup we consider, in which the familiar Maldacena-Wilson line \eqref{W} of the bulk $\mathcal{N}=4$ SYM theory is put on a 1/2-BPS boundary of the NS5-type, admits a major simplification at the large-$N$ limit. First, in addition to the free-theory description we have at small 't Hooft coupling for the bulk theory in terms of the SYM Lagrangian (while the boundary CFT is still strongly coupled on its own but weakly coupled to the bulk via gauging), there is another such description of the whole setup at strong coupling in terms of the supergravity solution we discussed in the previous section. Second, at weak coupling we need to only consider Wilson-line insertions transforming in the adjoint representation of the $SU(N)$ gauge group without involving traces, as these form the leading (in $N$) closed subsector of the theory. Indeed, correlators involving operators with traces are suppressed by powers of $N$ compared to those not involving traces at all (see \cite{Ferrero:2024wld1} for a recent discussion). We will therefore focus in what follows only on the leading planar limit, studying the spectrum of operators on the boundary Wilson line at the two regimes where a free-theory description is available. 

Before turning to this analysis, however, we should begin with identifying the symmetries preserved by the Wilson line, which would be useful in organizing the spectrum and in mapping operators between the two perturbative regimes. Since the boundary preserves half of the supersymmetry of the bulk theory (as well as the conformal group along the boundary directions), which amounts to a 3d $\mathcal{N}=4$ superconformal symmetry, the maximally supersymmetric Wilson line we insert on the boundary can be viewed as a 1/2-BPS line of a 3d $\mathcal{N}=4$ SCFT. The superalgebra preserved by such a line in general, as well as the possible multiplets and their properties, were analyzed in detail in \cite{Nathan:2020scd}, which we will next briefly review before specifying to our setup. 

\subsection{Algebra preserved by a 1/2-BPS line in a 3d $\ensuremath{\mathcal{N}=4}$ theory}\label{sec:1d-subalgebra}

We consider a generic superconformal straight line preserving half of the supercharges in a 3d $\mathcal{N}=4$ SCFT. The full algebra such a line preserves is given by \cite{Nathan:2020scd} 
\begin{equation}
    \mathfrak{psu}(1,1|2)\rtimes\mathfrak{u}(1)_{\mathcal{R}_2}\rtimes\mathfrak{u}(1)_b
\end{equation}
which has the bosonic subalgebra
\begin{equation}\label{eq:1d-SCFT-bosonic-subalgebra}
    \mathfrak{so}(2,1)_{\text{conf}}\oplus\mathfrak{su}(2)_{\mathcal{R}}\oplus\mathfrak{u}(1)_{\mathcal{R}_2}\oplus\mathfrak{u}(1)_b\,,
\end{equation}
where $\mathfrak{psu}(1,1|2)$ is the 1d $\mathcal{N}=4$ superconformal part with $\mathfrak{so}(2,1)_{\text{conf}}$ conformal group and $\mathfrak{su}(2)_{\mathcal{R}}$ $R$-symmetry. Meanwhile, $\mathfrak{u}(1)_{\mathcal{R}_2}$ and $\mathfrak{u}(1)_b$ are combinations of the $\mathfrak{so}(2)\simeq\mathfrak{u}(1)_{\text{rot}}$ rotation symmetry transverse to the line\footnote{Note that in our setup these transverse directions are inside the boundary and not in the bulk (such that there are two of them and not three).} and the $\mathfrak{u}(1)$ factor inside the $\mathfrak{so}(4)$ $R$-symmetry of the 3d theory that is preserved by the line and commutes with $\mathfrak{su}(2)_{\mathcal{R}}$, such that $\mathfrak{u}(1)_{\mathcal{R}_2}$ is central and $\mathfrak{u}(1)_b$ is a nontrivial outer-automorphism (see \cite{Nathan:2020scd} for more details). 

The multiplets formed by the operators of this 1d defect SCFT are composed of several conformal primaries and their towers of descendants, with the conformal primaries being related by the action of the  supercharges. Such conformal primaries are denoted by $\left[r\right]_\Delta^{(R)}$ where $r$ is the charge under $\mathfrak{u}(1)_{\mathcal{R}_2}$, $R$ is the $\mathfrak{su}(2)_{\mathcal{R}}$ Dynkin label and $\Delta$ the conformal scaling dimension. Meanwhile, the charge under $\mathfrak{u}(1)_b$ is suppressed in the notations of \cite{Nathan:2020scd} since it does not affect the unitarity bounds or the multiplet structure. As an example for this standard notation (though not for describing a primary), note that the 1d $\mathcal{N}=4$ superconformal algebra contains two sets of supercharges which transform as follows, 
\begin{equation}\label{eq:1d-SCFT-supercharges}
    \mathcal{Q}_a\in\left[0\right]_{\frac{1}{2}}^{(1)}\, ,\quad \bar{\mathcal{Q}}^a\in\left[0\right]_{\frac{1}{2}}^{(1)}
\end{equation}
where $a$ is an $\mathfrak{su}(2)_{\mathcal{R}}$ index. A multiplet is then described by specifying its conformal primaries, related by the action of the supercharges in (\ref{eq:1d-SCFT-supercharges}), with the rest of the multiplet given by their conformal descendants. The various possible multiplets were classified according to the shortening conditions that they satisfy in table 4 of \cite{Nathan:2020scd}, which we reproduce here in table \ref{tab:1d-SCFT-multiplets} for convenience. In this multiplet notation, each multiplet is denoted by the quantum numbers $\left[r\right]_{\Delta}^{\left(R\right)}$ of its superconformal primary (i.e. its highest weight) and by two letters, an unbarred one and a barred one, which denote whether or not the superconformal primary is annihilated by supercharges of the types $\mathcal{Q}$ and $\bar{\mathcal{Q}}$, respectively. In particular, the letter $A_1$ ($\bar{A}_1$) refers to an annihilation by some $\mathcal{Q}_{a}$ ($\bar{\mathcal{Q}}^{a}$), while the letter $L$ ($\bar{L}$) refers to a nontrivial action of all the $\mathcal{Q}_{a}$ ($\bar{\mathcal{Q}}^{a}$). 

\begin{table}
    \centering
    \begin{tblr}{colspec={|c|l|l|l|l|l|}, cell{3}{1}={r=2}{c}, cell{5}{1}={r=2}{c},hlines,vlines, cell{1}{1-5}={font=\bfseries}}
        Name & Primary & Unitary Bound & BPS & Null State & $\mathcal{Q}|\text{h.w.}\rangle = 0$ \\
        $L\bar{L}$ & $\left[r\right]_\Delta^{(R)}$ & $\Delta>|r|+\frac{1}{2}R$ & -- & -- & -- \\
        $A_1\bar{L}$ & $\left[r\right]_\Delta^{(R)}\ (r<0)$ & $\Delta=-r+\frac{1}{2}R$ & $\left(\frac{1}{4},0\right)$ & $\left[r\right]_{\Delta+\frac{1}{2}}^{(R+1)}$ & $\mathcal{Q}_1$ \\
        & $\left[r\right]_\Delta^{(0)}\ (r<0)$ & $\Delta = -r$ & $\left(\frac{1}{2},0\right)$ & $\left[r\right]_{\Delta+\frac{1}{2}}^{(1)}$ & $\mathcal{Q}_a$ \\
        $L\bar{A}_1$ & $\left[r\right]_\Delta^{(R)}\ (r>0)$ & $\Delta = r+\frac{1}{2}R$ & $\left(0,\frac{1}{4}\right)$ & $\left[r\right]_{\Delta+\frac{1}{2}}^{(R+1)}$ & $\bar{\mathcal{Q}}^2$ \\
        & $\left[r\right]_\Delta^{(0)}\ (r>0)$ & $\Delta = r$ & $\left(0,\frac{1}{2}\right)$ & $\left[r\right]_{\Delta+\frac{1}{2}}^{(1)}$ & $\bar{\mathcal{Q}}^a$ \\
        $A_1\bar{A}_1$ & $\left[0\right]_\Delta^{(R)}$ & $\Delta = \frac{1}{2}R$ & $\left(\frac{1}{4},\frac{1}{4}\right)$ & \raisebox{2pt}{$\left[0\right]_{\Delta+\frac{1}{2}}^{(R+1)}\oplus\left[0\right]_{\Delta+\frac{1}{2}}^{(R+1)}$} & $\mathcal{Q}_1,\ \bar{\mathcal{Q}}^2$ \\
    \end{tblr}
    \caption{The possible unitary multiplets of $\mathfrak{psu}(1,1|2)\rtimes\mathfrak{u}(1)_{\mathcal{R}_2}\rtimes\mathfrak{u}(1)_b$, where for each multiplet the unitarity bound, shortening condition and corresponding null state are specified. (Reproduction of table 4 of \cite{Nathan:2020scd}.)}
    \label{tab:1d-SCFT-multiplets}
\end{table}

As an example for a multiplet, let us consider the displacement multiplet $L\bar{A}_{1}\left[1\right]_{1}^{\left(0\right)}$ with conformal primaries 
\begin{equation}
    \label{D_multiplet_gen}L\bar{A}_{1}\left[1\right]_{1}^{\left(0\right)}:\quad\left[1\right]_{1}^{\left(0\right)}\overset{\mathcal{Q}}{\longrightarrow}\left[1\right]_{\frac{3}{2}}^{\left(1\right)}\overset{\mathcal{Q}}{\longrightarrow}\left[1\right]_{2}^{\left(0\right)}\,,
\end{equation}
where the scalar $\left[1\right]_{1}^{\left(0\right)}$ is associated with the $R$-symmetry broken by the line, the fermions $\left[1\right]_{\frac{3}{2}}^{\left(1\right)}$ with the broken supersymmetry and the scalar $\left[1\right]_{2}^{\left(0\right)}$ is the displacement operator. Note also that there is a conjugate multiplet $A_{1}\bar{L}\left[-1\right]_{1}^{\left(0\right)}$ which contains the conjugate operators. 

When a long multiplet hits the unitarity bound $\Delta_\star = -r+\frac{1}{2}R$ or $\bar{\Delta}_\star=r+\frac{1}{2}R$ (depending on whether $r$ is negative or positive, respectively) it decomposes into shorter multiplets according to the rules \cite{Nathan:2020scd} 
\begin{align}\label{eq:recombination_rules} &L\bar{L}\left[r\right]_{\Delta\rightarrow\bar{\Delta}_\star}^{(R)}=L\bar{A}_1\left[r\right]_{\bar{\Delta}_\star}^{(R)}\oplus L\bar{A}_1\left[r\right]_{\bar{\Delta}_\star+\frac{1}{2}}^{(R+1)},\nonumber\\ &L\bar{L}\left[r\right]_{\Delta\rightarrow\Delta_\star}^{(R)}=A_1\bar{L}\left[r\right]_{\Delta_\star}^{(R)}\oplus A_1\bar{L}\left[r\right]_{\Delta_\star + \frac{1}{2}}^{(R+1)},\\ &L\bar{L}\left[0\right]_{\Delta\rightarrow\Delta_\star}^{(R)}=A_1\bar{A}_1\left[0\right]_{\Delta_\star}^{(R)}\oplus2A_1\bar{A}_1\left[0\right]_{\Delta_\star+\frac{1}{2}}^{(R+1)}\oplus A_1\bar{A}_1\left[0\right]_{\Delta_\star + 1}^{(R+2)}.\nonumber 
\end{align}
This can happen for example in our setup when the 't Hooft coupling parameter is varied and the dimensions of long multiplets change (the dimensions of short and semi-short multiplets, on the other hand, are determined by $r$ and $R$ and are therefore protected). In the opposite direction, a collection of shorter multiplets can recombine at some value of the 't Hooft coupling as in \eqref{eq:recombination_rules} to form a long multiplet, whose dimension is no longer protected.

\subsection{Weak-coupling regime}\label{sec:weak-coupling-operators}
In order to adapt the general discussion of the previous subsection to our setup, let us begin by identifying the different $R$-symmetries we mentioned. In the weak coupling description of our setup, the bulk $SU(4)_{\mathcal{R}}$ $R$-symmetry is broken by the boundary to $SO(4)_\mathcal{R}\simeq SU(2)_X\times SU(2)_Y/\mathbb{Z}_{2}$, with the two $\mathfrak{su}(2)=\mathfrak{so}(3)$ factors acting on the scalar triplets $\vec{X}=(\Phi_1,\Phi_2,\Phi_3)$ and $\vec{Y}=(\Phi_4,\Phi_5,\Phi_6)$, respectively. Then, since the line on the boundary incorporates the scalar $\Phi_6$ in it (see \eqref{W}), the $\mathfrak{so}(3)_{Y}$ part is further broken to $\mathfrak{so}(2)_{Y}\simeq\mathfrak{u}(1)_{Y}$ while the $\mathfrak{so}(3)_{X}$ part is preserved. As a result, matching to the conventions of \cite{Nathan:2020scd} used in the previous subsection we identify\footnote{$\,$We here used Eq. (B.12) of \cite{Nathan:2020scd} which reads $\mathcal{R}_{2}=M_{+}^{\;+}-\mathtt{r}/2$, where $M_{+}^{\;+}$ is the generator of $\mathfrak{u}(1)_{\text{rot}}$ and we chose to identify $\mathtt{r}/2$ with minus the generator of $\mathfrak{u}(1)_Y$. The orthogonal combination then corresponds to $\mathfrak{u}(1)_{b}$. Note that we used here the symbol $\mathtt{r}$ rather than $r$ (as used in \cite{Nathan:2020scd}) in order to distinguish it from the charge under $\mathfrak{u}(1)_{\mathcal{R}_2}$ used in the multiplet notation.} 
\begin{equation}\label{u1b_u1R2_su2R}
\mathfrak{u}(1)_{\mathcal{R}_{2}}=\mathfrak{u}(1)_{\text{rot}}+\mathfrak{u}(1)_{Y}\,,\quad\mathfrak{u}(1)_{b}=\mathfrak{u}(1)_{\text{rot}}-\mathfrak{u}(1)_{Y}\,,\quad\mathfrak{su}(2)_{\mathcal{R}}=\mathfrak{su}(2)_{X}
\end{equation}
where recall that $\mathfrak{u}(1)_{\text{rot}}$ is the rotation symmetry transverse to the line. 

Turning next to discuss explicit multiplets, it would be useful to keep track of both $\mathfrak{u}(1)_{\mathcal{R}_{2}}$ and $\mathfrak{u}(1)_{b}$ (or alternatively $\mathfrak{u}(1)_{\text{rot}}$ and $\mathfrak{u}(1)_{Y}$) when labeling states, instead of only $\mathfrak{u}(1)_{\mathcal{R}_{2}}$ as in \cite{Nathan:2020scd}. We therefore further refine the notation $\left[r\right]_{\Delta}^{\left(R\right)}=\left[r_{\text{rot}}+r_Y\right]_{\Delta}^{\left(R\right)}$ used in the previous subsection to  
\begin{equation}
\label{new_notation}
    \left[r_Y,r_{\text{rot}}\right]_\Delta^{(R)},
\end{equation}
and use it from now on. Note that in this notation the supercharges transform as 
\begin{equation}\label{QQd}
\mathcal{Q}_{a}\in\left[-1/2,+1/2\right]_{1/2}^{(1)}\,,\quad\bar{\mathcal{Q}}^{a}\in\left[+1/2,-1/2\right]_{1/2}^{(1)}.
\end{equation}
We will next give an explicit example for each of the three types of multiplets in table \ref{tab:1d-SCFT-multiplets}, at the zero coupling point (i.e. $\lambda=0$) where the setup is basically given by the BPS pure Neumann boundary condition for the bulk theory plus a Maldacena-Wilson line and a decoupled $T_{[K^{N/K}]}[SU(N)]$ theory on the boundary. 

\begin{figure}
    \centering
    \begin{subfigure}{0.33\textwidth}
        \centering
        \begin{tikzpicture}
    \node[draw,fill=lightgray,align=center,minimum height=1.75cm,minimum width=2.5cm] (SP-LA) {\underline{$\left[+1,0\right]_1^{(0)}$}\\[6pt] $\Phi_+$};
    \node[draw,fill=lightgray,align=center,below of=SP-LA,yshift=-60pt,minimum height=1.75cm,minimum width=2.5cm] (LA-1) {\underline{$\left[+\frac{1}{2},+\frac{1}{2}\right]_\frac{3}{2}^{(1)}$}\\[6pt] $\lambda^{++}_a$};
    \node[draw,fill=lightgray,align=center,below of=LA-1,yshift=-60pt,minimum height=1.75cm,minimum width=2.5cm] (LA-2) {\underline{$\left[0,+1\right]_2^{(0)}$}\\[6pt] $\Tilde{\mathbb{F}}_+$};
    \draw[-{latex}] (SP-LA.south) -- (LA-1.north) node[midway,fill=white]{\small $\mathcal{Q}$};
    \draw[-{latex}] (LA-1.south) -- (LA-2.north) node[midway,fill=white]{\small $\mathcal{Q}$};
\end{tikzpicture}
        \caption{$L\bar{A}_1\left[+1,0\right]_1^{(0)}$}
    \end{subfigure}
    \begin{subfigure}{0.33\textwidth}
        \centering
        \begin{tikzpicture}
    \node[draw,fill=lightgray,align=center,minimum height=1.75cm,minimum width=2.5cm] (SP-AL) {\underline{$\left[-1,0\right]_1^{(0)}$}\\[6pt] $\Phi_-$};
    \node[draw,fill=lightgray,align=center,below of=SP-AL,yshift=-60pt,minimum height=1.75cm,minimum width=2.5cm] (AL-1) {\underline{$\left[-\frac{1}{2},-\frac{1}{2}\right]_\frac{3}{2}^{(1)}$}\\[6pt] $\lambda^{a--}$};
    \node[draw,fill=lightgray,align=center,below of=AL-1,yshift=-60pt,minimum height=1.75cm,minimum width=2.5cm] (AL-2) {\underline{$\left[0,-1\right]_2^{(0)}$}\\[6pt] $\Tilde{\mathbb{F}}_-$};
    \draw[-{latex}] (SP-AL.south) -- (AL-1.north) node[midway,fill=white]{\small $\bar{\mathcal{Q}}$};
    \draw[-{latex}] (AL-1.south) -- (AL-2.north) node[midway,fill=white]{\small $\bar{\mathcal{Q}}$};
\end{tikzpicture}
        \caption{$A_1\bar{L}\left[-1,0\right]_1^{(0)}$}
    \end{subfigure}
    \caption{The displacement multiplets $L\bar{A}_1\left[+1,0\right]_1^{(0)}$ and $A_1\bar{L}\left[-1,0\right]_1^{(0)}$, with the charges and representations of the different primaries denoted as in \eqref{new_notation}.}
    \label{fig:displacement-multiplets-weak}
\end{figure}
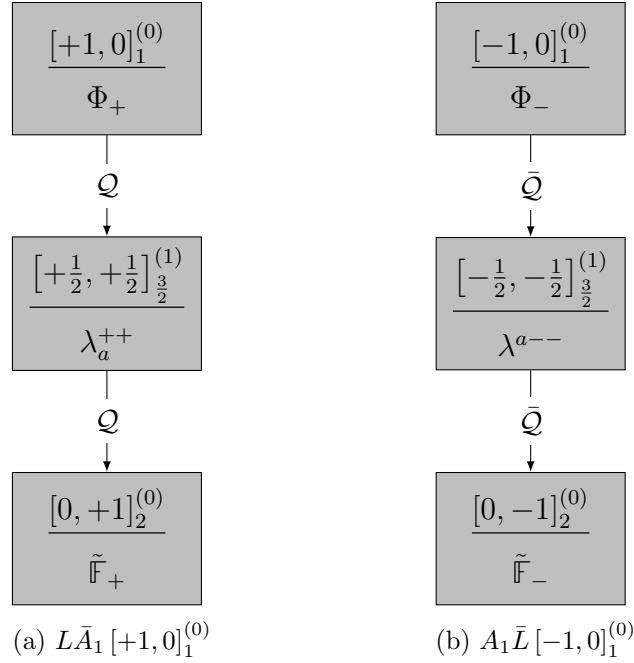

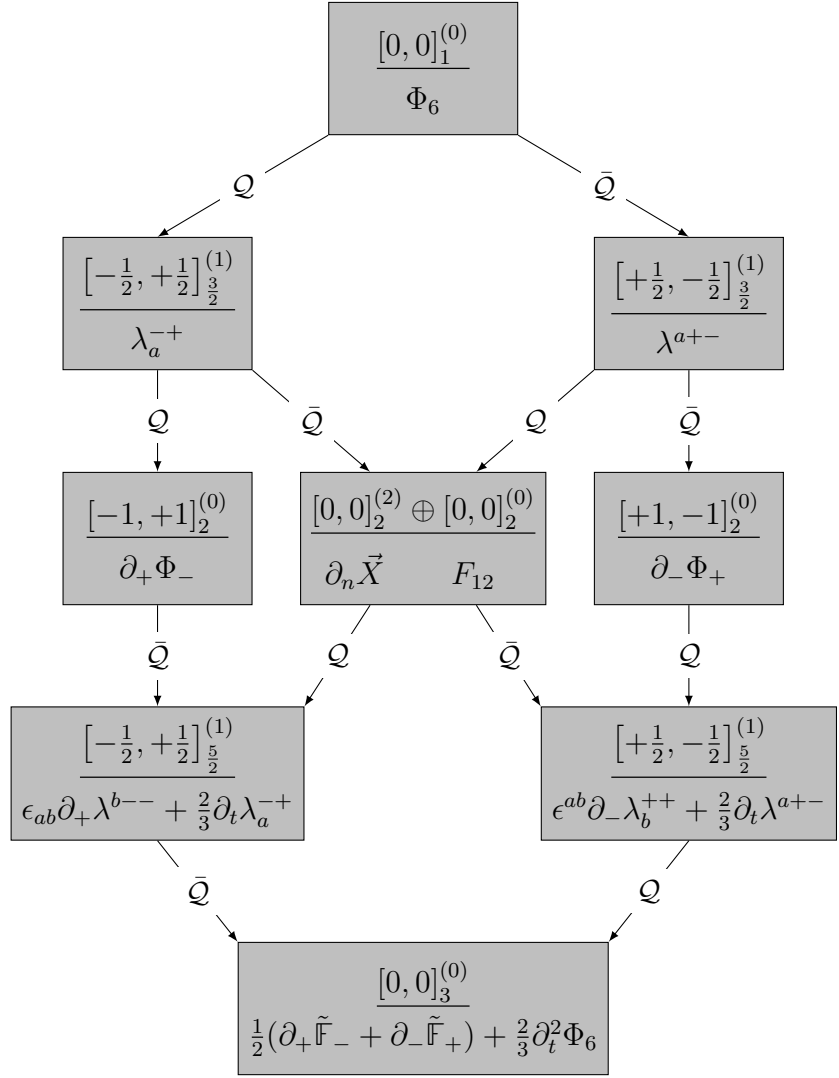
\begin{figure}
    \centering
    \begin{tikzpicture}
    \node[draw,fill=lightgray,align=center,minimum height=1.75cm,minimum width=2.5cm] (SP) {\underline{$\left[0,0\right]_1^{(0)}$}\\[6pt] $\Phi_6$};
    \node[draw,fill=lightgray,align=center,below of=SP,xshift=-100pt,yshift=-60pt,minimum height=1.75cm,minimum width=2.5cm] (l1-1) {\underline{$\left[-\frac{1}{2},+\frac{1}{2}\right]_\frac{3}{2}^{(1)}$}\\[4pt] $\lambda^{-+}_a$};
    \node[draw,fill=lightgray,align=center,below of=SP,xshift=100pt,yshift=-60pt,minimum height=1.75cm,minimum width=2.5cm] (l1-2) {\underline{$\left[+\frac{1}{2},-\frac{1}{2}\right]_\frac{3}{2}^{(1)}$}\\[4pt] $\lambda^{a+-}$};
    \node[draw,fill=lightgray,align=center,minimum height=1.75cm,minimum width=2.5cm,below of=l1-1, xshift=100pt,yshift=-60pt] (l2-2) {\underline{$\left[0,0\right]_2^{(2)}\oplus\left[0,0\right]_2^{(0)}$}\\[6pt] $\hspace{-10pt}\partial_n\Vec{X}\hspace{25pt} F_{12}$};
    \node[draw,fill=lightgray,align=center,minimum height=1.75cm,minimum width=2.5cm,below of=l1-1, yshift=-60pt] (l2-1) {\underline{$\left[-1,+1\right]_2^{(0)}$}\\[6pt] $\partial_+\Phi_-$};
    \node[draw,fill=lightgray,align=center,minimum height=1.75cm,minimum width=2.5cm,below of=l1-2,yshift=-60pt] (l2-3) {\underline{$\left[+1,-1\right]_2^{(0)}$}\\[6pt] $\partial_-\Phi_+$};
    \node[draw,fill=lightgray,align=center,minimum height=1.75cm,minimum width=2.5cm,below of=l2-1,yshift=-60pt] (l3-1) {\underline{$\left[-\frac{1}{2},+\frac{1}{2}\right]_{\frac{5}{2}}^{(1)}$}\\[4pt] $\epsilon_{ab}\partial_{+}\lambda^{b--}+\frac{2}{3}\partial_{t}\lambda_{a}^{-+}$}; 
    \node[draw,fill=lightgray,align=center,minimum height=1.75cm,minimum width=2.5cm,below of=l2-3,yshift=-60pt] (l3-2) {\underline{$\left[+\frac{1}{2},-\frac{1}{2}\right]_{\frac{5}{2}}^{(1)}$}\\[4pt] $\epsilon^{ab}\partial_{-}\lambda_{b}^{++}+\frac{2}{3}\partial_{t}\lambda^{a+-}$};
    \node[draw,fill=lightgray,align=center,minimum height=1.75cm,minimum width=2.5cm,below of=l3-1,xshift=100pt,yshift=-60pt] (l4) {\underline{$\left[0,0\right]_3^{(0)}$}\\[2pt] $\frac{1}{2}(\partial_{+}\mathbb{\tilde{F}}_{-}+\partial_{-}\mathbb{\tilde{F}}_{+})+\frac{2}{3}\partial_{t}^{2}\Phi_{6}$}; 
    \draw[-{latex}] (SP.south west) -- (l1-1.north) node[midway,fill=white]{\small $\mathcal{Q}$};
    \draw[-{latex}] (SP.south east) -- (l1-2.north) node[midway,fill=white]{\small $\bar{\mathcal{Q}}$};
    \draw[-{latex}] (l1-1.south) -- (l2-1.north) node[midway,fill=white]{\small $\mathcal{Q}$};
    \draw[-{latex}] (l1-2.south) -- (l2-3.north) node[midway,fill=white]{\small $\bar{\mathcal{Q}}$};
    \draw[-{latex}] (l1-1.south east) -- ($(l2-2.north)-(20pt,0pt)$) node[midway,fill=white]{\small $\bar{\mathcal{Q}}$};
    \draw[-{latex}] (l1-2.south west) -- ($(l2-2.north)+(20pt,0pt)$) node[midway,fill=white]{\small $\mathcal{Q}$};
    \draw[-{latex}] (l2-1.south) -- (l3-1.north) node[midway,fill=white]{\small $\bar{\mathcal{Q}}$};
    \draw[-{latex}] (l2-3.south) -- (l3-2.north) node[midway,fill=white]{\small $\mathcal{Q}$};
    \draw[-{latex}] ($(l2-2.south)-(20pt,0pt)$) -- (l3-1.north east) node[midway,fill=white]{\small $\mathcal{Q}$};
    \draw[-{latex}] ($(l2-2.south)+(20pt,0pt)$) -- (l3-2.north west) node[midway,fill=white]{\small $\bar{\mathcal{Q}}$};
    \draw[-{latex}] (l3-1.south) -- (l4.north west) node[midway,fill=white]{\small $\bar{\mathcal{Q}}$};
    \draw[-{latex}] (l3-2.south) -- (l4.north east) node[midway,fill=white]{\small $\mathcal{Q}$};
\end{tikzpicture}
    \caption{The long multiplet $L\bar{L}\left[0,0\right]_1^{(0)}$ containing the operator $\Phi_6$ as its bottom component. The fermionic descendants at level one are denoted by $\lambda_{a}^{-+}$ and $\lambda^{a+-}$ according to their charges under $U(1)_Y\times U(1)_{\text{rot}}$.}
    \label{fig:LL-multiplet-weak}
\end{figure}

\begin{itemize}
    \item \textit{Displacement Multiplets}: These correspond to the two multiplets $L\bar{A}_{1}\left[1\right]_{1}^{\left(0\right)}$ and $A_{1}\bar{L}\left[-1\right]_{1}^{\left(0\right)}$ discussed in \eqref{D_multiplet_gen}, which in the notation \eqref{new_notation} are denoted by $L\bar{A}_1\left[+1,0\right]_1^{(0)}$ and $A_1\bar{L}\left[-1,0\right]_1^{(0)}$. As in the case of $\mathcal{N}=4$ SYM theory without a boundary (see e.g. \cite{Cooke:2017wavy,Ferrero:2024wld1,Julius:2026rosetta}), it is easy to show that the displacement operators on the line are given by
\begin{equation}\label{displ_operator}
\Tilde{\mathbb{F}}_{\pm}=iF_{t\pm}+\mathcal{\partial}_{\pm}\Phi_{6}\in\left[0,\pm1\right]_{2}^{(0)},
\end{equation}
where $t$ is the direction along the line (see \eqref{W}) and $\pm$ the two directions transverse to it on the boundary, $x_{\pm}=x_{1}\pm ix_{2}$, with $\mathfrak{u}(1)_{\text{rot}}$ charges $\pm 1$. Similarly, the scalars associated with the $R$-symmetry breaking by the line, which sit at the bottom of the displacement multiplets, are given by 
\begin{equation}
\Phi_{\pm}=\Phi_{4}\pm i\Phi_{5}\in\left[\pm1,0\right]_{1}^{\left(0\right)},\end{equation}
while the fermions which are their super-descendants (and are associated with the supersymmetry breaking) have charges $\left[\pm1/2,\pm1/2\right]_{3/2}^{(1)}$ and are denoted by $\lambda_{a}^{++}$ and $\lambda^{a--}$. See figure \ref{fig:displacement-multiplets-weak} for a depiction of these multiplets. 

\item \textit{The long multiplet containing $\Phi_6$}: The operator $\Phi_6$, which corresponds to $\left[0,0\right]_1^{(0)}$, lies at the bottom of the long multiplet $L\bar{L}\left[0,0\right]_1^{(0)}$. The operator $F_{ij}$, which is associated with the worldsheet axion (see section \ref{sec:map} for a more detailed discussion), corresponds to $\left[0,0\right]_2^{(0)}$ and is a super-descendant of it at level two, as can be seen in figure \ref{fig:LL-multiplet-weak} which displays the entire (primary part of the) multiplet.\footnote{This multiplet is obtained by branching the corresponding $\Phi_6$ multiplet in the no-boundary case \cite{Cooke:2017wavy} under the symmetry preserved by the boundary, imposing the boundary conditions, using the free bulk equations of motion to eliminate dependent components, and requiring annihilation by the one-dimensional special conformal generator to fix the primary representatives.} Note that in the case of the Maldacena-Wilson line in $\mathcal{N}=4$ SYM without a boundary (as in e.g. \cite{Cooke:2017wavy,Ferrero:2024wld1,Julius:2026rosetta}), the operator $\Phi_6$ is the bottom component of a (semi) short multiplet instead of a long one. However, as one increases the value of the 't Hooft coupling this short multiplet combines with another such multiplet to form a long one, whose dimension is not protected. 

\item \textit{The multiplet containing the moment-map operator of the boundary theory}: On the boundary, we have in addition to the restriction of bulk fields also the 3d SCFT $T_{[K^{N/K}]}[SU(N)]$, whose operators can also be inserted on the Wilson line. As we discussed, our focus is on the leading planar limit, at which we consider only operators transforming in the adjoint representation of the bulk $SU(N)$ gauge group, without involving traces. From the perspective of the decoupled $T_{[K^{N/K}]}[SU(N)]$ theory at the zero coupling point (i.e. $\lambda=0$), this $SU(N)$ is its global Higgs-branch symmetry, and the simplest (lowest dimension) operator transforming in its adjoint is the corresponding moment-map operator $\vec{\mu}$ which is the bottom component
of the 3d $\mathcal{N}=4$ conserved-current multiplet of this symmetry. It has dimension $\Delta=1$ and it transforms in the adjoint of the $\mathfrak{su}(2)_{H}=\mathfrak{su}(2)_{X}$ part of the $R$-symmetry of the boundary theory \cite{cordova:2016multiplets}.\footnote{We identified $\mathfrak{su}(2)_{X}$ with $\mathfrak{su}(2)_{H}$ since when the bulk 4d $\mathcal{N}=4$ vector multiplet is written in terms of 3d $\mathcal{N}=4$ multiplets, the $\mathfrak{su}(2)_{X}$ part acts on the scalars of the hypermultiplet. In contrast, $\mathfrak{su}(2)_{Y}$ acts on the scalars of the 3d $\mathcal{N}=4$ vector multiplet and so is identified with $\mathfrak{su}(2)_{C}$. Note also that the coupling of the 3d vector multiplet to the 3d boundary SCFT is such that the gauge field couples to its Higgs-branch symmetry current while $\vec{\mu}$ couples to the $D$-term of the vector multiplet, which in turn transforms in the adjoint of $\mathfrak{su}(2)_{X}$ (and therefore so is $\vec{\mu}$).} As a result, since $\mathfrak{su}(2)_{X}$ is preserved by the line on the boundary and is identified with $\mathfrak{su}(2)_{\mathcal{R}}$ of its 1d $\mathcal{N}=4$ algebra, this moment map operator corresponds to $\left[0,0\right]_1^{(2)}$ and so is the superconformal primary of the short multiplet $A_1\bar{A}_1\left[0,0\right]_1^{(2)}$. 
\end{itemize}

In order to obtain the full multiplet content on the line at the weak-coupling regime, one should construct the corresponding spectrum systematically and decompose it in terms of multiplets. Since the $T_{[K^{N/K}]}[SU(N)]$ theory on the boundary is a strongly coupled SCFT whose full unprotected spectrum is not known, we will next focus on the part of the spectrum resulting from the bulk fields and describe its contribution to the partition function (in analogy to the case with no boundary, see \cite{Ferrero:2024wld1,Julius:2026rosetta}). Note that for pure BPS Neumann boundary condition, this computation yields the full planar partition function of the line at $\lambda=0$. 

\subsubsection*{Bulk contribution to the planar partition function of the line}\label{sec:character-neumann-BCs}

Our starting point is the single-letter character of the 4d bulk fields restricted to the line in the case with no boundary, which was discussed at length in \cite{Ferrero:2024wld1,Julius:2026rosetta} and from which the no-boundary line partition function at $\lambda=0$ can easily be computed. It is given explicitly in Eqs. (4.8) and (4.9) of \cite{Ferrero:2024wld1},\footnote{Note that the fugacity $q$ we use here is the square of the one used in \cite{Ferrero:2024wld1}.} 
\begin{equation}
\label{chi_gen}
\chi=\left(1+\chi_{\left[0,1\right]}^{\mathfrak{sp}(4)}\right)\chi\left(\Phi\right)+2\chi_{\left[1,0\right]}^{\mathfrak{sp}(4)}\chi\left(\Psi\right)+2\chi\left(F\right)
\end{equation}
where
\begin{equation}
\label{chis}
    \begin{aligned}
\chi\left(\Phi\right)=q\left(1+\sum_{n\geq1}q^{n}\chi_{n}^{\mathfrak{su}(2)}\left(z\right)^{2}\right)&\,,\quad\chi\left(\Psi\right)=q^{3/2}\left(\chi_{1}^{\mathfrak{su}(2)}\left(z\right)+\sum_{n\geq1}q^{n}\chi_{n}^{\mathfrak{su}(2)}\left(z\right)\chi_{n+1}^{\mathfrak{su}(2)}\left(z\right)\right),\\  
\chi\left(F\right)=q^{2}&\left(\chi_{2}^{\mathfrak{su}(2)}\left(z\right)+\sum_{n\geq1}q^{n}\chi_{n}^{\mathfrak{su}(2)}\left(z\right)\chi_{n+2}^{\mathfrak{su}(2)}\left(z\right)\right)
    \end{aligned}
\end{equation}
with $q$ being the fugacity that counts the conformal dimension via $q^{\Delta}$ and with $z$ being the fugacity for the $\mathfrak{su}(2)\simeq\mathfrak{so}(3)$ rotation symmetry preserved by a line in 4d. The infinite sums above count the four-dimensional derivative towers, with the free bulk equations of motion imposed and the Lorentz representations branched to the rotation symmetry preserved by the line. They include both transverse derivatives and the derivative $\partial_t$ along the line, which generates
one-dimensional conformal descendants. Note also that since the Maldacena-Wilson line in 4d picks one scalar (i.e. $\Phi_6$), the bulk $\mathfrak{so}(6)_\mathcal{R}$ $R$-symmetry algebra is broken by it to $\mathfrak{sp}(4)\simeq\mathfrak{so}(5)$, which is why the representations in \eqref{chi_gen} are written this way. 

Now, in order to obtain the analogous character in our setup, where the same line is put on a boundary, we decompose the representations of the line in 4d to those preserved by the line on a boundary and impose the boundary conditions. To do this, we begin by introducing the fugacities for the symmetries preserved in our setup, 
\begin{equation}\label{fugacity_def}
x:\textrm{fugacity for \ensuremath{\mathfrak{su}(2)_{X}=\mathfrak{su}(2)_{\mathcal{R}}}}\,,\;y:\textrm{fugacity for \ensuremath{\mathfrak{u}(1)_{Y}}}\,,\;w:\textrm{fugacity for \ensuremath{\ensuremath{\mathfrak{u}(1)_{\text{rot}}}}}
\end{equation}
such that the $\mathfrak{sp}(4)\simeq\mathfrak{so}(5)$ $R$-symmetry preserved by the line in 4d is broken by the addition of the boundary to $\mathfrak{su}(2)_{X}\times\mathfrak{u}(1)_{Y}$, and the $\mathfrak{su}(2)\simeq\mathfrak{so}(3)$ rotation symmetry around the line to $\mathfrak{u}(1)_{\text{rot}}$ (which is its Cartan). The relevant branchings of the $R$-symmetry representations appearing in \eqref{chi_gen} under the preserved subgroup are 
\begin{equation}
    \mathfrak{sp}(4)\rightarrow\mathfrak{su}(2)_{X}\times\mathfrak{u}(1)_{Y}:\quad\left[0,1\right]\rightarrow\mathbf{3}_{0}\oplus\mathbf{1}_{+1}\oplus\mathbf{1}_{-1}\,,\quad\left[1,0\right]\rightarrow\mathbf{2}_{+1/2}\oplus\mathbf{2}_{-1/2}
\end{equation}
or 
\begin{equation}
\label{R_branch_chi}
1+\chi_{\left[0,1\right]}^{\mathfrak{sp}(4)}\rightarrow1+\chi_{2}^{\mathfrak{su}(2)}\left(x\right)+y+y^{-1}\,,\quad\chi_{\left[1,0\right]}^{\mathfrak{sp}(4)}\rightarrow\chi_{1}^{\mathfrak{su}(2)}\left(x\right)\left(y^{1/2}+y^{-1/2}\right),
\end{equation}
where the interpretation of the different terms in the first equation in \eqref{R_branch_chi} (corresponding to the scalar fields) is as follows, 
\begin{equation}
1\leftrightarrow\Phi_{6}\,,\quad\chi_{2}^{\mathfrak{su}(2)}\left(x\right)\leftrightarrow\vec{X}\,,\quad y^{\pm1}\leftrightarrow\Phi_{\pm}\,.
\end{equation}
Meanwhile, the transverse-spin branching for $\mathfrak{so}(3)_{\textrm{rot}}\rightarrow\mathfrak{so}(2)_{\textrm{rot}}$ is done by simply replacing the $z$ fugacity in the characters $\chi_{n}^{\mathfrak{su}(2)}\left(z\right)$ in \eqref{chis} by the fugacity $w$ such that $z^{2}=w$ (we could of course have chosen the normalization such that nothing changes in this branching), e.g. 
\begin{equation}
    \chi_{1}^{\mathfrak{su}(2)}\left(z\right)=z+z^{-1}\rightarrow w^{1/2}+w^{-1/2}\,.
\end{equation}

We next turn to impose the boundary conditions, which at the zero-coupling point we consider amount to \eqref{eq:pure-neumann-boundary-conditions}. In order to do it in a transparent way, we first make a digression and discuss the origin of the characters $\chi_{n}^{\mathfrak{su}(2)}\left(z\right)$ in \eqref{chis}, which as we mentioned represent the derivative tower on the line in the case with no boundary, following appendix C.4 of \cite{Ferrero:2024wld1}. Considering for simplicity a single free scalar $\phi$ of dimension one (such that on its own it contributes a $q$ factor to a character) that satisfies the free bulk equation of motion $\Delta\phi=0$, the four-dimensional character of its conformal multiplet is given by 
\begin{equation}
\chi^{\left(4d\right)}\left(\phi\right)=\frac{q\left(1-q^{2}\right)}{\left(1-qx\bar{x}\right)\left(1-qx^{-1}\bar{x}\right)\left(1-qx\bar{x}^{-1}\right)\left(1-qx^{-1}\bar{x}^{-1}\right)}
\end{equation}
where the denominator factors correspond to the four-dimensional derivatives $\partial_{\alpha\dot{\alpha}}$ with fugacities $x$ and $\bar{x}$ for $\mathfrak{su}(2)_{L}\times\mathfrak{su}(2)_{R}$, and the $(1-q^{2})$ factor in the numerator accounts for the equation of motion. The restriction to a line then preserves the diagonal subalgebra $\mathfrak{su}(2)_{\textrm{rot}}\subset\mathfrak{su}(2)_{L}\times\mathfrak{su}(2)_{R}$ and the branching is achieved at the character level by setting $x=\bar{x}=z$, with $z$ the same fugacity as before. We then obtain the following 1d character on the line, 
\begin{equation}
\label{chi_phi_1d}
\chi^{\left(1d\right)}\left(\phi\right)=\frac{q\left(1-q^{2}\right)}{\left(1-qz^{2}\right)\left(1-q\right)\left(1-q\right)\left(1-qz^{-2}\right)}=\frac{q\left(1+q\right)}{\left(1-qz^{2}\right)\left(1-q\right)\left(1-qz^{-2}\right)}
\end{equation}
which can equivalently be written as $\chi\left(\Phi\right)$ in Eq. \eqref{chis}. This is what is needed for a line in 4d and what appears in appendix C.4 of \cite{Ferrero:2024wld1}, but let us now rederive \eqref{chi_phi_1d} (and interpret the different factors in it) using a language that can be straightforwardly adapted to our setup and that will be very useful for imposing boundary conditions when adding a boundary. We begin by writing \eqref{chi_phi_1d} using the fugacity $w$ of $\mathfrak{u}(1)_{\textrm{rot}}$, 
\begin{align}
        \chi^{\left(1d\right)}\left(\phi\right)&=\frac{q\left(1+q\right)}{\left(1-qw\right)\left(1-q\right)\left(1-qw^{-1}\right)} \nonumber\\
&=\frac{q}{\left(1-qw\right)\left(1-q\right)\left(1-qw^{-1}\right)}+\frac{q^{2}}{\left(1-qw\right)\left(1-q\right)\left(1-qw^{-1}\right)}\label{chi_phi_1d_w}
\end{align}
and note that there are four derivatives with the following charges under $\mathfrak{u}(1)_{\textrm{rot}}$,
\begin{equation}
\textrm{Charge \ensuremath{0}}:\,\partial_{t}\,,\,\partial_{n}\,,\quad\textrm{Charge \ensuremath{\pm1}}:\,\partial_{\pm}
\end{equation}
where we use notations that are suitable for our setup even though we are here still in 4d without a boundary. In particular, $\partial_{t}$ is the derivative along the line, while $\partial_{n}$ and $\partial_{\pm}$ are the derivatives in the three transverse directions that form the triplet of $\mathfrak{su}(2)_{\textrm{rot}}$, written according to their charges under the Cartan $\mathfrak{u}(1)_{\textrm{rot}}$. Then, to obtain the character of the conformal multiplet of $\phi$, we need to act on it with the different derivatives while imposing the constraint $\Delta\phi=0$, and collect the contributions. To do this, we notice that at level $q^2$ (i.e. one derivative acting on $\phi$) all the four derivatives contribute independently, but that starting at level $q^3$ (i.e. two derivatives) all the higher derivatives in the $n$ direction $(\partial_{n})^{>1}\phi$ can be replaced using the equation of motion by the $\partial_{t}$ and $\partial_{\pm}$ derivatives acting on $\phi$ and $\partial_{n}\phi$. For example, we have 
\begin{equation}
\partial_{n}^{2}\phi=-\partial_{t}^{2}\phi-\partial_{+}\partial_{-}\phi\,,\quad\partial_{n}^{3}\phi=-\partial_{t}^{2}\left(\partial_{n}\phi\right)-\partial_{+}\partial_{-}\left(\partial_{n}\phi\right)
\end{equation}
and similarly (recursively) for higher-order derivatives. We therefore see that the net effect of the equation of motion is to eliminate the $(\partial_{n})^{>1}\phi$ derivatives from the counting, i.e. the independent operators in the character are generated by freely acting with the $\partial_{t}$ and $\partial_{\pm}$ derivatives on the two seed operators (or letters) $\phi$ and $\partial_{n}\phi$. This exactly corresponds to the two terms on the right hand side of \eqref{chi_phi_1d_w}, with the denominators corresponding to the three derivatives $\partial_{t}$ and $\partial_{\pm}$ (with charges $0$, $\pm1$) and the numerators $q$ and $q^2$ to $\phi$ and $\partial_{n}\phi$, respectively. 

We would now like to turn to our setup and consider the line on the boundary. Here the six scalars transform under the preserved symmetries according to Eq. \eqref{R_branch_chi}, i.e. they split into $\vec{X}$, which transforms in the adjoint of $\mathfrak{su}(2)_{X}$, and $\vec{Y}$, which includes the scalars $\Phi_{6}$ and $\Phi_{\pm}$ with charges $0$ and $\pm1$ under $\mathfrak{u}(1)_{Y}$. As we saw above, the character of each scalar in the bulk is generated by the derivatives $\partial_{t}$, $\partial_{\pm}$ acting on the scalar and its first normal derivative, where now $\partial_{n}$ has the meaning of a derivative in the normal direction to the boundary and $\partial_{\pm}$ are the two derivatives along the boundary directions which are transverse to the line. It is then easy to see the effect of the boundary conditions \eqref{eq:pure-neumann-boundary-conditions} on the scalar characters: the character of $\vec{X}$ is obtained by acting with the derivatives on only $\partial_{n}\vec{X}$ (since $\vec{X}=0$) while the character of $\vec{Y}$ is obtained by acting with the derivatives on only $\vec{Y}$ (since $\partial_{n}\vec{Y}=0$). We therefore get (see \eqref{chi_phi_1d_w}) 
\begin{equation}
\chi_{\vec{Y}}^{\textrm{bdy}}=\frac{q\left(1+y+y^{-1}\right)}{\left(1-q\right)\left(1-qw\right)\left(1-qw^{-1}\right)}\,,\quad\chi_{\partial_{n}\vec{X}}^{\textrm{bdy}}=\frac{q^{2}\chi_{2}^{\mathfrak{su}(2)}\left(x\right)}{\left(1-q\right)\left(1-qw\right)\left(1-qw^{-1}\right)}
\end{equation}
such that the total scalar contribution is
\begin{equation}
\chi_{\textrm{scal}}^{\textrm{bdy}}=\frac{q\left(1+y+y^{-1}\right)+q^{2}\chi_{2}^{\mathfrak{su}(2)}\left(x\right)}{\left(1-q\right)\left(1-qw\right)\left(1-qw^{-1}\right)}\,.
\end{equation}

Turning next to the fermions, the effect of the boundary conditions is to simply remove the factor of $2$ in the fermionic part of \eqref{chi_gen}, and using the branchings of the symmetries, we obtain the fermionic contribution 
\begin{equation}
\chi_{\textrm{ferm}}^{\textrm{bdy}}=\frac{q^{3/2}\chi_{1}^{\mathfrak{su}(2)}\left(x\right)\left(y^{1/2}+y^{-1/2}\right)\left(w^{1/2}+w^{-1/2}\right)}{\left(1-q\right)\left(1-qw\right)\left(1-qw^{-1}\right)}\,.
\end{equation}
Similarly, the boundary conditions remove the factor of $2$ from the gauge part of the character in \eqref{chi_gen}. To see this, note that the field-strength components divide into two sets with $\mathfrak{u}(1)_{\textrm{rot}}$ character $1+w+w^{-1}$, the first is $(F_{nt},F_{ni})$ which is set to zero by the boundary conditions and the second is $(F_{12},F_{ti})$ which is nontrivial and serves as the seed operators for the conformal character. This is what is responsible for the factor of $2$ turning into $1$. In addition to this, the derivative tower appears as follows. There are two bulk equations of motion, the Bianchi identity $\partial_{[\mu}F_{\nu\rho]}=0$ and $\partial_{\mu}\partial^{\mu}F_{\nu\rho}=0$ (which follows from the Bianchi identity and $\partial^{\mu}F_{\mu\rho}=0$), which can be used to eliminate all the normal derivatives of the seed field-strength components from the counting of independent operators. Explicitly, 
\begin{equation}
0=3\,\partial_{[n}F_{ti]}=\partial_{n}F_{ti}+\partial_{t}F_{in}+\partial_{i}F_{nt}=\partial_{n}F_{ti}
\end{equation}
(where we have used the boundary conditions $F_{in}=F_{nt}=0$) and similarly $\partial_{n}F_{12}=0$, while for the second derivative $\partial_{n}^{2}F_{\mu\nu}$ we have
\begin{equation}
\partial_{n}^{2}F_{\mu\nu}=-\partial_{t}^{2}F_{\mu\nu}-\partial_{+}\partial_{-}F_{\mu\nu}
\end{equation}
and similarly (recursively) for the higher derivatives. Moreover, the Bianchi identity for the choice of components $\partial_{[t}F_{12]}=0$ eliminates all the $\partial_t$ derivatives acting on $F_{12}$, 
\begin{equation}
\partial_tF_{12} = \frac{i}{2}(\partial_+F_{t-}-\partial_-F_{t+})\,,
\end{equation}
such that the total gauge contribution to the character is given by 
\begin{equation}
\chi_{F}^{\textrm{bdy}}=\frac{q^{2}\left(1-q+w+w^{-1}\right)}{\left(1-q\right)\left(1-qw\right)\left(1-qw^{-1}\right)}
\end{equation}
with the $1-q$ part in the numerator canceling the $\partial_t$ derivative tower of $F_{12}$.

Overall, the full single-letter line character of the bulk fields is therefore given by
\begin{equation}
\label{eq:superconformal-character-w/-boundary}
\chi_{\textrm{bulk}}^{\textrm{bdy}}\hspace{-1.5pt}=\hspace{-1.5pt}\frac{q(1+y+y^{-1})+q^{\frac{3}{2}}\chi_{1}^{\mathfrak{su}(2)}\hspace{-1.5pt}(x)(y^{\frac{1}{2}}+y^{-\frac{1}{2}})(w^{\frac{1}{2}}+w^{-\frac{1}{2}})+q^{2}(\chi_{2}^{\mathfrak{su}(2)}\hspace{-1.5pt}(x)+1-q+w+w^{-1})}{\left(1-q\right)\left(1-qw\right)\left(1-qw^{-1}\right)}
\end{equation}
and is the analog of \eqref{chi_gen} for our setup. Obtaining the corresponding contribution to the partition function is now very simple. Since all the bulk fields contributing to \eqref{eq:superconformal-character-w/-boundary} are in the adjoint of the $SU(N)$ gauge group, and since the results of products of such fields should also be taken in the adjoint representation in order to serve as a valid insertion on the Wilson line, the order of the fields in each product matters and no symmetrization is needed when computing the partition function. As a result, we obtain the following contribution of the bulk to the planar partition function of the line at the zero-coupling point, 
\begin{equation}\label{Z_weak_bulk}
Z_{\textrm{weak}}^{\textrm{bulk}}=\sum_{k=0}^{\infty}\left(\chi_{\textrm{bulk}}^{\textrm{bdy}}\right)^{k}=\frac{1}{1-\chi_{\textrm{bulk}}^{\textrm{bdy}}}\,.
\end{equation}

Now that we have discussed the free-field description at small 't Hooft coupling and examined some of the multiplets and part of the spectrum on the line, we turn to strong coupling and consider its holographic description.

\subsection{Strong-coupling regime}\label{sec:strong-coupling-operators}
The second regime in which we have a perturbative description is given by taking the 't Hooft coupling to be very large. In this case, operator insertions on the Wilson line admit a dual description in terms of the fluctuations of a bulk fundamental string ending on the boundary along the Wilson line. As reviewed in the previous section, the supergravity solution dual to our setup involves the space $AdS_{4}\times S_{1}^{2}\times S_{2}^{2}\times\Sigma$ with metric \eqref{eq:supergravity-solution}, and the string dual to the Wilson line lies along an $AdS_{2}$ subspace inside the $AdS_{4}$ part. More explicitly, writing the $AdS_{4}$ metric in terms of $AdS_{2}$ slicing, we have 
\begin{equation}
\label{AdS4_AdS2}
ds_{AdS_{4}}^2=\frac{\left(1+\frac{1}{4}\bm{\bar{x}}^2\right)^2}{\left(1-\frac{1}{4}\bm{\bar{x}}^2\right)^2}ds_{AdS_{2}}^2+\frac{d\bm{\bar{x}}^2}{\left(1-\frac{1}{4}\bm{\bar{x}}^2\right)^2}\,\,,\quad ds_{AdS_{2}}^2=\frac{1}{\sigma^2}\left(-dt^2+d\sigma^2\right)
\end{equation}
with $t\in\mathbb{R}$, $\sigma\in(0,\infty)$ and $\bm{\bar{x}}^2 = \bar{x}^i\bar{x}^i<1$, $i=1,2$, and the string is located at $\bm{\bar{x}}=0$. Meanwhile, its position in the $\Sigma$ plane is fixed by the requirement that it is 1/2-BPS with respect to the 3d boundary supersymmetry, and is given by \eqref{z_string}. 

In order to obtain the effective worldsheet theory of this string, we should start from its action and, as usual, perturb its location around this embedding, thereby getting an effective action for these small perturbations. We therefore start from the action of a fundamental string, whose  bosonic part is given in string frame by 
\begin{equation}\label{eq:nambu-goto-action}
    S_{\text{F1}} = -\frac{1}{2\pi\alpha'}\int\mathrm{d}^2\xi\left(\sqrt{-\det\left(G_{\mu\nu}(X)\partial_aX^\mu\partial_bX^\nu\right)}+\frac{1}{2}\epsilon^{ab}B_{\mu\nu}\partial_aX^\mu\partial_bX^\nu\right)
\end{equation}
where $X^\mu=X^\mu(\xi)$, $\mu = 0,\dots,9$ are the bulk (target space) coordinates 
\begin{equation}
    X^\mu = \left(\underbrace{t, \sigma, \bar{x}^1, \bar{x}^2}{\tikz{\node[inner sep=0pt] (AdS){};}}, \underbrace{\theta_1, \phi_1}{\tikz{\node[inner sep=0pt] (S_1){};}}, \underbrace{\theta_2, \phi_2}{\tikz{\node[inner sep=0pt] (S_2){};}}, \underbrace{x, y}{\tikz{\node[inner sep=0pt] (Sigma){};}}\right)
\end{equation}
\begin{tikzpicture}[overlay]
    \node[below of=AdS,align=center,yshift=8pt,xshift=-27.5pt] {$AdS_{4}$};
    \node[below of=S_1,align=center,yshift=8pt,xshift=-16pt] {$S^2_1$};
    \node[below of=S_2,align=center,yshift=8pt,xshift=-16pt] {$S^2_2$};
    \node[below of=Sigma,align=center,yshift=8pt,xshift=-13pt] {$\Sigma$};
\end{tikzpicture} \\
which in turn are functions of the worldsheet coordinates $\xi^a$, $a=0,1$, and $B_{\mu\nu}$ is the $B$-field given in \eqref{eq:B_field}. Fixing the static gauge $(\xi^0,\xi^1)=(t,\sigma)$, we can now expand \eqref{eq:nambu-goto-action} in small perturbations of the coordinates transverse to the worldsheet around the embedding location, 
\begin{gather}\label{eq:worldsheet-fluctuations}
    \bar{x}^i = \delta\bar{x}^i\ ;\ \theta_1=\delta\theta_1\ ;\ \phi_1 = \delta\phi_1\ ;\ \theta_2=\frac{\pi}{2}+\delta\theta_2\ ;\ \phi_2 = \delta\phi_2\ ;\ x = \delta x\ (\geq 0)\ ;\ y = y_{\text{str}} + \delta y
\end{gather}
and get an effective action for them. Note that $y_{\text{str}}$ here is the location of the string on the $y$ axis of the $\Sigma$ plane (given in \eqref{z_string}), and that we chose to expand $\theta_2$ around $\theta_2=\pi/2$ for convenience (all choices for the string location in $S_1^2\times S_2^2$ are of course equivalent).  
Applying this to the supergravity solution corresponding to our setup, which we recall describes $N$ D3-branes ending on $N_{\textrm{NS5}}=N^{1-\alpha}/\kappa$ NS5 branes, results in the following worldsheet action to cubic order, 
\begin{equation}\label{SF1}
S_{\mathrm{F1}}
=-\int d^2\xi\,\sqrt{-\widehat g}\,
\left[T_{\mathrm F}+\mathcal L_2+\mathcal L_3
+\mathcal O(\delta^4)\right].
\end{equation}
Here $T_{\mathrm F}=1/(2\pi\alpha')$ is the fundamental-string
tension, and $\widehat g_{ab}$ is the induced string-frame metric
on the unperturbed worldsheet. It is related to the unit-radius
$AdS_2$ metric $g_{ab}$ in \eqref{AdS4_AdS2} by
\begin{equation}
\widehat g_{ab}=L^2g_{ab}\,,
\qquad
\widehat g_{ab}\,d\xi^a d\xi^b
=\frac{L^{2}}{\sigma^{2}}\left(-dt^{2}+d\sigma^{2}\right)\,,
\end{equation}
where the physical $AdS_2$ radius $L$ is given by (see \eqref{eq:supergravity-solution})\footnote{The expression for $L^2$ at $u=0$ is understood by continuity, giving $L^2=\alpha'\sqrt{2\lambda}$. For general $u$, the corresponding effective tension $T_{\mathrm F}L^2$ agrees with that extracted from the circular Wilson-loop expectation value in \cite[Eq.~(60)]{He:2024djr}, through $\log\langle W_{\mathrm f}\rangle=2\pi T_{\mathrm F}L^2$ at leading order.}
\begin{equation}
L^2=f_{4,\textrm{(str)}}^{2}|_{z_{\textrm{string}}}=\frac{\alpha'\sqrt{\lambda}}{4}
\left(
\sqrt{u+8}
+\frac{8}{\sqrt u}\operatorname{arsinh}\sqrt{\frac{u}{8}}
\right)
\end{equation}
with
\begin{equation}\label{u}
u\equiv\lim_{N\to\infty}\lambda\kappa^2N^{-2(1-\alpha)}
=\begin{cases}
0\,, & 0\leq\alpha<1\,,\\[2pt]
v=\lambda\kappa^2\,, & \alpha=1\,.
\end{cases}
\end{equation}
The quadratic and cubic parts of the action are 
\begin{equation}\label{L2}
\begin{aligned}
\mathcal L_2={}&
\frac12\widehat g^{ab}
\partial_a\delta\widetilde{\bar x}^{\,i}
\partial_b\delta\widetilde{\bar x}^{\,i}
+\frac{1}{L^2}
\delta\widetilde{\bar x}^{\,i}\delta\widetilde{\bar x}^{\,i}
+\frac12\widehat g^{ab}
\partial_a\delta\widetilde r^{\,m}
\partial_b\delta\widetilde r^{\,m}
\\
&+\frac12\widehat g^{ab}
\partial_a\delta\widetilde\theta_2
\partial_b\delta\widetilde\theta_2
+\frac12\widehat g^{ab}
\partial_a\delta\widetilde\varphi_2
\partial_b\delta\widetilde\varphi_2
+\frac12\widehat g^{ab}
\partial_a\delta\widetilde y\,
\partial_b\delta\widetilde y
+\frac{1}{L^2}(\delta\widetilde y)^2
\end{aligned}
\end{equation}
and
\begin{equation}\label{L3}
\begin{aligned}
\mathcal L_3=c(u)\Bigg[&
\frac{1}{L^2}\left(
3\,\delta\widetilde y\,
\delta\widetilde r^{\,m}\delta\widetilde r^{\,m}
-(\delta\widetilde y)^3\right)
-3\,\delta\widetilde y\,\widehat g^{ab}
\partial_a\delta\widetilde r^{\,m}
\partial_b\delta\widetilde r^{\,m}
\\
&-3\,\delta\widetilde y\,\widehat g^{ab}
\partial_a\delta\widetilde y\,
\partial_b\delta\widetilde y
-\frac{\epsilon^{ab}}{\sqrt{-\widehat g}}\,
\epsilon^{lmn}\delta\widetilde r^{\,l}
\partial_a\delta\widetilde r^{\,m}
\partial_b\delta\widetilde r^{\,n}
\Bigg],
\end{aligned}
\end{equation}
with
\begin{equation}\label{c}
c(u)=\frac{\sqrt{2\pi}}{3\lambda^{1/4}}
\frac{u+6}{(u+8)^{3/4}}\,.
\end{equation}
Here, $x$, $\theta_{1}$ and $\phi_{1}$ combined to form the coordinates $r^m$ which form a triplet of $\mathfrak{so}(3)$, 
\begin{equation}
r^{m}=x\left(\sin\theta_{1}\cos\phi_{1},\sin\theta_{1}\sin\phi_{1},\cos\theta_{1}\right)\,,\quad m=1,2,3\,,
\end{equation}
and the following rescaling was done in order to get canonically normalized worldsheet fields, 
\begin{equation}
(\delta\widetilde{\bar{x}}^{\,i},\delta\widetilde{\theta}_{2},\delta\widetilde{\phi}_{2})=\sqrt{T_{\mathrm F}L^2}\,(\delta\bar{x}^{i},\delta\theta_{2},\delta\phi_{2})\,,\quad(\delta\widetilde{r}^{\,m},\delta\widetilde{y})=\frac{\lambda^{1/4}}{2\sqrt{2}}(u+8)^{1/4}(\delta r^{m},\delta y)\,.
\end{equation}

Notice that the uniform parameter $u$ in \eqref{u} interpolates smoothly between the $0\leq\alpha<1$ and $\alpha=1$ cases, with the $0\leq\alpha<1$ case obtained as the small-$u$ limit of the $\alpha=1$ case. This includes the worldsheet action, the string and pole locations and the expressions for the dilaton and the curvature at the string location (see Eqs. \eqref{z_string}-\eqref{R_dil_string_alpha1}), which can all be written in terms of $u$. Moreover, note that the curvature at the string location can be written in a simple way in terms of the coupling \eqref{c} of cubic interactions, 
\begin{equation}\label{eq:curvature-coupling-relation}
\alpha'\mathcal{R}^{\left(\textrm{str}\right)}|_{z_{\text{string}}}=\frac{27}{\pi}\,c(u)^{2}\,,
\end{equation}
see Eqs. \eqref{R_dil_string_alpha}-\eqref{R_dil_string_alpha1}. We can therefore clearly see that the validity criterion of the supergravity approximation in studying the string, given in Eq. \eqref{sugra_criterion} by $N_{\textrm{NS5}}\gg1$, corresponds to the condition for having a weakly interacting worldsheet theory. Let us also mention that even when the supergravity description breaks down (i.e. when $N_{\textrm{NS5}}=\mathcal{O}(1)$), the curvature turns out to be of order one only in the $S^2_1\times \Sigma$ directions (or the $\mathbb{R}\times S^{3}$ directions transverse to the NS5 brane near the pole), while still being small in the $AdS_4 \times S^2_2$ directions. This, in turn, corresponds again to the cubic interactions in \eqref{L3}, which are between perturbations of the string in the $S^2_1\times \Sigma$ directions, becoming large (i.e. of order one). More generally, at quartic order, we begin to also have interactions between the perturbations in the $AdS_4 \times S^2_2$ and $S^2_1\times \Sigma$ directions, as well as self interactions between the $AdS_4 \times S^2_2$ perturbations, but only the self interactions between the $S^2_1\times \Sigma$ perturbations become large for $N_{\textrm{NS5}}=\mathcal{O}(1)$ (while all the interactions are small for $N_{\textrm{NS5}}\gg1$). A further simplification occurs as $\lambda\to\infty$ at fixed $N_{\mathrm{NS5}}$ (with the planar limit taken first), where the string-frame metric at the string position takes the form 
\begin{equation}\label{eq:CHS-bg}
ds_{\mathrm{(str)}}^{2}|_{z_{\textrm{string}}}\simeq\frac{\alpha'\lambda}{4N_{\mathrm{NS5}}}\left(ds_{AdS_{4}}^{2}+ds_{S_{2}^{2}}^{2}\right)+N_{\mathrm{NS5}}\alpha'\left(\frac{dR^{2}}{R^{2}}+ds_{S^{3}}^{2}\right)
\end{equation}
with $R=|z-z_{\mathrm{pole}}|$.\footnote{Note that the near-pole expansion in (2.27) is taken at fixed $\lambda$, is written in Einstein frame, and is not uniform in $\lambda$. Here we instead evaluate the string-frame metric at the $\lambda$-dependent string position before taking $\lambda\to\infty$. Although
$R_{\mathrm{string}}\sim 2N_{\mathrm{NS5}}/\sqrt{\pi\lambda}\to0$, the regular term in $h_2$ proportional to $\lambda/N_{\mathrm{NS5}}$ dominates its logarithmic term at this position.} We then see that the $AdS_4\times S^2_2$ directions become locally flat (with their radius scaling like $\sqrt{\lambda}$), while the transverse directions retain the NS5 throat geometry. This, in fact, is the local type-IIB Callan-Harvey-Strominger throat (with the remaining background fields also approaching their CHS form)  \cite{Callan:1991dj,Callan:1991at,Aharony:1998ub}, in which the sector of the worldsheet theory associated with the directions transverse to the NS5-brane worldvolume consists of a linear-dilaton superfield and a supersymmetric $SU(2)$ WZW model at level $N_{\mathrm{NS5}}$.\footnote{The conventional supersymmetric WZW description applies for $N_{\mathrm{NS5}}\geq2$, with its bosonic $SU(2)$ level being $N_{\mathrm{NS5}}-2$. The single-NS5 case requires a separate treatment, see \cite{Dei:2025ilx} for a recent discussion. We will comment about this case in the next subsection, where we discuss what the superconformal index suggests about the spectrum in our setup.} 

Let us next examine the spectrum and some of the multiplet types on the dual Wilson line at the infinite-coupling point $\lambda=\infty$, as follows from the free worldsheet action \eqref{L2}. Note that even though the validity of the supergravity approximation for the string only requires $N_{\textrm{NS5}}\gg1$, to have the free worldsheet-theory description at $\lambda=\infty$ we should instead require the stronger condition that $N_{\textrm{NS5}}$ is infinite, which corresponds to the $\alpha<1$ case and which we will hence assume here. This, in turn, ensures that the worldsheet interactions vanish at the $\lambda=\infty$ point, and that the spectrum is freely generated by the fields in \eqref{L2}.\footnote{On the other hand, the case $\alpha=1$ with fixed $\kappa\ll1$ (corresponding to fixed $N_{\textrm{NS5}}\gg1$), while still compatible with the supergravity description, results in the finite cubic coupling $c=\sqrt{2\pi\kappa}/3>0$ at $\lambda=\infty$ and in a non freely generated spectrum. In particular, for highly enough excited states (with $\sim c^{-2}\sim\kappa^{-1}$ excitations) the size of $\mathcal{L}_3$
is of the same order as that of $\mathcal{L}_2$.\label{foot:free_gen}} We begin then by identifying the multiplets that the operators dual to the fundamental worldsheet fields in \eqref{L2} reside in, using the same notations as in the previous subsection (see \eqref{new_notation}). Note that in doing so, the identification of symmetries in the notation $\left[r_{Y},r_{\text{rot}}\right]_{\Delta}^{(R)}$ is such that the remaining $\mathfrak{so}(2)$ isometry of $S^2_2$ (left from the breaking of its $\mathfrak{so}(3)$ isometry by the location of the string in it) corresponds to $\mathfrak{u}(1)_Y$, the $\mathfrak{so}(2)$ rotating the two directions transverse to $AdS_{2}$ inside $AdS_{4}$ (acting on $\delta\widetilde{\bar{x}}^{\,i}$ as a doublet) corresponds to $\mathfrak{u}(1)_{\text{rot}}$, and the $\mathfrak{so}(3)$ associated with the shrunken $S^2_1$ (acting on $\delta\widetilde{r}^{\,m}$ as a triplet) corresponds to $\mathfrak{su}(2)_{\mathcal{R}}$. Moreover, in extracting the spectrum, we use standard
quantization for the worldsheet fields, imposing fixed-source boundary conditions at the $AdS_2$ boundary $\sigma=0$ and keeping the Wilson-line contour and its internal embedding fixed. The allowed normalizable fluctuations therefore behave as $O(\sigma)$ for the massless scalars and $O(\sigma^2)$ for those with $m^2L^2=2$, giving $\Delta=1$ and $\Delta=2$, respectively. Meanwhile, the fermionic partners obey the supersymmetry-compatible
source-free boundary conditions: their $\sigma^{-1/2}$
source coefficients vanish, while the normalizable modes behave as $\sigma^{3/2}$, giving $\Delta=3/2$.\footnote{Notice that changing the worldsheet boundary conditions can instead define a different line operator, as illustrated by the ordinary Wilson line without scalar coupling
\cite{Polchinski:2011im,Beccaria:2017rbe}.}
\begin{itemize}
    \item \textit{Displacement Multiplets}: Using the identification of symmetries we did above and the familiar relation $m^2L^2=\Delta(\Delta-d)$ for $d=1$, where $m^2$ is the mass of a field living on the $AdS_2$ worldsheet and $\Delta$ the dimension of the dual operator on the Wilson line, we see that the massless scalars $\delta\widetilde{\theta}_{2}$ and $\delta\widetilde{\phi}_{2}$ in \eqref{L2} are dual to operators with the quantum numbers $\left[\pm 1,0\right]_1^{(0)}$, which we denote by $\mathsf{\Theta}_\pm$. Meanwhile, $\delta\widetilde{\bar{x}}^{\,i}$ in \eqref{L2} have $m^2L^2=2$ and are dual to operators transforming as $\left[0,\pm 1\right]_2^{(0)}$, which we denote by $\mathsf{\bar{X}}_\pm$. Together with the fermionic operators with quantum numbers $\left[\pm1/2,\pm1/2\right]_{3/2}^{(1)}$ at level one, these form the displacement multiplets $L\bar{A}_1\left[+1,0\right]_1^{(0)}$ and $A_1\bar{L}\left[-1,0\right]_1^{(0)}$ we encountered before (see figure \ref{fig:displacement-multiplets-weak}). Here, the operators $\mathsf{\Theta}_\pm$ are the bottom components of these multiplets and are associated as expected with the $R$-symmetry broken by the line (which here is the $\mathfrak{so}(3)$ isometry of $S^2_2$ broken by the string to $\mathfrak{so}(2)$), while $\mathsf{\bar{X}}_\pm$  are their super-descendants at level 2 and are the displacement operators associated with the broken translations (here the translations transverse to $AdS_{2}$ inside $AdS_{4}$). 

    \item \textit{The Multiplet $A_1\bar{A}_1\left[0,0\right]_1^{(2)}$}: The symmetry identifications and the mass-dimension relation $m^2L^2=\Delta(\Delta-1)$ imply that the three massless scalars $\delta\widetilde{r}^{\,m}$ are dual to an operator with the quantum numbers $\left[0,0\right]_1^{(2)}$, which we denote by $\mathsf{R}^m$ (or $\vec{\mathsf{R}}$), while the remaining $m^2L^2=2$ scalar $\delta\widetilde{y}$ is dual to an operator $\left[0,0\right]_2^{(0)}$ which is denoted by $\mathsf{Y}$. Together with two fermionic operators with quantum numbers $\left[-1/2,+1/2\right]_{3/2}^{(1)}\oplus\left[+1/2,-1/2\right]_{3/2}^{(1)}$ at level one, they form the short multiplet $A_1\bar{A}_1\left[0,0\right]_1^{(2)}$, such that $\vec{\mathsf{R}}$ is the superconformal primary, while $\mathsf{Y}$ is its super-descendant at level 2. 
\end{itemize}

Having identified the quantum numbers and multiplet structure of the basic operators on the Wilson line that are dual to the fundamental worldsheet fields, the rest of the spectrum at $\lambda=\infty$ is easily obtained since it is freely generated by these operators. More explicitly, the total single-letter character is given in this case by 
\begin{equation}\label{chi_total_strong}
\begin{aligned}
\chi(y,w,x;q)
&=
\chi_{L\bar{A}_{1}\left[+1,0\right]_{1}^{(0)}}
+\chi_{A_{1}\bar{L}\left[-1,0\right]_{1}^{(0)}}
+\chi_{A_{1}\bar{A}_{1}\left[0,0\right]_{1}^{(2)}}
\\
&=
\frac{1}{1-q}\Bigg(
q\left(\chi_{2}^{\mathfrak{su}(2)}(x)+y+y^{-1}\right)
+q^{2}\left(1+w+w^{-1}\right)
\\
&\phantom{=\frac{1}{1-q}\Bigg(}
+q^{3/2}\chi_{1}^{\mathfrak{su}(2)}(x)
\left(y^{1/2}+y^{-1/2}\right)
\left(w^{1/2}+w^{-1/2}\right)
\Bigg)
\end{aligned}
\end{equation}
where the characters of the individual multiplets are 
\begin{align}
\chi_{L\bar{A}_{1}\left[+1,0\right]_{1}^{(0)}}
&=
\frac{1}{1-q}
\left(
qy+q^{3/2}\chi_{1}^{\mathfrak{su}(2)}(x)y^{1/2}w^{1/2}+q^{2}w
\right),\nonumber
\\
\chi_{A_{1}\bar{L}\left[-1,0\right]_{1}^{(0)}}
&=
\frac{1}{1-q}
\left(
qy^{-1}+q^{3/2}\chi_{1}^{\mathfrak{su}(2)}(x)y^{-1/2}w^{-1/2}
+q^{2}w^{-1}
\right),\label{chis_mult}
\\
\chi_{A_{1}\bar{A}_{1}\left[0,0\right]_{1}^{(2)}}
&=
\frac{1}{1-q}
\Big(
q\chi_{2}^{\mathfrak{su}(2)}(x)+q^{3/2}\chi_{1}^{\mathfrak{su}(2)}(x)
\left(y^{-1/2}w^{1/2}+y^{1/2}w^{-1/2}\right)+q^{2}\Big).\nonumber
\end{align}
These, in turn, are obtained from the content of the multiplets described in the bullets above, while employing the character conventions used in the previous subsection (see the discussion around \eqref{chis} and \eqref{fugacity_def}). Note in particular that the terms in the parenthesis in Eq. \eqref{chis_mult} correspond to the primaries of the multiplets, while the $(1-q)$ factors in the denominators correspond to the $\partial_t$ derivative along the line and generate the infinite towers of descendants. 

Given the single-letter character \eqref{chi_total_strong}, the partition function is obtained by using appropriately the plethystic exponential. This is in contrast to the computation we did at the zero-coupling point $\lambda=0$ (see \eqref{Z_weak_bulk}), where the basic letters were adjoint valued (and their products should have been taken to be in the adjoint as well) and we did not need to apply any specific symmetrization. Here, on the other hand, the letters are gauge singlets and therefore appropriate symmetrizations are required for bosons and fermions when constructing composite operators. Specifically, the partition function is computed by first splitting the character \eqref{chi_total_strong} to bosonic and fermionic parts, 
\begin{equation}
\begin{aligned}
\chi_{\mathrm{B}}(y,w,x;q)
&=
\frac{
q(\chi_{2}^{\mathfrak{su}(2)}(x)+y+y^{-1})
+q^{2}\left(1+w+w^{-1}\right)
}{1-q}\,,
\\[2mm]
\chi_{\mathrm{F}}(y,w,x;q)
&=
\frac{
q^{3/2}\chi_{1}^{\mathfrak{su}(2)}(x)
\left(y^{1/2}+y^{-1/2}\right)
\left(w^{1/2}+w^{-1/2}\right)
}{1-q}
\end{aligned}
\end{equation}
and then taking their plethystic exponentials as follows, 
\begin{equation}
Z_{\mathrm{strong}}
=
\exp\left[
\sum_{n=1}^{\infty}\frac{1}{n}
\left(
\chi_{\mathrm{B}}(y^{n},w^{n},x^{n};q^{n})
+(-1)^{n+1}
\chi_{\mathrm{F}}(y^{n},w^{n},x^{n};q^{n})
\right)
\right].
\end{equation}
The difference between bosons and fermions manifests itself here in the factor $(-1)^{n+1}$ in front of $\chi_{\mathrm{F}}$, which is responsible for accounting correctly for the anti-symmetrization appearing in products of fermions.

\subsection{Superconformal index}
\label{subsec:Index}

In the previous two subsections, we have discussed two descriptions of the Wilson line in the setup we consider, one at the weak-coupling regime $\lambda\rightarrow0$ and the other at the strong-coupling regime $\lambda\rightarrow \infty$. It would hence be useful to examine a protected quantity that is invariant under changing $\lambda$, and check how it matches between the two descriptions. For this purpose, we consider the superconformal index of the Wilson line at the leading large-$N$ limit, which is indeed independent of $\lambda$. For completeness, we also use an exactly analogous approach in appendix \ref{app:no-boundary-index} for computing the line index in the more familiar case where there is no boundary, reproducing known results. 

Beginning with the general construction of the index (see \cite{Gadde:2020yah} for a review and \cite{Gang:2012yr} for a related computation in the case with no boundary), we choose a nilpotent supercharge $Q$ preserved by the Wilson line and denote by
$Q^{\dagger}$ its adjoint in radial quantization. The operator
\begin{equation}\label{delta}
    \delta\equiv\left\{Q,Q^{\dagger}\right\}
\end{equation}
is then positive semidefinite, and the line-operator superconformal index is defined by
\begin{equation}\label{I}
\mathcal{I}=\textrm{Tr}_{\mathcal{H}_{\textrm{line}}}(-1)^{F}e^{-\beta\delta}\prod_{i}z_{i}^{J_{i}}
\end{equation}
where $\mathcal{H}_{\textrm{line}}$ is the line Hilbert space, $J_i$ are charges that commute with $Q$ and $Q^{\dagger}$ and $z_i$ are the corresponding fugacities. In such a trace, every state with
$\delta>0$ is paired by $Q$ with a state of opposite fermion parity such that their
contributions to the index cancel due to the $(-1)^F$ factor. As a result, only states satisfying
$\delta=0$ (or equivalently representatives of the $Q$-cohomology) contribute, and the index is independent of $\beta$. Such states belong to short or semi-short superconformal multiplets (see table \ref{tab:1d-SCFT-multiplets}),
while long multiplets have vanishing index. This, in particular, means that even though individual short and semi-short multiplets can contribute nontrivially to the index, the total contribution of any collection of such multiplets that can recombine to form a long multiplet (see \eqref{eq:recombination_rules}) vanishes. Now since the only way in which the content of short and semi-short multiplets can change upon varying $\lambda$ is either through such recombinations into long multiplets or (in the opposite direction) through long multiplets hitting the unitarity bound and decomposing into shorter multiplets, to both of which the index is blind, we see that it is indeed independent of $\lambda$. 

To construct a specific index for our setup, we then start by choosing one of the supercharges in \eqref{QQd} preserved by the Wilson line. We choose $Q=\mathcal{Q}_{1}$ and using 
\begin{equation}\label{QSb}
\left\{ \mathcal{Q}_{a},\mathcal{\bar{S}}^{b}\right\} =\delta_{a}^{b}\left(D+\mathcal{R}_{2}\right)-\mathcal{R}_{a}^{\;b}\,,\quad\left(\mathcal{Q}_{a}\right)^{\dagger}=\mathcal{\bar{S}}^{a}
\end{equation}
following appendix B.1 of \cite{Nathan:2020scd}, where $D$ is the dilatation generator and $\mathcal{R}_{2}$, $\mathcal{R}_{a}^{\;b}$ the generators of $\mathfrak{u}(1)_{\mathcal{R}_{2}}$ and $\mathfrak{su}(2)_{\mathcal{R}}$ respectively, we find 
\begin{equation}
\delta=\left\{ Q,Q^{\dagger}\right\} =\left\{ \mathcal{Q}_{1},\left(\mathcal{Q}_{1}\right)^{\dagger}\right\} =\left\{ \mathcal{Q}_{1},\mathcal{\bar{S}}^{1}\right\} =D+\mathcal{R}_{2}-\mathfrak{R}
\end{equation}
where we denoted by $\mathfrak{R}\equiv\mathcal{R}_{1}^{\;1}$ the Cartan of $\mathfrak{su}(2)_{\mathcal{R}}$. We would next like to find generators that commute with both $Q$ and $Q^{\dagger}$ in order to refine the index, corresponding to $J_i$ in \eqref{I}. Since $\mathcal{R}_{2}$ is central (recall our discussion below \eqref{eq:1d-SCFT-bosonic-subalgebra}), it clearly qualifies as a first such generator. In addition to it, one can easily show using the commutation relations in appendix B.1 of \cite{Nathan:2020scd} that 
\begin{equation}
\left[2D-b,Q\right]=\left[2D-b,Q^{\dagger}\right]=0\,,
\end{equation}
where $b$ is the generator of $\mathfrak{u}(1)_{b}$ (see \eqref{u1b_u1R2_su2R}). Hence, $2D-b$ is another charge we can refine the index by, and is in fact the only independent such charge left. Overall, we obtain the following explicit expression for the index, 
\begin{equation}
\mathcal{I}\left(\eta,\xi\right)=\textrm{Tr}_{D=\mathfrak{R}-\mathcal{R}_{2}}\left(-1\right)^{F}\eta^{\mathcal{R}_{2}}\xi^{2D-b}
\end{equation}
where $\eta$ and $\xi$ are fugacities for the two generators we found and the trace is performed only over operators that satisfy the condition $D=\mathfrak{R}-\mathcal{R}_{2}$ (i.e. we wrote \eqref{I} only for the $\delta=0$ sector which is the one that can contribute nontrivially). Using \eqref{u1b_u1R2_su2R}, the index can also be written alternatively in terms of the $\mathfrak{u}(1)_{\text{rot}}$ and $\mathfrak{u}(1)_{Y}$ charges (instead of the $\mathfrak{u}(1)_{\mathcal{R}_{2}}$ and $\mathfrak{u}(1)_b$ charges), 
\begin{equation}\label{I_expl}
\mathcal{I}(\eta,\xi)=\textrm{Tr}_{D=\mathfrak{R}-r_{\mathrm{rot}}-r_{Y}}(-1)^{F}\,\eta^{r_{\mathrm{rot}}+r_{Y}}\,\xi^{\,2D-r_{\mathrm{rot}}+r_{Y}}\,.
\end{equation}
It evaluates on the different multiplets in table \ref{tab:1d-SCFT-multiplets} (using the notation \eqref{new_notation})\footnote{In order to distinguish between the operators $r_{Y}$, $r_{\text{rot}}$ and the corresponding charges of the superconformal primary of a multiplet (which is used in its notation), we denote such charges by $r_{Y}^{(0)}$ and $r_{\text{rot}}^{(0)}$.} as follows,  

\begin{itemize}
    \item \textit{Multiplets of the type $A_{1}\bar{L}[r_{Y}^{(0)},r_{\text{rot}}^{(0)}]_{\Delta}^{(R)}$}: For these multiplets $\Delta=-r_{Y}^{(0)}-r_{\text{rot}}^{(0)}+R/2$ (with $r_{Y}^{(0)}+r_{\text{rot}}^{(0)}<0$) such that the operators that satisfy the condition $\delta=D+r_{Y}+r_{\text{rot}}-\mathfrak{R}=0$ to contribute to the index are the $su(2)_\mathcal{R}$ highest-weight component of the superconformal primary (for which $\mathfrak{R}=R/2$) and its $\bar{\mathcal{Q}}^{2}$ descendant, since $[\delta,\bar{\mathcal{Q}}^{2}]=0$ with $\bar{\mathcal{Q}}^{2}$ acting nontrivially on the super-primary. Their total contribution to the index is then (see \eqref{QQd} for the charges of $\bar{\mathcal{Q}}^{2}$) 
    \begin{equation}\label{I_ALb}
    \mathcal{I}_{A_{1}\bar{L}[r_{Y}^{(0)},r_{\mathrm{rot}}^{(0)}]_{\Delta}^{(R)}}=(-1)^{2r_{\mathrm{rot}}^{(0)}}\,\eta^{\,r_{Y}^{(0)}+r_{\mathrm{rot}}^{(0)}}\xi^{\,R-r_{Y}^{(0)}-3r_{\mathrm{rot}}^{(0)}}\left(1-\xi^{2}\right)
    \end{equation}
    where the factor $(-1)^{2r_{\mathrm{rot}}^{(0)}}$ captures the fermion parity of the super-primary and is equal to $(-1)^{F_0}$. 
    \item \textit{Multiplets of the type $A_{1}\bar{A}_{1}[r_{Y}^{(0)},r_{\mathrm{rot}}^{(0)}]_{\Delta}^{(R)}$}: For these multiplets $\Delta=R/2$ with $r_{Y}^{(0)}+r_{\mathrm{rot}}^{(0)}=0$ and with both $\bar{\mathcal{Q}}^{2}$ and $\mathcal{Q}_1$ (for which $[\delta,\bar{\mathcal{Q}}^{2}]=[\delta,\mathcal{Q}_{1}]=0$) annihilating the superconformal primary. As a result, only the $su(2)_\mathcal{R}$ highest-weight component of the superconformal primary contributes, with the result being 
    \begin{equation}\label{I_AAb}
    \mathcal{I}_{A_{1}\bar{A}_{1}[r_{Y}^{(0)},r_{\mathrm{rot}}^{(0)}]_{\Delta}^{(R)}}=(-1)^{2r_{\mathrm{rot}}^{(0)}}\,\xi^{\,R-r_{\mathrm{rot}}^{(0)}+r_{Y}^{(0)}}\,.
    \end{equation}
    \item \textit{Multiplets of the types $L\bar{A}_{1}[r_{Y}^{(0)},r_{\mathrm{rot}}^{(0)}]_{\Delta}^{(R)}$ and $L\bar{L}[r_{Y}^{(0)},r_{\mathrm{rot}}^{(0)}]_{\Delta}^{(R)}$}: In this case no operator satisfies the condition $\delta=D+r_{Y}+r_{\text{rot}}-\mathfrak{R}=0$ such that the result is trivial, 
    \begin{equation}
    \mathcal{I}_{L\bar{A}_{1}[r_{Y}^{(0)},r_{\mathrm{rot}}^{(0)}]_{\Delta}^{(R)}}=\mathcal{I}_{L\bar{L}[r_{Y}^{(0)},r_{\mathrm{rot}}^{(0)}]_{\Delta}^{(R)}}=0\,.
    \end{equation}
\end{itemize}

We can now turn to the evaluation of the index at the weak and strong coupling regimes we studied, focusing on the $\alpha<1$ case where $N_{\textrm{NS5}}$ is infinite at the planar limit (see footnote \ref{foot:free_gen} and the discussion around it). Starting with strong coupling, as we saw in the previous subsection the spectrum on the Wilson line at $\lambda=\infty$ is freely generated by the operators of the displacement multiplets $L\bar{A}_{1}\left[+1,0\right]_{1}^{(0)}$ and $A_{1}\bar{L}\left[-1,0\right]_{1}^{(0)}$, along with those of the multiplet $A_{1}\bar{A}_{1}\left[0,0\right]_{1}^{(2)}$. Treating them as the basic letters (as in \eqref{chi_total_strong}) we then obtain the single-letter index 
\begin{equation}\label{i_s.l.}
i_{\textrm{s.l.}}^{\textrm{strong}}(\eta,\xi)=\mathcal{I}_{A_{1}\bar{L}\left[-1,0\right]_{1}^{(0)}}+\mathcal{I}_{A_{1}\bar{A}_{1}\left[0,0\right]_{1}^{(2)}}=\eta^{-1}\xi\left(1-\xi^{2}\right)+\xi^{2}
\end{equation}
where we have used \eqref{I_ALb} and \eqref{I_AAb}. Note also that there is no derivative contribution since $\partial_t$ does not satisfy $\delta=0$. The full index is then given by taking the (standard) plethystic exponential of \eqref{i_s.l.},
\begin{align}\label{I_result}
\mathcal{I}(\eta,\xi)=\operatorname{PE}\!\left[i_{\textrm{s.l.}}^{\textrm{strong}}(\eta,\xi)\right]=\exp\left[\sum_{n=1}^{\infty}\frac{1}{n}i_{\textrm{s.l.}}^{\textrm{strong}}\!\left(\eta^n,\xi^n\right)\right]=\frac{1-\eta^{-1}\xi^3}{\left(1-\eta^{-1}\xi\right)\left(1-\xi^2\right)}\,.
\end{align}
Let us next turn to weak coupling. Here we basically have at $\lambda=0$ pure BPS Neumann boundary conditions for the bulk $\mathcal{N}=4$ SYM fields plus a Maldacena-Wilson line and a decoupled $T_{[K^{N/K}]}[SU(N)]$ theory on the boundary. As a result, there are two types of adjoint insertions we can consider on the line, one from the restriction of bulk fields to the boundary and the other from operators of the boundary theory transforming in the adjoint of its Higgs-branch $SU(N)$ symmetry (which is the one we gauge with the bulk gauge field). Recall also that we consider the leading large-$N$ limit such that no traces are allowed. Starting from the contribution of the bulk fields to the line index, we list in table \ref{tab:bulk-field-index-charges} the field components that are not set to zero by the boundary condition, along with their charges under the generators that appear in the expression \eqref{I_expl} for the index. 
\begin{table}[t]
    \centering
    \renewcommand{\arraystretch}{1.35}
    \resizebox{\textwidth}{!}{%
    \begin{tabular}{c|cccccccccccc}
        \hline
        \textbf{Field/charge}
        & $\Phi_{6}$
        & $\Phi_{\pm}$
        & $\partial_{n}\vec{X}$
        & $\lambda_{a}^{++}$
        & $\lambda^{a--}$
        & $\lambda_{a}^{-+}$
        & $\lambda^{a+-}$
        & $F_{12}$
        & $F_{t\pm}$
        & $\partial_{t}$
        & $\partial_{n}$
        & $\partial_{\pm}$
        \\
        \hline
        $\Delta$
        & $1$
        & $1$
        & $2$
        & $\frac{3}{2}$
        & $\frac{3}{2}$
        & $\frac{3}{2}$
        & $\frac{3}{2}$
        & $2$
        & $2$
        & $1$
        & $1$
        & $1$
        \\
        $\mathfrak{R}$
        & $0$
        & $0$
        & $0,\pm1$
        & $\pm\frac{1}{2}$
        & $\pm\frac{1}{2}$
        & $\pm\frac{1}{2}$
        & $\pm\frac{1}{2}$
        & $0$
        & $0$
        & $0$
        & $0$
        & $0$
        \\
        $r_Y$
        & $0$
        & $\pm1$
        & $0$
        & $+\frac{1}{2}$
        & $-\frac{1}{2}$
        & $-\frac{1}{2}$
        & $+\frac{1}{2}$
        & $0$
        & $0$
        & $0$
        & $0$
        & $0$
        \\
        $r_{\mathrm{rot}}$
        & $0$
        & $0$
        & $0$
        & $+\frac{1}{2}$
        & $-\frac{1}{2}$
        & $+\frac{1}{2}$
        & $-\frac{1}{2}$
        & $0$
        & $\pm1$
        & $0$
        & $0$
        & $\pm1$
        \\
        \hline
    \end{tabular}%
    }
    \caption{Bulk fields restricted to the boundary and their charges under
    the generators entering the definition of the index.}
    \label{tab:bulk-field-index-charges}
\end{table}
We can then see that the letters with $\delta=D+r_{Y}+r_{\text{rot}}-\mathfrak{R}=0$, which therefore contribute nontrivially to the index, are $\Phi_{-}$ and $\lambda^{2--}$ that contribute $\eta^{-1}\xi$ and $-\eta^{-1}\xi^{3}$, respectively. In addition, the derivative $\partial_{-}$ also has $\delta=0$ and can therefore be used to generate new contributing operators by acting on those built from $\Phi_{-}$ and $\lambda^{2--}$. We therefore end up with the following single-letter index corresponding to the bulk fields, 
\begin{equation}\label{i_s.l._weak}
i_{\textrm{s.l.}}^{\textrm{weak,bulk}}(\eta,\xi)=\frac{\eta^{-1}\xi-\eta^{-1}\xi^{3}}{1-\eta^{-1}\xi^{3}}\,,
\end{equation}
where the denominator captures the contribution of $\partial_{-}$. Denoting by $i_{\textrm{s.l.}}^{\textrm{weak,\ensuremath{T}}}(\eta,\xi)$ the single-letter index corresponding to the contribution of basic seed operators from the boundary theory, the total single-letter index is then given by\footnote{Note that since we consider here only adjoint-valued operators such that their ordering matters, we must take into account the contribution of boundary-theory seed operators in the single-letter index together with that of the bulk fields, rather than computing the full index of each type of contribution separately and multiplying.} 
\begin{equation}\label{i_sl_weak_total}
i_{\textrm{s.l.}}^{\textrm{weak}}(\eta,\xi)=\frac{\eta^{-1}\xi-\eta^{-1}\xi^{3}}{1-\eta^{-1}\xi^{3}}+i_{\textrm{s.l.}}^{\textrm{weak,\ensuremath{T}}}(\eta,\xi)\,.
\end{equation}
As we discussed above Eq. \eqref{Z_weak_bulk}, in this weakly coupled description (of the bulk part and the Wilson line) no symmetrization, and hence no plethystic exponential, is needed when constructing operators and computing their contribution to the index (or partition function). As a result, the index obtained from \eqref{i_sl_weak_total} is given by 
\begin{equation}\label{I_weak}
\mathcal{I}\left(\eta,\xi\right)=\sum_{k=0}^{\infty}\left(i_{\textrm{s.l.}}^{\textrm{weak}}(\eta,\xi)\right)^{k}=\frac{1}{1-i_{\textrm{s.l.}}^{\textrm{weak}}(\eta,\xi)}\,,
\end{equation}
where we have used the same letter $\mathcal{I}$ as in \eqref{I_result} since the index is independent of $\lambda$. Comparing then \eqref{I_weak} to \eqref{I_result} and using \eqref{i_sl_weak_total}, we find that in order for the two expressions for the index to match, the contribution of the $T_{[K^{N/K}]}[SU(N)]$ theory to the single-letter index at $\lambda=0$ should be given by 
\begin{equation}\label{iT}
i_{\textrm{s.l.}}^{\textrm{weak,\ensuremath{T}}}(\eta,\xi)=\frac{\xi^{2}-\eta^{-1}\xi^{3}}{1-\eta^{-1}\xi^{3}}\,.
\end{equation}
To check this, let us first examine the contribution of operators residing in the 3d $\mathcal{N}=4$ conserved-current multiplet of the $SU(N)$ Higgs-branch global symmetry of the $T_{[K^{N/K}]}[SU(N)]$ theory, which is the most natural place to start searching for operators with a nontrivial index contribution since all the operators lying in this multiplet are in the adjoint of $SU(N)$ and so are nontrivial Wilson-line insertions at the planar limit. The simplest such operator is the superconformal primary of this 3d multiplet, the moment map operator $\vec{\mu}$ we already encountered before. Recall that in terms of 1d line multiplets (to which we can decompose any 3d multiplet), $\vec{\mu}$ is the super-primary of $A_{1}\bar{A}_{1}\left[0,0\right]_{1}^{(2)}$ and hence, as we showed in the bullets above, its $su(2)_\mathcal{R}$ highest-weight component $\mu_{\textrm{hw}}$ contributes to the index a factor of $\xi^{2}$. Meanwhile, the rest of this $A_{1}\bar{A}_{1}\left[0,0\right]_{1}^{(2)}$ multiplet, obtained by acting on $\vec{\mu}$ with the supercharges $\mathcal{Q}_{a}$, $\bar{\mathcal{Q}}^{a}$ preserved by the line (see \eqref{QQd}), does not contribute to the index. In order to obtain other operators in the 3d multiplet and check whether they contribute or not, we should act on $\vec{\mu}$ with the supercharges broken by the line. Denoting the 3d supercharges by $Q_{I\beta}$ (with $\beta=\pm$ and $I=1,\dots,4$), the preserved supercharges are given by \cite{Nathan:2020scd} 
\begin{equation}
\mathcal{Q}_{a}=Q_{+}^{2a-1}+iQ_{+}^{2a}\,,\quad\bar{\mathcal{Q}}^{a}=Q_{-}^{2a-1}-iQ_{-}^{2a}
\end{equation}
while the broken ones by 
\begin{equation}
\mathcal{\mathbf{Q}}^{a}=Q_{+}^{2a-1}-iQ_{+}^{2a}\,,\quad\bar{\mathcal{\mathbf{Q}}}_{a}=Q_{-}^{2a-1}+iQ_{-}^{2a}\,.
\end{equation}
Among these, the only broken supercharge  with $\delta=D+r_{Y}+r_{\text{rot}}-\mathfrak{R}=0$, which therefore yields an operator contributing nontrivially to the index when acting on $\mu_{\textrm{hw}}$, is $\bar{\mathcal{\mathbf{Q}}}_2$ whose quantum numbers are 
\begin{equation}
\bar{\mathcal{\mathbf{Q}}}_{2}:\quad\Delta=\frac{1}{2}\,,\quad\mathfrak{R}=-\frac{1}{2}\,,\quad(r_{Y},r_{\text{rot}})=\left(-\frac{1}{2},-\frac{1}{2}\right).
\end{equation}
As a result, in addition to $\mu_{\textrm{hw}}$ we have in the 3d current multiplet also the fermionic operator $\Psi\equiv\bar{\mathcal{\mathbf{Q}}}_{2}\mu_{\textrm{hw}}$ with quantum numbers  
\begin{equation}
\Psi:\quad\Delta=\frac{3}{2}\,,\quad\mathfrak{R}=\frac{1}{2}\,,\quad(r_{Y},r_{\text{rot}})=\left(-\frac{1}{2},-\frac{1}{2}\right),
\end{equation}
which contributes $-\eta^{-1}\xi^{3}$ to the index. Even though this operator is a super-descendant in the 3d multiplet, it is the $su(2)_\mathcal{R}$ highest-weight component of the superconformal primary of a 1d line multiplet $A_{1}\bar{L}[-1/2,-1/2]_{3/2}^{(1)}$. Now, as we discussed in the bullets above, also the $\bar{\mathcal{Q}}^{2}$ descendant of such an operator contributes to the index. However, in this case this results in the operator 
\begin{equation}
\bar{\mathcal{Q}}^{2}\Psi=\bar{\mathcal{Q}}^{2}\bar{\mathcal{\mathbf{Q}}}_{2}\mu_{\textrm{hw}}=\left\{ \bar{\mathcal{Q}}^{2},\bar{\mathcal{\mathbf{Q}}}_{2}\right\} \mu_{\textrm{hw}}\propto\partial_{-}\mu_{\textrm{hw}}\end{equation}
which we already account for in the single-letter index by adding the infinite tower of $\partial_{-}$ derivatives acting on $\mu_{\textrm{hw}}$ and $\Psi$. There are no other independent seed operators in the 3d current multiplet that can contribute to the index, and adding the contributions of $\mu_{\textrm{hw}}$ and $\Psi$ together with the corresponding towers of operators generated by $\partial_{-}$, we get exactly the single-letter index \eqref{iT}. Therefore, even if operators in other 3d multiplets of the $T_{[K^{N/K}]}[SU(N)]$ theory can contribute to the planar line index, their total contribution to $i_{\textrm{s.l.}}^{\textrm{weak,\ensuremath{T}}}$ vanishes. We will not investigate here whether there are such additional canceling contributions or not, but it would be interesting to check this explicitly elsewhere. 

Having studied the index in the two extreme regimes $\lambda=0$ and $\lambda=\infty$, let us next consider the other direction and check what we can learn from it about the multiplet content for generic $\lambda$. Writing the index \eqref{I_result} in the form 
\begin{equation}\label{I_expan}
\mathcal I=\frac{1-\eta^{-1}\xi^{3}}
{\left(1-\eta^{-1}\xi\right)\left(1-\xi^{2}\right)}
=1+\sum_{n=1}^{\infty}\xi^{2n}
+\sum_{m=1}^{\infty}\sum_{n=0}^{\infty}
\eta^{-m}\xi^{m+2n}\left(1-\xi^{2}\right),
\end{equation}
we begin by examining the terms $\xi^{2n}$. Using \eqref{I_ALb}-\eqref{I_AAb} it is easy to see that the only multiplets with such a contribution are 
\begin{equation}
A_{1}\bar{A}_{1}\left[-k,k\right]_{n+k}^{(2n+2k)}\,,\quad k\in\frac{1}{2}\mathbb{Z}\,,\quad k\geq\frac{1}{2}-n\,,\quad\mathcal{I}_{A_{1}\bar{A}_{1}\left[-k,k\right]_{n+k}^{(2n+2k)}}=(-1)^{2k}\xi^{2n}\,,
\end{equation}
where the two conditions on $k$ in the two middle equations follow from $R\in\mathbb{Z}_{+}$ and $\Delta>0$. We then obtain that the coefficient $+1$ of $\xi^{2n}$ in \eqref{I_expan} corresponds to the following conditions on the $A_{1}\bar{A}_{1}$ multiplet content for any $\lambda$, 
\begin{equation}
\sum_{\substack{
k\in\frac{1}{2}\mathbb Z\\
k\geq \frac{1}{2}-n
}}
(-1)^{2k}\,
\mathsf N\!\left(
A_{1}\bar A_{1}
\left[-k,k\right]_{n+k}^{(2n+2k)}
\right)
=1\,,
\quad n\geq1\,,
\label{AA_multiplicity_constraint}
\end{equation}
where $\mathsf N$ denotes the number of multiplets of the corresponding type (see \cite{Dorey:2018klg} for similar alternating sums in the context of 1d indices). Similarly, the absence of $\xi^{2n+1}$ terms in the index yields the conditions 
\begin{equation}
\sum_{\substack{
k\in\frac{1}{2}\mathbb Z\\
k\geq -n
}}
(-1)^{2k}\,
\mathsf N\!\left(
A_{1}\bar A_{1}
\left[-k,k\right]_{n+\frac12+k}^{(2n+1+2k)}
\right)
=0\,,
\quad n\geq0\,.
\label{AA_odd_multiplicity_constraint}
\end{equation}
Turning next to the terms $\eta^{-m}\xi^{m+2n}\left(1-\xi^{2}\right)$ in \eqref{I_expan} and using again \eqref{I_ALb}-\eqref{I_AAb}, the multiplets that contribute such terms are 
\begin{equation}
A_{1}\bar{L}\left[-m-k,k\right]_{m+n+k}^{(2n+2k)}\,,\quad m\geq1\,,\quad n\geq0\,,\quad k\in\frac{1}{2}\mathbb{Z}\,,\quad k\geq-n\,
\end{equation}
with 
\begin{equation}
\mathcal{I}_{A_{1}\bar{L}\left[-m-k,k\right]_{m+n+k}^{(2n+2k)}}=(-1)^{2k}\eta^{-m}\xi^{m+2n}\left(1-\xi^{2}\right)\,.
\end{equation}
This, in turn, results in the conditions 
\begin{equation}
\sum_{\substack{ k\in\frac{1}{2}\mathbb Z\\ k\geq-n }} (-1)^{2k}\, \mathsf N\!\left( A_{1}\bar L \left[-m-k,k\right]_{m+n+k}^{(2n+2k)} \right) =1\,, \quad m\geq1\,,\quad n\geq0\,.
\label{AL_multiplicity_constraint}
\end{equation}
Meanwhile, the absence of $\eta^{-m}\xi^{m+2n+1}\left(1-\xi^{2}\right)$ terms in the index results in 
\begin{equation}\label{AL0_multiplicity_constraint}
\sum_{\substack{k\in\frac{1}{2}\mathbb{Z}\\
k\geq-n-\frac{1}{2}
}
}(-1)^{2k}\,\mathsf{N}\!\left(A_{1}\bar{L}\left[-m-k,k\right]_{m+n+\frac{1}{2}+k}^{(2n+1+2k)}\right)=0\,,\quad m\geq1\,,\quad n\geq0\,.
\end{equation}
Notice that the minimal solution to all of the conditions above (compatible with our analysis at $\lambda=\infty$) is given by 
\begin{equation}\label{N_min_sol}
\mathsf{N}\!\left(A_{1}\bar{A}_{1}\left[0,0\right]_{n}^{(2n)}\right)=1\,,\quad\mathsf{N}\!\left(A_{1}\bar{L}\left[-m,0\right]_{m+n}^{(2n)}\right)=1\,,\quad m\geq1\,,\quad n\geq0
\end{equation}
with all the rest of the multiplicities vanishing, which corresponds to the simple index form\footnote{In \eqref{I_ISum} it is understood that the contribution of any short or semi-short multiplet with a nonzero multiplicity is written explicitly, even if some of them would cancel together.} 
\begin{equation}\label{I_ISum}
\mathcal{I}=1+\sum_{n=1}^{\infty}\mathcal{I}_{A_{1}\bar{A}_{1}\left[0,0\right]_{n}^{(2n)}}+\sum_{m=1}^{\infty}\sum_{n=0}^{\infty}\mathcal{I}_{A_{1}\bar{L}[-m,0]_{m+n}^{(2n)}}
\end{equation}
as can also be readily seen by using 
\begin{equation}
\mathcal{I}_{A_{1}\bar{A}_{1}\left[0,0\right]_{n}^{(2n)}}=\xi^{2n}\,,\quad\mathcal{I}_{A_{1}\bar{L}[-m,0]_{m+n}^{(2n)}}=\eta^{-m}\xi^{m+2n}\left(1-\xi^{2}\right)
\end{equation}
in \eqref{I_expan}. A possible scenario is that the multiplets in \eqref{N_min_sol} remain protected as $\lambda$ is varied, while additional short or semi-short multiplets present at $\lambda=0$ recombine into long multiplets at generic nonzero coupling (as also expected in the case without a boundary, see appendix \ref{app:no-boundary-index}). This scenario cannot, however, be inferred from the index alone, which determines only the alternating multiplicity constraints. Nevertheless, the spectra analyzed in the previous subsections give a stronger conclusion near the two endpoints. In the free strong-coupling limit, each of the multiplets in \eqref{N_min_sol} lacks the partners required by the recombination rules \eqref{eq:recombination_rules}, and therefore remains protected under sufficiently small deformations away from $\lambda=\infty$.
The same conclusion similarly holds near $\lambda=0$ for the $A_1\bar A_1[0,0]_n^{(2n)}$ and $A_1\bar L[-m,0]_m^{(0)}$ towers. As an example, let us consider the multiplet $A_1\bar A_1[0,0]_1^{(2)}$, whose superconformal primary is the moment map operator $\vec{\mu}$ of the boundary theory at $\lambda=0$ and the operator $\vec{\mathsf{R}}$ at $\lambda=\infty$. Specializing \eqref{eq:recombination_rules}
to this case (with our extra charge refinement) then gives 
\begin{equation}\label{eq:threshold-LL-A1A1}
\begin{aligned}
L\bar L[0,0]_{\Delta\to1}^{(2)}
={}& A_1\bar A_1[0,0]_1^{(2)}
\\
&\oplus A_1\bar A_1
  \left[-\tfrac12,+\tfrac12\right]_{3/2}^{(3)}
 \oplus A_1\bar A_1
  \left[+\tfrac12,-\tfrac12\right]_{3/2}^{(3)}
\\
&\oplus A_1\bar A_1[0,0]_2^{(4)}.
\end{aligned}
\end{equation}
The first and last multiplets are present at both endpoints, with the last having superconformal primary $\vec{\mu}^{\,\otimes_{\mathrm{s.t.}}2}$ at $\lambda=0$ and $\vec{\mathsf{R}}^{\,\otimes_{\mathrm{s.t.}}2}$ at $\lambda=\infty$ (where $\otimes_{\mathrm{s.t.}}2$ denotes the symmetric traceless square in the $SU(2)_X$ indices). However, the two middle multiplets, whose superconformal primaries are fermionic and have $R=3$, are absent at both endpoints. Thus the moment-map multiplet remains protected under sufficiently small deformations away from either endpoint. Note that this does not, however, establish protection throughout the full range of $\lambda$. 

Let us finally mention that even though the pure BPS Neumann boundary condition does not admit a controlled supergravity description for the boundary Wilson line at strong coupling (since the curvature is large at the location of the dual string), the above index discussion might suggest a simple worldsheet theory for the dual string. To see this, note that for pure Neumann boundary condition (i.e. no boundary theory is present) the total single-letter line index at $\lambda=0$ is given by \eqref{i_s.l._weak}, such that the full index is\footnote{This result also follows from the large-$N$ limit of the fundamental Wilson-line half-index computed in \cite{Hatsuda:2025yzp}, normalized by the half-index without line insertions.} 
\begin{equation}\label{I_N}
\mathcal{I}_{\textrm{Neumann}}\left(\eta,\xi\right)=\frac{1}{1-i_{\textrm{s.l.}}^{\textrm{weak,bulk}}(\eta,\xi)}=\frac{1-\eta^{-1}\xi^{3}}{1-\eta^{-1}\xi}\,.
\end{equation}
Comparing with \eqref{i_s.l.}-\eqref{I_result}, we observe that this result is equal to the index obtained for a spectrum which is freely generated by the displacement multiplets only, i.e. 
\begin{equation}\label{I_N_displ}
\mathcal{I}_{\textrm{Neumann}}\left(\eta,\xi\right)=\textrm{PE}\left[\mathcal{I}_{A_{1}\bar{L}\left[-1,0\right]_{1}^{(0)}}\left(\eta,\xi\right)\right].
\end{equation}
This index result is therefore consistent with the Wilson line on the Neumann boundary being described at $\lambda=\infty$ by a free bulk worldsheet theory corresponding to perturbations of a dual string wrapping an $AdS_{2}\subset AdS_{4}$ along the $AdS_{4}\times S^{2}$ directions only. This, in turn, corresponds to the free theory in \eqref{L2} written for the $(\delta\widetilde{\bar{x}}^{\,i},\delta\widetilde{\theta}_{2},\delta\widetilde{\phi}_{2})$ fields only. This description is of course only suggestive by the index result \eqref{I_N_displ}, which however serves as a nontrivial piece of evidence for it. Note, in particular, that given the line index \eqref{I_N}, the free-theory description in terms of $(\delta\widetilde{\bar{x}}^{\,i},\delta\widetilde{\theta}_{2},\delta\widetilde{\phi}_{2})$ is the minimal one needed in order to match this index (with any additional sector necessarily having a vanishing contribution to the index). Moreover, this description is the exact analogy of the free theory associated with the same Wilson line in the case with no boundary. Indeed, as we review in appendix \ref{app:no-boundary-index}, the index of the Maldacena-Wilson line in $\mathcal{N}=4$ SYM with no boundary can similarly be written as the plethystic exponential of the index of the displacement multiplet, which corresponds to the familiar holographic description at strong coupling in terms of a free worldsheet theory of string perturbations along the $AdS_{5}\times S^{5}$ directions. 

Let us also comment briefly on the supersymmetry of such a fundamental string which is assumed to propagate in a six-dimensional spacetime and contain the displacement multiplets as the only generators of its spectrum, as suggested by the index computation. For simplicity, we consider the limit $\lambda\rightarrow\infty$ (or alternatively take the flat-space limit) such that the fluctuations $\delta\tilde{\bar{x}}^i$ have vanishing mass, resulting in all four bosons $(\delta\tilde{\bar{x}}^i,\delta\tilde{\theta}_2,\delta\tilde{\phi}_2)$ becoming identical. In this case, the rotation symmetry transverse to the worldsheet is further enhanced from $U(1)_Y\times U(1)_{\text{rot}}$ to $SO(4)$. 
As such, we introduce the six-dimensional gamma matrices in a basis suited for a straight string in static gauge $(\xi^0,\xi^1) = (X^0,X^1)$, 
\begin{equation}
    \Gamma^a = \rho^a\otimes\mathbb{1}_4\quad a=0,1\,,\quad \Gamma^i = \rho^{(2)}_\star\otimes \gamma^i\,,\quad i=2,\dots,5\,,
\end{equation}
where $\rho^a$ and $\gamma^i$ are the $SO(1,1)$ and $SO(4)$ gamma matrices, respectively, and $\rho^{(2)}_\star = \rho^0\rho^1$ corresponds to the 2d chirality matrix.\footnote{For even spacetime dimensions $d=2m$, the Lorentzian chirality matrix in mostly-plus signature is given by $\Gamma^{(L)}_\star = (-i)^{m-1}\Gamma^0\cdots\Gamma^{d-1}$ and the Euclidean chirality matrix by $\Gamma^{(E)}_\star = (-i)^m\Gamma_{(E)}^0\cdots\Gamma_{(E)}^{d-1}$.} Then, the generators of $U(1)_{\text{rot}}$ and $U(1)_Y$ on spinors are given by 
\begin{gather}
    J_{\text{rot}} = \frac{i}{2}\Gamma^2\Gamma^3 = \frac{i}{2}\big(\mathbb{1}_2\otimes\gamma^2\gamma^3\big)\,,\quad J_Y=\frac{i}{2}\Gamma^4\Gamma^5 = \frac{i}{2}\big(\mathbb{1}_2\otimes\gamma^4\gamma^5\big)\, ,\\
    J_{\text{rot}}J_Y  = \frac{1}{4}\big(\mathbb{1}_2\otimes\gamma^{(4)}_\star\big)
\end{gather}
where $\gamma^{(4)}_\star = -\gamma^2\cdots\gamma^5$ is the $SO(4)$ chirality matrix, and recall that the CFT$_1$ fermions $\mathsf{\Psi}_a^{++}$ and $\mathsf{\Psi}^{a --}$ in the displacement multiplets have charges $\left[+1/2,+1/2\right]$ and $\left[-1/2,-1/2\right]$ under $U(1)_Y\times U(1)_{\text{rot}}$. In the Green-Schwarz formalism (see \cite{GSW:2012superstring} for more details), this means that the corresponding six-dimensional spacetime fermions, which inherit opposite worldsheet chiralities after $\kappa$-symmetry gauge-fixing and are denoted by $S_1$ and $S_2$, satisfy
\begin{equation}
    \big(\mathbb{1}_2\otimes\gamma^{(4)}_\star\big)S_{1,2} = +S_{1,2}\, .
\end{equation}
Then, together with $\big(\rho^{(2)}_\star\otimes\mathbb{1}_4\big)S_1 = -S_1$ and $\big(\rho^{(2)}_\star\otimes\mathbb{1}_4\big)S_2 = +S_2$ this implies that their 6d chirality is
\begin{equation}
    \left\{
    \begin{IEEEeqnarraybox}[\IEEEeqnarraystrutmode][c]{l}
         \Gamma^{(6)}_\star S_1 = -S_1 \\[2pt]
         \Gamma^{(6)}_\star S_2 = +S_2
    \end{IEEEeqnarraybox}
    \right.
   \quad ,\quad \Gamma^{(6)}_\star = -\Gamma^0\dots\Gamma^5 = \rho^{(2)}_\star\otimes\gamma^{(4)}_\star\, , 
\end{equation}
i.e. that they arise from a superstring with $\mathcal{N}=(1,1)_6$.

\section{Mapping operators from weak to strong coupling}
\label{sec:map}

In the previous section, we have discussed the two very different descriptions we have for our setup at strong and weak coupling, and illustrated how various multiplets of the symmetry preserved by the line on the boundary are realized. In this section, assuming no level-crossing takes place as $\lambda$ is varied, we would find how some of the operators in these two regimes are mapped to each other. 
Since at strong coupling, the supergravity description at the location of the dual string is only valid for $N_{\textrm{NS5}}\gg1$, and since a free worldsheet theory at $\lambda=\infty$ is only obtained for infinite $N_{\textrm{NS5}}$ (corresponding to $\alpha<1$), we will focus on the latter case in what follows. We will in addition comment throughout on the case where the boundary condition is the pure BPS Neumann one (corresponding to $N_{\textrm{NS5}}=1$), following our discussion below Eq. \eqref{I_N_displ}. 

\begin{figure}
    \centering
    \begin{subfigure}{0.495\textwidth}
        \centering
        \begin{tikzpicture}[anchor=south,baseline]
    \coordinate (O) at (0,0);
    \coordinate (X) at (4,0);
    \coordinate (Y) at (0,4.75);
    \coordinate (Z) at (4,5);
    \node at (2,5) {\underline{$r=0$}};
    \foreach \y in {1.5,3,4.5}
        \draw[thick] (-0.1,\y) -- (0.1,\y) (3.9,\y) -- (4.1,\y);
    \draw[thick] (Y) node[fill=white,yshift=-0.25cm]{$\Delta$} -- (O) -- node[midway,below]{$\lambda$} (X) -- (Z);
        
    \node[align=right,anchor=east] at (0,1.5) {\footnotesize \textcolor{myblue}{$\vec{X}$}\ \textcolor{myred}{$\Phi_6$}};
    \node[align=left,anchor=west] at (4,1.5) {\footnotesize \textcolor{myblue}{$\vec{\mathsf{R}}$}};
    
    \begin{pgfonlayer}{bg}

        \draw[myblue,thick] (0,1.5) -- (0.75,1.5);
        \draw[myblue,thick] (3.25,1.5) -- (4,1.5);
        \draw[myblue,thick,dashed] (0.75,1.5) to [curve through={(1.25,1.4) (2.25,1.6)}] (3.25,1.5);
    
    \end{pgfonlayer}

    \node[align=right,anchor=east] at (0,3) {\footnotesize \textcolor{myblue}{$\Tilde{\mathbb{F}}_n$}\ \textcolor{myorange}{$\vec{X}^{\otimes_{\text{s.t.}} 2}$}\\
    \footnotesize \textcolor{myred}{$F_{12}$}};
    \node[align=left,anchor=west] at (4,3) {\footnotesize \textcolor{myblue}{$\mathsf{Y}$}\ \textcolor{myorange}{$\vec{\mathsf{R}}^{\otimes_{\text{s.t.}} 2}$}\\
    \footnotesize \textcolor{myred}{$b_1\mathsf{\Theta}_+\mathsf{\Theta}_- + b_2\mathsf{R}^n\mathsf{R}^n$}};
    \node[align=left,anchor=west] at (4,4.5) {\footnotesize \textcolor{myred}{$\mathsf{A}$}};
    
    \begin{pgfonlayer}{bg}
     
        \draw[myblue,thick,dash pattern=on 6pt off 6pt] (0,3) -- (0.75,3);
        \draw[myblue,thick,dash pattern=on 6pt off 6pt] (3.25,3) -- (4,3);
        \draw[myblue,thick,dashed] (0.75,3) to [curve through={(1.25,2.9) (2.25,3.1)}] (3.25,3);
        \draw[myorange,thick,dash pattern=on 6pt off 6pt,dash phase=6pt] (0,3) -- (0.75,3);
        \draw[myorange,thick,dash pattern=on 6pt off 6pt,dash phase=6pt] (3.25,3) -- (4,3);
        \draw[myorange,thick,dashed] (0.75,3) to [curve through={(2,3.15) (3,3)}] (3.25,3); 
        
        \path[draw,myred,thick,dashed]
            (0,1.5) to [curve through={(1.25,2) (1.5,2.25) (3,2.75)}] (4,3)
            (0,3) to [curve through={(1.25,3.5) (1.5,3.75) (3,4.25)}] (4,4.5);

    \end{pgfonlayer}

    \draw[myred,thick] (3.5,6.5) -- (3.75,6.5)node[right]{\footnotesize \textcolor{myred}{$L\bar{L}\left[0,0\right]_1^{(0)}$}};
    \draw[myblue,thick] (3.5,6) -- (3.75,6)node[right]{\footnotesize \textcolor{myblue}{$A_1\bar{A}_1\left[0,0\right]_1^{(2)}$}};
    \draw[myorange,thick] (3.5,5.5) -- (3.75,5.5)node[right]{\footnotesize \textcolor{myorange}{$A_1\bar{A}_1\left[0,0\right]_2^{(4)}$}};
    
\end{tikzpicture}
        \caption{}
    \end{subfigure}
    \begin{subfigure}{0.495\textwidth}
        \centering
        \begin{tikzpicture}[anchor=south,baseline]
    \coordinate (O) at (0,0);
    \coordinate (X) at (4,0);
    \coordinate (Y) at (0,4.75);
    \coordinate (Z) at (4,5);
    \node at (2,5) {\underline{$|r|=1$}};
    \foreach \y in {1.5,3,4.5}
        \draw[thick] (-0.1,\y) -- (0.1,\y) (3.9,\y) -- (4.1,\y);
    \draw[thick] (Y) node[fill=white,yshift=-0.25cm]{$\Delta$} -- (O) -- node[midway,below]{$\lambda$} (X) -- (Z);
        
    \node[align=right,anchor=east] at (0,1.5) {\footnotesize \textcolor{myblue}{$\Phi_\pm$}};
    \node[align=left,anchor=west] at (4,1.5) {\footnotesize \textcolor{myblue}{$\mathsf{\Theta}_\pm$}};
    
    \begin{pgfonlayer}{bg}

        \draw[myblue,thick] (0,1.5) -- (4,1.5);
    
    \end{pgfonlayer}

    \node[align=right,anchor=east] at (0,3) {\footnotesize \textcolor{myblue}{$\Tilde{\mathbb{F}}_\pm$}};
    \node[align=left,anchor=west] at (4,3) {\footnotesize \textcolor{myblue}{$\mathsf{\bar{X}}_\pm$}};
    
    \begin{pgfonlayer}{bg}
     
        \draw[myblue,thick] (0,3) -- (4,3);

    \end{pgfonlayer}

    \draw[myblue,thick] (3.5,6) -- (3.75,6)node[right,text width=2cm]{\footnotesize \textcolor{myblue}{$L\bar{A}_1\left[+1,0\right]_1^{(0)} \big/$}\\
    \footnotesize \textcolor{myblue}{$A_1\bar{L}\left[-1,0\right]_1^{(0)}$}};
    
\end{tikzpicture}
        \caption{}
    \end{subfigure}
    \begin{subfigure}{0.495\textwidth}
        \centering
        \begin{tikzpicture}
    \coordinate (O) at (0,0);
    \coordinate (X) at (4,0);
    \coordinate (Y) at (0,4.75);
    \coordinate (Z) at (4,5);
    \node at (2,5) {\underline{$|r|=2$}};
    \foreach \y in {1.5,3,4.5}
        \draw[thick] (-0.1,\y) -- (0.1,\y) (3.9,\y) -- (4.1,\y);
    \draw[thick] (Y) node[fill=white,yshift=-0.25cm]{$\Delta$} -- (O) -- node[midway,below]{$\lambda$} (X) -- (Z);

    \node[align=right,anchor=east] at (0,3) {\footnotesize \textcolor{myblue}{$\Phi_\pm^2$}\ \textcolor{myred}{$\partial_\pm\Phi_\pm$}};
    \node[align=left,anchor=west] at (4,3) {\footnotesize \textcolor{myblue}{$\Theta_\pm^2$}};
    \node[align=left,anchor=west] at (4,4.5) {\footnotesize \textcolor{myred}{$\mathsf{\Theta}_\pm\mathsf{\bar{X}}_\pm - \frac{3}{4}\epsilon^{ab}\mathsf{\Psi}_a^{\pm\pm}\mathsf{\Psi}_b^{\pm\pm}$}};

    \begin{pgfonlayer}{bg}

        \draw[myblue,thick] (0,3) -- (0.75,3);
        \draw[myblue,thick] (3.25,3) -- (4,3);
        \draw[myblue,thick,dashed] (0.75,3) to [curve through={(2,3.05) (3,2.9)}] (3.25,3);
        
        \draw[myred,thick,dashed] (0,3) to [bend left,in=170,out=0,looseness=1] (4,4.5);

    \end{pgfonlayer}

    \draw[myred,thick] (3.5,6.25) -- (3.75,6.25)node[right,text width=2cm]{\footnotesize \textcolor{myred}{$L\bar{A}_1\left[+1,+1\right]_2^{(0)} \big/A_1\bar{L}\left[-1,-1\right]_2^{(0)}$}};
    \draw[myblue,thick,text width=2cm] (3.5,5.75) -- (3.75,5.75)node[right]{\footnotesize \textcolor{myblue}{$L\bar{A}_1\left[+2,0\right]_2^{(0)} \big/ A_1\bar{L}\left[-2,0\right]_2^{(0)}$}};
    
\end{tikzpicture}
        \caption{}
    \end{subfigure} 
    \caption{The dimensions $\Delta$ of the conformal primary operators of $U(1)_{\mathcal{R}_2}$ charge $r\in\{0;\pm1;\pm2\}$ discussed in this section as a function of the 't Hooft coupling $\lambda$. Operators of the same color within a plot belong to the same supermultiplet, which is indicated in the corresponding legend. A horizontal line segment signals protectedness of an operator for some range of $\lambda$, while a dashed curve represents a possibly unprotected domain.}
    \label{fig:operator-mapping}
\end{figure}
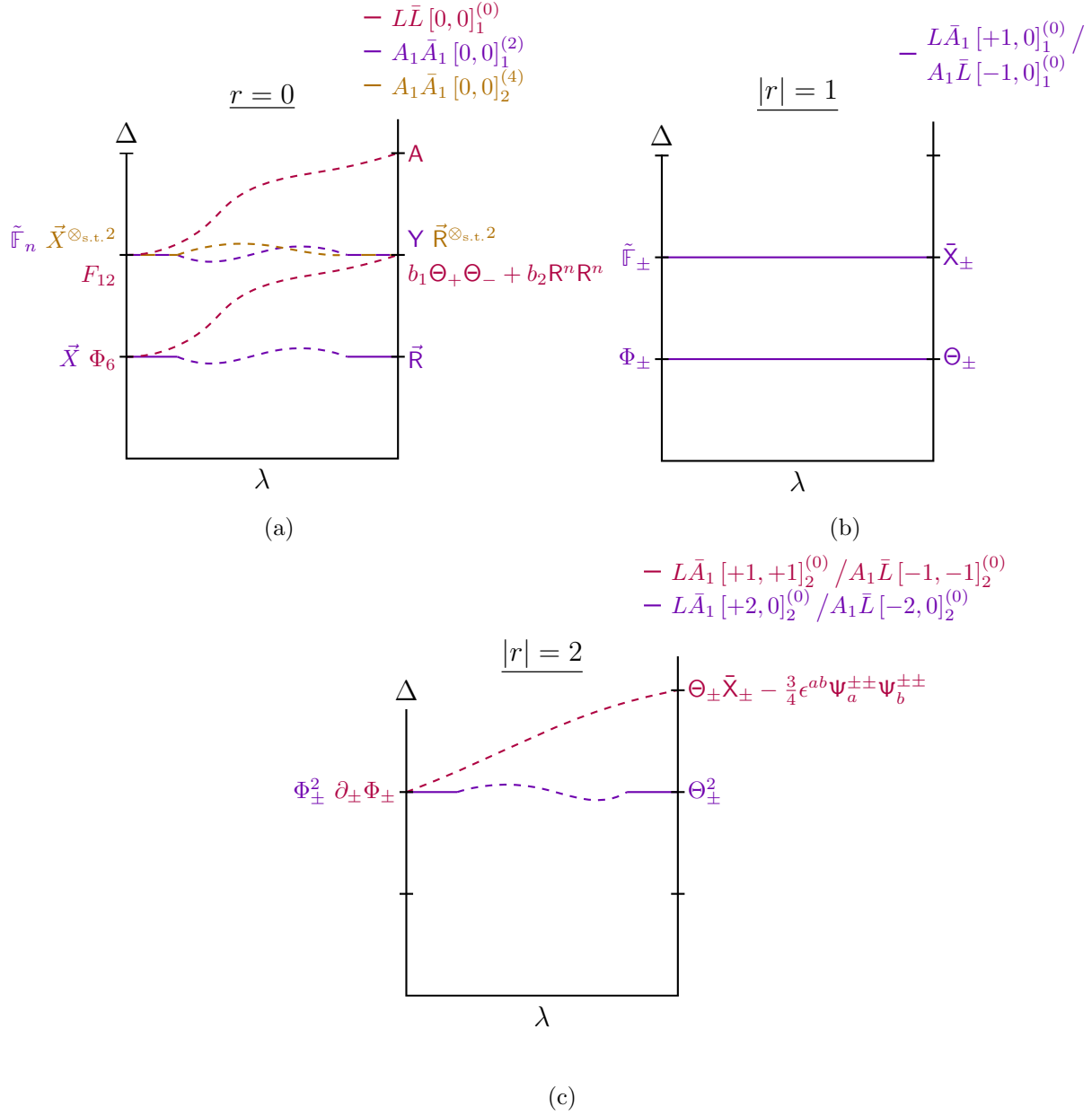

The simplest starting point is given by considering the displacement multiplets $L\bar{A}_1\left[+1,0\right]_1^{(0)}$ and $A_1\bar{L}\left[-1,0\right]_1^{(0)}$, which are associated with currents broken by the line and are necessarily protected upon varying $\lambda$. As a result, the multiplets at $\lambda=0$ are straightforwardly mapped to the corresponding ones at $\lambda=\infty$. In particular, their bottom components (which are the tilt operators for the $R$-symmetry broken by the line) and top components (which are the displacement operators associated with broken translations) map as 
\begin{equation}\label{displ_map}
\Phi_{\pm}\longrightarrow\mathsf{\Theta}_{\pm}\,,\quad\mathbb{\tilde{F}}_{\pm}\longrightarrow\mathsf{\bar{X}}_{\pm}\,,
\end{equation}
where recall that $\mathbb{\tilde{F}}_{\pm}=iF_{t\pm}+\partial_{\pm}\Phi_{6}$ (see \eqref{displ_operator}) while $\mathsf{\Theta}_{\pm}$ and $\mathsf{\bar{X}}_{\pm}$ are the line operators dual to the worldsheet fields $(\delta\widetilde{\theta}_{2},\delta\widetilde{\phi}_{2})$ and $\delta\widetilde{\bar{x}}^{\,i}$, respectively. Note also that in the strong-coupling description suggested for the line on a pure Neumann boundary below Eq. \eqref{I_N_displ} (i.e. for $N_{\textrm{NS5}}=1$), the same map applies. The behavior of the scaling dimensions of these operators as a function of $\lambda$, although trivial, is depicted in figure \ref{fig:operator-mapping}.b.

The next simplest multiplet to consider is $A_1\bar{A}_1\left[0,0\right]_1^{(2)}$, which at strong coupling generates the spectrum together with the displacement multiplets (see \eqref{chi_total_strong} and the discussion around it). At weak coupling, its bottom component is the so-called moment-map operator $\vec{\mu}$ associated with the Higgs-branch $SU(N)$ global symmetry of the boundary $T_{[K^{N/K}]}[SU(N)]$ theory, which is defined as the super-primary of the corresponding 3d $\mathcal{N}=4$ current multiplet. Its suggested map to strong coupling is therefore 
\begin{equation}\label{mu_map}
\vec{\mu}\longrightarrow\vec{\mathsf{R}}\,,
\end{equation}
where recall that $\vec{\mathsf{R}}$ is the line operator dual to the worldsheet fields $\delta\vec{\widetilde{r}}$, and is the bottom component of $A_1\bar{A}_1\left[0,0\right]_1^{(2)}$ at $\lambda=\infty$. Note that the dimensions of both operators in \eqref{mu_map} are the same (equal to $1$), which is consistent with the $A_1\bar{A}_1\left[0,0\right]_1^{(2)}$ multiplet being protected along the flow. However, we did not assume such protectedness in \eqref{mu_map}. Notice also that at weak coupling the boundary condition for the bulk scalar $\vec{X}$, which is $\vec{X}|_{x^3=0}=0$ at $\lambda=0$, turns into $\vec{X}|_{x^3=0}=-\vec{\mu}$ outside this point \cite{Gaiotto:2009sbc}. As a result, the map \eqref{mu_map} can also be viewed as $\vec{X}\longrightarrow\vec{\mathsf{R}}$ such that $\delta\vec{\widetilde{r}}$ is essentially the bulk worldsheet dual of $\vec{X}$ at strong coupling. Similarly, the line operator $\mathsf{Y}$ (lying at the top of $A_1\bar{A}_1\left[0,0\right]_1^{(2)}$ at $\lambda=\infty$) dual to the worldsheet field $\delta\widetilde{y}$ corresponds at weak coupling to the level-two descendant of $\vec{\mu}$ in $A_1\bar{A}_1\left[0,0\right]_1^{(2)}$, which for $\lambda>0$ can be viewed as $\mathbb{\tilde{F}}_{n}|_{x^3=0}$ where $\mathbb{\tilde{F}}_{n}=iF_{tn}+D_{n}\Phi_{6}$ (as follows from the boundary condition).\footnote{At $\lambda=0$ this level-two descendant is given by $\mathcal{O}_{Y}^{(3)}-iJ_{t}$, where $J_t$ is the component along the line of the $SU(N)$ current of the boundary theory and $\mathcal{O}_{Y}^{(3)}$ the $\mathfrak{u}(1)_{Y}$-neutral component of the $\mathfrak{su}(2)_{Y}$ triplet $\mathcal{O}_{Y}^A$ appearing with $J_i$ at level two in the 3d $\mathcal{N}=4$ current multiplet. At $\lambda>0$ the boundary conditions $F_{nt}/g^2_{YM}=J_t$, $D_{n}\Phi_{6}/g^2_{YM}=\mathcal{O}_{Y}^{(3)}$ can then be used to write this operator as $\mathbb{\tilde{F}}_{n}/g^2_{YM}$.}
Let us finally comment that in the case of the pure Neumann boundary condition, no boundary theory exists (as well as no fields $\delta\widetilde{y}$ and $\delta\vec{\widetilde{r}}$ in the suggested strong-coupling description below Eq. \eqref{I_N_displ}) such that the map \eqref{mu_map} is absent. The scaling dimensions of the bosonic operators in the $A_1\bar{A}_1\left[0,0\right]_1^{(0)}$ multiplet are presented schematically as a function of $\lambda$ in figure \ref{fig:operator-mapping}.a by the purple curves. Following the discussion around Eq. \eqref{eq:threshold-LL-A1A1}, these dimensions are constant near the endpoints $\lambda=0$ and $\lambda=\infty$ due to the absence of some of the semi-short multiplets required for the corresponding recombination rule.

We can generalize the maps \eqref{displ_map} and \eqref{mu_map} by taking their powers in such a way that the operator on each side of the resulting maps continues to be the unique lightest operator in the corresponding symmetry channel. For the map \eqref{displ_map} this results in 
\begin{equation}\label{Phi_n_map}
\left(\Phi_{\pm}\right)^{n}\longrightarrow\left(\mathsf{\Theta}_{\pm}\right)^{n}\,,\quad n\geq1\,,
\end{equation}
which are the superconformal primaries of the multiplets $A_{1}\bar{L}[-n,0]_{n}^{(0)}$ and $L\bar{A}_{1}[n,0]_{n}^{(0)}$ at weak and strong coupling, i.e. 
\begin{equation}\label{Phi_pm_2}
\left(\Phi_{+}\right)^{n},\left(\mathsf{\Theta}_{+}\right)^{n}\in L\bar{A}_{1}[n,0]_{n}^{(0)}\,,\quad\left(\Phi_{-}\right)^{n},\left(\mathsf{\Theta}_{-}\right)^{n}\in A_{1}\bar{L}[-n,0]_{n}^{(0)}
\end{equation}
such that in fact the whole multiplets map in the same way. Note again that the dimensions of the two operators in \eqref{Phi_n_map} are equal, which is consistent with the multiplets $A_{1}\bar{L}[-n,0]_{n}^{(0)}$ and $L\bar{A}_{1}[n,0]_{n}^{(0)}$ continuing to be protected along the flow (i.e. not recombining with other semi-short multiplets to form long ones). This, in turn, is reminiscent of the protectedness of the $B_{1}\left[0\right]_{n}^{\left(0,n\right)}$ multiplets (in the notations of \cite{Nathan:2020scd}), among which $n=1$ corresponds to the displacement multiplet, in the case of a 1/2-BPS Wilson line in $\mathcal{N}=4$ SYM without a boundary (see appendix \ref{app:no-boundary-index}). In the latter case, though, the multiplets $B_{1}\left[0\right]_{n}^{\left(0,n\right)}$ are fully protected by representation theory, in the sense that they do not appear in any recombination rule. The case $n=2$ of (\ref{Phi_n_map}) is shown by the purple lines in figure \ref{fig:operator-mapping}.c where, once again, the dimension is constant as a function of $\lambda$ at the weak and strong coupling regimes.

Similarly, the generalization of the map \eqref{mu_map} is given by\footnote{In the boundary $T_{[K^{N/K}]}[SU(N)]$ theory, a complex highest-weight component $\mu$ of the Higgs-branch
moment map obeys $\mu^{N/K}=\mu^{N_{\textrm{NS5}}}=0$
in the chiral ring \cite{Gaiotto:2009sdua}, with powers
understood as matrix multiplication in flavor indices.
This nilpotency relation does not truncate the tower above at any fixed operator degree in the limit
$N_{\textrm{NS5}}\to\infty$ considered here.} 
\begin{equation}\label{mu_n_map}
\vec{\mu}^{\,\otimes_{\textrm{s.t.}}m}\longrightarrow\vec{\mathsf{R}}^{\,\otimes_{\textrm{s.t.}}m}\,,\quad m\geq1
\end{equation}
where $\otimes_{\textrm{s.t.}}m$ stands for the $m$-th symmetric traceless power. These operators are the bottom components of the multiplets $A_{1}\bar{A}_{1}\left[0,0\right]_{m}^{(2m)}$, again with the same dimension at weak and strong coupling (which is consistent with them being protected). See the orange line in figure \ref{fig:operator-mapping}.a for a depiction of the $\lambda$ dependence of the dimension in the $m=2$ case. 

Let us next consider an example in which the dimensions of the operators on the two sides of the map are different (i.e. the operator at $\lambda=0$ develops an anomalous dimension when $\lambda$ is increased). Let us examine $\partial_{\pm}\Phi_{\pm}$ (with the signs correlated), which are the superconformal primaries of the short multiplets $L\bar{A}_{1}\left[+1,+1\right]_{2}^{\left(0\right)}$ and $A_{1}\bar{L}\left[-1,-1\right]_{2}^{\left(0\right)}$, i.e. 
\begin{equation}
\partial_{+}\Phi_{+}\in L\bar{A}_{1}\left[+1,+1\right]_{2}^{\left(0\right)}\,,\quad\partial_{-}\Phi_{-}\in A_{1}\bar{L}\left[-1,-1\right]_{2}^{\left(0\right)}\,.
\end{equation}
At $\lambda=\infty$ there are no operators of dimension two in the same symmetry channels (i.e. $\left[r_{Y},r_{\text{rot}}\right]^{(R)}=\left[\pm1,\pm1\right]^{(0)}$), and hence these multiplets cannot be protected for all $\lambda$. Instead, they must recombine into long multiplets. The lightest operators at $\lambda=\infty$ in these symmetry channels are the dimension-three operators $\mathsf{\bar{X}}_{\pm}\mathsf{\Theta}_{\pm}$ and $\epsilon^{ab}\mathsf{\Psi}_{a}^{\pm\pm}\mathsf{\Psi}_{b}^{\pm\pm}$, where $\mathsf{\Psi}_{a}^{\pm\pm}\in\left[\pm1/2,\pm1/2\right]_{3/2}^{(1)}$ are the level-one super-descendants of $\Theta_{\pm}$ in the displacement multiplets, 
\begin{equation}\label{Psi_pmpm}
\mathsf{\Psi}_{a}^{++}=\mathcal{Q}_{a}\Theta_{+}\,,\quad\mathsf{\Psi}^{a--}=\mathcal{\bar{Q}}^{a}\Theta_{-}\,,\quad\mathsf{\bar{X}}_{+}=\mathcal{Q}_{1}\mathcal{Q}_{2}\Theta_{+}\,,\quad\mathsf{\bar{X}}_{-}=\mathcal{\bar{Q}}^{1}\mathcal{\bar{Q}}^{2}\Theta_{-}\,.
\end{equation}
As a result, assuming no level-crossing takes place, $\partial_{\pm}\Phi_{\pm}$ flow at strong coupling to some linear combination of $\mathsf{\bar{X}}_{\pm}\mathsf{\Theta}_{\pm}$ and $\epsilon^{ab}\mathsf{\Psi}_{a}^{\pm\pm}\mathsf{\Psi}_{b}^{\pm\pm}$, which we denote by $\mathcal{O}_\pm$. To find the specific combination, we use the fact that $\mathcal{O}_\pm$ are superconformal primaries (of long multiplets) and are therefore annihilated by the conformal supercharges $\mathcal{{S}}_{a}$, $\mathcal{\bar{S}}^{a}$. Using the anti-commutators in \eqref{QSb} and their conjugates, as well as the explicit relations \eqref{Psi_pmpm} it is then easy to check that the combination that satisfies these conditions is given by 
\begin{equation}
\mathcal{O}_{\pm}=\Theta_{\pm}\mathsf{\bar{X}}_{\pm}-\frac{3}{4}\epsilon^{ab}\mathsf{\Psi}_{a}^{\pm\pm}\mathsf{\Psi}_{b}^{\pm\pm}\,,\quad\mathcal{S}_{a}\mathcal{O}_{\pm}=\mathcal{\bar{S}}^{a}\mathcal{O}_{\pm}=0\,.
\end{equation}
Meanwhile, the orthogonal combination corresponds to the level-two descendants of $(\Theta_{\pm})^2$ in $L\bar{A}_{1}\left[+2,0\right]_{2}^{\left(0\right)}$ and $A_{1}\bar{L}\left[-2,0\right]_{2}^{\left(0\right)}$ (see \eqref{Phi_pm_2}), 
\begin{equation}
\mathcal{Q}_{1}\mathcal{Q}_{2}\left(\Theta_{+}^{2}\right)=2\Theta_{+}\mathsf{\bar{X}}_{+}+\epsilon^{ab}\mathsf{\Psi}_{a}^{++}\mathsf{\Psi}_{b}^{++}\,,\quad\mathcal{\bar{Q}}^{1}\mathcal{\bar{Q}}^{2}\left(\Theta_{-}^{2}\right)=2\Theta_{-}\mathsf{\bar{X}}_{-}+\epsilon^{ab}\mathsf{\Psi}_{a}^{--}\mathsf{\Psi}_{b}^{--}\,.
\end{equation}
We therefore see that the operators $\partial_{\pm}\Phi_{\pm}$ begin at $\lambda=0$ as the super-primaries of short multiplets and end up at $\lambda=\infty$ as the super-primaries of long multiplets, with the map given by 
\begin{equation}\label{dPhi_map}
\partial_{\pm}\Phi_{\pm}\longrightarrow\Theta_{\pm}\mathsf{\bar{X}}_{\pm}-\frac{3}{4}\epsilon^{ab}\mathsf{\Psi}_{a}^{\pm\pm}\mathsf{\Psi}_{b}^{\pm\pm}
\end{equation}
and with the dimensions increasing from $\Delta=2$ to $\Delta=3$ (see figure \ref{fig:operator-mapping}.c). In the pure Neumann case and its conjectured strong-coupling description of the line, the map \eqref{dPhi_map} remains unchanged. 

Let us finally consider the example of the simple operator $\Phi_6\in L\bar{L}\left[0,0\right]_1^{(0)}$, which is a superconformal primary of a long multiplet already at $\lambda=0$. Here again, there is no operator of dimension one in the singlet symmetry channel $\left[0,0\right]^{(0)}$ at $\lambda=\infty$, and the next lightest operators are $\mathsf{\Theta}_+\mathsf{\Theta}_-$ and $\mathsf{R}^n\mathsf{R}^n$, of dimension two. The corresponding map is therefore given by 
\begin{equation}\label{Phi6_map}
\Phi_{6}\longrightarrow b_{1}\,\mathsf{\Theta}_{+}\mathsf{\Theta}_{-}+b_{2}\,\mathsf{R}^{n}\mathsf{R}^{n}\,,
\end{equation}
for some coefficients $b_{1,2}$, across which the dimension increases from $\Delta=1$ to $\Delta=2$. We will not determine $b_{1,2}$ here, but let us mention a natural possibility based on the analogy with the Wilson line without a boundary. In the no-boundary case, $\Phi_{6}$ maps to $z^Az^A$, where $z^A$ ($A=1,\dots,5$) is the bottom component of the displacement multiplet at $\lambda=\infty$ (of dimension one) \cite{Giombi:2017cqn,Julius:2026rosetta}. At weak-coupling, $z^A$ corresponds to $\Phi_{1,\dots,5}$, which upon imposing the boundary conditions of our setup turns (at $\lambda>0$) $\Phi_{1,2,3}$ to $\vec{\mu}$ while keeping $\Phi_{4,5}$ unchanged as $\Phi_{\pm}$. At strong coupling, $\vec{\mu}$ and $\Phi_{\pm}$ flow to $\vec{\mathsf{R}}$ and $\Theta_{\pm}$, respectively (see \eqref{displ_map}-\eqref{mu_map}), both of the same dimension $\Delta=1$. Identifying them as an $SO(5)$ vector $\mathsf{Z}^{A}$, i.e. setting 
\begin{equation}
\Theta_{\pm}=\mathsf{Z}^{1}\pm i\,\mathsf{Z}^{2}\,,\quad\mathsf{R}^{n}=\mathsf{Z}^{n+2}\,,
\end{equation}
and treating $\mathsf{Z}^{A}$ as the analog of $z^A$ (or its boundary restriction), suggests that $\Phi_{6}$ maps to the $SO(5)$ singlet $\mathsf{Z}^{A}\mathsf{Z}^{A}$ as in the no-boundary case. This then results in  
\begin{equation}
\Phi_{6}\longrightarrow\mathsf{Z}^{A}\mathsf{Z}^{A}=\mathsf{\Theta}_{+}\mathsf{\Theta}_{-}+\mathsf{R}^{n}\mathsf{R}^{n}\,,
\end{equation}
or $b_1=b_2$ in \eqref{Phi6_map}, as the map motivated by the analogy to the no-boundary case. The scaling dimension of $\Phi_6$ as a function of $\lambda$ is displayed in figure \ref{fig:operator-mapping}.a, alongside that  of the following level-two descendant (see figure \ref{fig:LL-multiplet-weak} for a recollection of the structure of the supermultiplet),\footnote{The other $\mathcal{Q}\mathcal{\bar{Q}}$ combination is $\{\mathcal{Q}_{a},\mathcal{\bar{Q}}^{b}\}=\delta_{a}^{b}P_{t}$, which does not yield a conformal primary.}
\begin{equation}\label{F12_Phi6}
F_{12}=\frac{1}{2}\left[\mathcal{Q}_{a},\mathcal{\bar{Q}}^{a}\right]\Phi_{6}\,.
\end{equation}
In four-dimensional confining gauge theories, lattice spectra provide evidence for a massive pseudoscalar
excitation on the QCD flux tube which is odd under both transverse and longitudinal parity, conventionally denoted by $0^{--}$ and referred to
as the worldsheet axion \cite{Dubovsky:2013axion,Dubovsky:2015axion,Dubovsky:2025axion}. In pure Yang-Mills, transverse parity $\mathbf{P}$ consists of a reflection of one of the coordinates transverse to the worldsheet swept by the flux tube, here chosen to be $x^1$, also sending the gauge field component $A_1$ to $-A_1$. On the other hand, longitudinal parity $\mathbf{CT}$ combines a time reversal transformation $t\rightarrow -t$, $A_0\rightarrow -A_0$ with charge conjugation $A_\mu\rightarrow -A_\mu^T$. For $\mathcal{N}=4$ SYM, these parities also carry onto the other members of the vector multiplet, such that they correspond to discrete symmetries of the BPS Wilson line on the 3d boundary. The supersymmetric extensions of $\mathbf{P}$ and $\mathbf{CT}$, as well as their holographic dual transformations, can be found in appendix \ref{app:discrete-symmetries}. Evidently, $F_{12}$ has the same pseudoscalar quantum numbers as the axion under the discrete symmetries of the line, so that it is its natural weak-coupling candidate. Applying the operator in \eqref{F12_Phi6} to \eqref{Phi6_map}, we then obtain its map to strong coupling, 
\begin{equation}\label{F12_map}
F_{12}\longrightarrow \mathsf{A}:=b_{1}\,\left(\mathsf{\Psi}_{a}^{++}\mathsf{\Psi}^{a--}+\mathsf{\Theta}_{+}\overleftrightarrow{\partial_{t}}\mathsf{\Theta}_{-}\right)+b_{2}\,\bm{\chi}_{a}^{-+}\bm{\chi}^{a+-}\ , 
\end{equation}
where $\bm{\chi}_{a}^{-+}\in\left[-1/2,+1/2\right]_{3/2}^{\left(1\right)}$ and $\bm{\chi}^{a+-}\in\left[+1/2,-1/2\right]_{3/2}^{\left(1\right)}$ are the level-one super-descendants of $\vec{\mathsf{R}}$ in $A_{1}\bar{A}_{1}\left[0,0\right]_{1}^{\left(2\right)}$. In order to check the oddness of $\mathsf{A}$ under $\mathbf{P}$ and $\mathbf{CT}$, let us first consider the $\mathsf{\Theta}_+\overleftrightarrow{\partial_t}\mathsf{\Theta}_-$ term. For longitudinal parity, this combination picks up a minus sign from the transformation $\partial_t\rightarrow -\partial_t$. The case of transverse parity is less intuitive in that the holographic dual transformation involves flipping one of the coordinates on the sphere $S_2^2$, effectively sending $\mathsf{\Theta}_\pm$ to $-\mathsf{\Theta}_\mp$ such that again we obtain a net minus sign. As to the fermion bilinears in (\ref{F12_map}), one can study the transformation properties of the supercharges in (\ref{F12_Phi6}) and obtain the same minus signs, see appendix \ref{app:discrete-symmetries} for the explicit details.

Let us finally mention that in the conjectured strong-coupling description of the line in the pure Neumann case, the operators $\vec{\mathsf{R}}$ in \eqref{Phi6_map} and $\bm{\chi}_{a}^{-+}$, $\bm{\chi}^{a+-}$ in \eqref{F12_map} are absent such that the maps are $\Phi_{6}\longrightarrow\mathsf{\Theta}_{+}\mathsf{\Theta}_{-}$ and $F_{12}\longrightarrow\mathsf{\Psi}_{a}^{++}\mathsf{\Psi}^{a--}+\mathsf{\Theta}_{+}\overleftrightarrow{\partial_{t}}\mathsf{\Theta}_{-}$, without undetermined coefficients.

\section{Conclusions}
In this paper we considered $\mathcal{N}=4$ SYM theory in AdS$_4$ with a Wilson line placed on its boundary. At any coupling this line corresponds to a conformal defect, which at strong coupling can also be described as a long fundamental string. We focused on the leading order computations at weak and strong coupling as well as on exact supersymmetric observables. A natural next step is a more detailed study of perturbative corrections to non-protected operators in the two perturbative regimes. It would also be interesting to compare the behavior of the defect CFT observables to those of a flux tube of pure large-$N$ Yang-Mills in AdS. Since the defect we study is not integrable and 4d conformal symmetry is broken we expect to see more similarities, as compared to the $\mathcal{N}=4$ case without a boundary. Nevertheless, holography still supports strongly that the CFT data changes smoothly from weak to strong coupling. At the same time, in pure Yang-Mills, or in other confining theories, smoothness of the corresponding interpolation is still a conjecture. We believe that our present study can help with settling this important question.

Another interesting direction is to break supersymmetry by compactifying a spatial direction parallel to the boundary on a circle with antiperiodic boundary conditions for the fermions. In the absence of a boundary, such a compactification of $\mathcal{N}=4$ SYM yields pure three-dimensional Yang–Mills theory in an appropriate small-circle limit \cite{Witten:1998confinement,Aharony:2000stg}. Extending this construction to the present setup would allow us to investigate confinement and the dynamics of boundary Wilson lines in the resulting three-dimensional theory.

We only briefly discussed in this paper the regime of a small number of NS5 branes associated with the boundary conditions, and in particular the case of a single brane corresponding to pure Neumann boundary conditions. While this regime is complicated from the point of view of perturbation theory at strong coupling, it is also  potentially very interesting. One can hope that the worldsheet theory simplifies and that some of the bulk directions "disappear", as suggested by our discussion around Eq. \eqref{I_N_displ} and as has recently been discussed in a similar setting in \cite{Dei:2025ilx} and in AdS$_3$ holography in \cite{Eberhardt:2018worldsheet,Eberhardt:2019deriving,Eberhardt:2025localising}.

\section*{Acknowledgements}

We thank Ofer Aharony, Sergei Dubovsky, Barak Gabai, Julius Julius, Carlos Nunes and Bendeguz Offertaler for useful discussions. 
This work was supported by the Simons Foundation grant 994310 (Simons Collaboration on Confinement and QCD Strings), and by a grant of access to OpenAI models through the ChatGPT for Academic Researchers program. The models were used to a limited extent, primarily to cross-check computations.

\appendix

\section{The superconformal index of a 1/2-BPS Wilson line in $\mathcal{N}=4$ SYM}
\label{app:no-boundary-index}

Considering $\mathcal{N}=4$ SYM on a flat space with no boundaries, we will here compute the superconformal index of a 1/2-BPS Maldacena-Wilson line at the leading large-$N$ limit using the same approach and methods as in subsection \ref{subsec:Index}, reproducing the result obtained in \cite{Gang:2012yr}. 

We begin by identifying the symmetries preserved by the line, following the conventions of \cite{Nathan:2020scd}. The superconformal algebra of the bulk theory is broken by the line to the 1d $\mathcal{N}=8$ superconformal algebra $\mathfrak{osp}(4^{*}|4)$, whose bosonic subalgebra is 
\begin{equation}
\mathfrak{so}(2,1)_{\textrm{conf}}\oplus\mathfrak{su}(2)_{\textrm{rot}}\oplus\mathfrak{usp}(4)_{\mathcal{R}}\,.
\end{equation}
The states are denoted by $\left[j\right]_{\Delta}^{\left(m,n\right)}$ where $\Delta$ is the conformal dimension, $j$ the Dynkin label of the $\mathfrak{su}(2)_{\textrm{rot}}$ transverse-rotation representation and $\left(m,n\right)$ the Dynkin labels of the preserved $\mathfrak{usp}(4)_{\mathcal{R}}$ $R$-symmetry. For example, the preserved supercharges transform as follows, 
\begin{equation}
\mathcal{Q}_{a\alpha}\in\left[1\right]_{\frac{1}{2}}^{\left(1,0\right)}
\end{equation}
where $a$ is a $\mathfrak{usp}(4)_{\mathcal{R}}$ fundamental index and $\alpha$ an $\mathfrak{su}(2)_{\textrm{rot}}$ fundamental index. 

In order to construct an index we need to choose a specific preserved supercharge $Q$, compute the resulting operator $\delta=\{Q,Q^{\dagger}\}$ and find charges that commute with both $Q$ and $Q^{\dagger}$ in order to refine it by them, see the discussion around Eqs. \eqref{delta}-\eqref{I}. We choose the supercharge 
\begin{equation}
Q=\mathcal{Q}_{12}
\end{equation}
which has $\mathfrak{su}(2)_{\textrm{rot}}$ weight $-1/2$ and $\mathfrak{usp}(4)_{\mathcal{R}}$ weight $\left(1,0\right)$. Then, in order to compute $\delta$ we use \cite{Nathan:2020scd} 
\begin{equation}
\left\{ \mathcal{Q}_{a\alpha},\mathcal{S}_{b}^{\beta}\right\} =\Omega_{ab}\left(\mathcal{M}_{\alpha}^{\;\beta}+\delta_{\alpha}^{\beta}D\right)+\delta_{\alpha}^{\beta}\mathcal{R}_{ab}\,,\quad\left(\mathcal{Q}_{a\alpha}\right)^{\dagger}=\Omega_{ab}\mathcal{S}_{b}^{\alpha}
\end{equation}
where $D$ is the dilatation generator, $\mathcal{M}_{\alpha}^{\;\beta}$ are the generators of $\mathfrak{su}(2)_{\textrm{rot}}$, $\mathcal{R}_{ab}$ the generators of $\mathfrak{usp}(4)_{\mathcal{R}}$ and $\Omega_{ab}$ is the $\mathfrak{usp}(4)_{\mathcal{R}}$ symplectic form with the nonzero entries being $\Omega_{14}=\Omega_{23}=1$. We obtain 
\begin{equation}
\delta=\left\{ \mathcal{Q}_{12},\left(\mathcal{Q}_{12}\right)^{\dagger}\right\} =\left\{ \mathcal{Q}_{12},\mathcal{S}_{4}^{2}\right\} =\mathcal{M}_{2}^{\;2}+D+\mathcal{R}_{14}=D-\mathcal{M}_{1}^{\;1}+\mathcal{R}_{14}
\end{equation}
where in the last step we have used the fact that $\mathcal{M}_{\alpha}^{\;\beta}$ is traceless. Denoting 
\begin{equation}
j\equiv\mathcal{M}_{1}^{\;1}\,,\quad r_{1}\equiv-\mathcal{R}_{14}\,,\quad r_{3}\equiv-\mathcal{R}_{23}
\end{equation}
to match the notations of \cite{Gang:2012yr}, we then get the expression
\begin{equation}
\delta=D-j-r_{1}\,.
\end{equation}
We can now turn to find generators that commute with both $Q$ and $Q^{\dagger}$. Using the commutators \cite{Nathan:2020scd} 
\begin{equation}
\begin{gathered}
\left[D,\mathcal{Q}_{a\alpha}\right]
 =\frac{1}{2}\mathcal{Q}_{a\alpha}\,,\quad
\left[\mathcal{M}_{\alpha}^{\;\beta},\mathcal{Q}_{a\gamma}\right]
 =\delta_{\gamma}^{\beta}\mathcal{Q}_{a\alpha}
 -\frac{1}{2}\delta_{\alpha}^{\beta}\mathcal{Q}_{a\gamma}\,,\quad
\left[\mathcal{R}_{ab},\mathcal{Q}_{c\alpha}\right]
 =\Omega_{ac}\mathcal{Q}_{b\alpha}
 +\Omega_{bc}\mathcal{Q}_{a\alpha}\,,
\\[3pt]
\left[D,\mathcal{S}_{a}^{\alpha}\right]
 =-\frac{1}{2}\mathcal{S}_{a}^{\alpha}\,,\quad
\left[\mathcal{M}_{\alpha}^{\;\beta},\mathcal{S}_{a}^{\gamma}\right]
 =-\delta_{\alpha}^{\gamma}\mathcal{S}_{a}^{\beta}
 +\frac{1}{2}\delta_{\alpha}^{\beta}\mathcal{S}_{a}^{\gamma}\,,\quad
\left[\mathcal{R}_{ab},\mathcal{S}_{c}^{\alpha}\right]
 =\Omega_{ac}\mathcal{S}_{b}^{\alpha}
 +\Omega_{bc}\mathcal{S}_{a}^{\alpha}
\end{gathered}
\end{equation}
one can readily show that 
\begin{equation}
\left[D+j,Q\right]=\left[D+j,Q^{\dagger}\right]=0\,,\quad\left[r_{3},Q\right]=\left[r_{3},Q^{\dagger}\right]=0
\end{equation}
such that we get the following index considered in \cite{Gang:2012yr}, 
\begin{equation}\label{I_N4SYM}
\mathcal{I}\left(\eta,\xi\right)=\textrm{Tr}_{D=j+r_{1}}\left(-1\right)^{F}\xi^{D+j}\eta^{r_{3}}
\end{equation}
where the trace is performed only over operators that satisfy the condition $\delta=D-j-r_{1}=0$. 

Now that we got the explicit index expression \eqref{I_N4SYM}, we can turn to its evaluation at weak ($\lambda=0$) and strong ($\lambda=\infty$) coupling. 

\begin{itemize}
    \item \textit{Strong coupling}: Here we have the familiar holographic description in which the Wilson line corresponds to a fundamental string in $AdS_{5}\times S^5$. The basic fields on its worldsheet are the massless bosonic modes corresponding to fluctuations along $S^{5}$, the massive bosonic modes corresponding to fluctuations of the string wrapping $AdS_{2}\subset AdS_{5}$ along the transverse $AdS_{3}$ directions, and the fermionic partners. On the dual Maldacena-Wilson line, these correspond to the following operators that serve as basic letters, 
    \begin{equation}\label{B1_content}
\left[0\right]_{1}^{\left(0,1\right)}\oplus\left[1\right]_{\frac{3}{2}}^{\left(1,0\right)}\oplus\left[2\right]_{2}^{\left(0,0\right)}
    \end{equation}
    and which form together the absolutely-protected displacement multiplet $B_{1}\left[0\right]_{1}^{\left(0,1\right)}$ (we here use the multiplet notation of \cite{Nathan:2020scd}). To check which of these letters contributes to the index and how, we note that the $(r_{1},r_{3})$ charges of the different states in the representations $\mathbf{4}_{\mathfrak{usp}(4)_{\mathcal{R}}}$ and $\mathbf{5}_{\mathfrak{usp}(4)_{\mathcal{R}}}$ (i.e. $\left(1,0\right)$ and $\left(0,1\right)$) are 
    \begin{equation}
        \mathbf{4}_{\mathfrak{usp}(4)_{\mathcal{R}}}:\;\left(\pm1,0\right)\,,\,\left(0,\pm1\right)\,,\quad\mathbf{5}_{\mathfrak{usp}(4)_{\mathcal{R}}}:\;\left(\pm1,\pm1\right)\,,\,\left(0,0\right)
    \end{equation}
    such that the letters with $D=j+r_{1}$ are the ones with $(r_{1},r_{3})=(1,\pm1)$ in $\left[0\right]_{1}^{\left(0,1\right)}$ and the one with $(r_{1},r_{3})=(1,0)$ in $\left[1\right]_{\frac{3}{2}}^{\left(1,0\right)}$ (while $\left[2\right]_{2}^{\left(0,0\right)}$ does not contribute). Their contribution to the index \eqref{I_N4SYM} is 
    \begin{equation}\label{i_sl_N4SYM}
i_{\textrm{s.l.}}^{\textrm{strong}}\left(\eta,\xi\right)=\left(\eta+\eta^{-1}\right)\xi-\xi^{2}\,,
    \end{equation}
    which constitutes the single-letter index at strong coupling. The full index is then given by taking the plethystic exponential of \eqref{i_sl_N4SYM}, resulting in 
    \begin{equation}\label{I_expl_N4SYM}
\mathcal{I}\left(\eta,\xi\right)=\textrm{PE}\left[i_{\textrm{s.l.}}^{\textrm{strong}}\left(\eta,\xi\right)\right]=\exp\left[\sum_{n=1}^{\infty}\frac{1}{n}i_{\textrm{s.l.}}^{\textrm{strong}}\left(\eta^{n},\xi^{n}\right)\right]=\frac{1-\xi^{2}}{\left(1-\eta\xi\right)\left(1-\eta^{-1}\xi\right)}\,.          \end{equation}
    Note that since the single-letter index \eqref{i_sl_N4SYM} is the index of the displacement multiplet \eqref{B1_content} which generates the spectrum, i.e. $i_{\textrm{s.l.}}^{\textrm{strong}}=\mathcal{I}_{B_{1}\left[0\right]_{1}^{\left(0,1\right)}}$, the full index \eqref{I_expl_N4SYM} can be written as 
    \begin{equation}
\mathcal{I}\left(\eta,\xi\right)=\textrm{PE}\left[\mathcal{I}_{B_{1}\left[0\right]_{1}^{\left(0,1\right)}}\left(\eta,\xi\right)\right],
    \end{equation}
    exactly like in the case with a pure BPS Neumann boundary condition (see Eq. \eqref{I_N_displ}). 
    \item \textit{Weak coupling}: Here we have the free field-theory description, and we show in table \ref{tab:no-boundary-field-index-charges} the basic fields and the various derivatives along with their charges under the generators that appear in the expression \eqref{I_N4SYM} for the index. 
    \begin{table}
    \centering
    \renewcommand{\arraystretch}{1.35}
    \resizebox{\textwidth}{!}{%
    \begin{tabular}{c|ccccccc}
        \hline
        \textbf{Field/charge}
        & $\Phi_{6}$
        & $\Phi_{1},\ldots,\Phi_{5}$
        & $\Psi,\overline{\Psi}$
        & $F,\overline{F}$
        & $\partial_{11}$
        & $\partial_{22}$
        & $\partial_{12},\partial_{21}$
        \\
        \hline
        $\Delta$
        & $1$
        & $1$
        & $\frac{3}{2}$
        & $2$
        & $1$
        & $1$
        & $1$
        \\
        $j$
        & $0$
        & $0$
        & $\pm\frac{1}{2}$
        & $0,\pm1$
        & $1$
        & $-1$
        & $0$
        \\
        $(r_{1},r_{3})$
        & $(0,0)$
        & $(\pm1,\pm1),(0,0)$
        & $(\pm1,0),(0,\pm1)$
        & $(0,0)$
        & $(0,0)$
        & $(0,0)$
        & $(0,0)$
        \\
        \hline
    \end{tabular}%
    }
    \caption{Fundamental fields of $\mathcal{N}=4$ SYM and derivatives along with their charges under the generators entering the
    expression for the index. Note that the line breaks the bulk rotation algebra $\mathfrak{su}(2)_{L}\oplus\mathfrak{su}(2)_{R}$ to the diagonal $\mathfrak{su}(2)_{\textrm{rot}}$ (with Cartan generator $j$), such that dotted and undotted spinor indices are identified as
    $\mathfrak{su}(2)_{\mathrm{rot}}$ indices and the dots are suppressed (for example, $\partial_{\alpha\dot{\beta}}\rightarrow\partial_{\alpha\beta}$).}
    \label{tab:no-boundary-field-index-charges}
    \end{table}
    The letters with $\Delta=j+r_{1}$ that contribute to the index are the ones corresponding to $(r_{1},r_{3})=(1,\pm1)$ in $\Phi_{1,\ldots,5}$, to $(r_{1},r_{3})=(1,0)$ in $\Psi$ and $\bar{\Psi}$ and the derivative $\partial_{11}$. The resulting single-letter index at weak coupling is then 
    \begin{equation}\label{i_sl_N4SYM_weak}
i_{\textrm{s.l.}}^{\textrm{weak}}\left(\eta,\xi\right)=\frac{\left(\eta+\eta^{-1}\right)\xi-2\xi^{2}}{1-\xi^{2}}
    \end{equation}
    where the denominator corresponds to the derivative contribution. As we discussed above Eq. \eqref{Z_weak_bulk}, in such a description, no symmetrization, and hence no plethystic exponential, is needed when constructing operators and computing their contribution to the index. As a result, the index obtained from \eqref{i_sl_N4SYM_weak} is given by 
    \begin{equation}
\mathcal{I}\left(\eta,\xi\right)=\sum_{k=0}^{\infty}\left(i_{\textrm{s.l.}}^{\textrm{weak}}\left(\eta,\xi\right)\right)^{k}=\frac{1}{1-i_{\textrm{s.l.}}^{\textrm{weak}}\left(\eta,\xi\right)}=\frac{1-\xi^{2}}{\left(1-\eta\xi\right)\left(1-\eta^{-1}\xi\right)}\,,
    \end{equation}
    matching the strong-coupling result \eqref{I_expl_N4SYM} as expected. 
\end{itemize}

As in the case with a boundary we considered in the main text, also here the index $\mathcal{I}\left(\eta,\xi\right)$ gives us alternating-sum constraints on the short and semi-short multiplet content for any $\lambda$ (recall \eqref{AA_multiplicity_constraint}-\eqref{AL0_multiplicity_constraint}). At $\lambda=\infty$, the minimal solution for these constraints is realized, for which there is a single absolutely protected short multiplet $B_{1}\left[0\right]_{k}^{\left(0,k\right)}$ (which cannot recombine) for any $k\geq1$, along with only long multiplets (see \cite{Ferrero:2024wld1} for a recent discussion). In this case, the index decomposes as 
\begin{equation}
\mathcal{I}\left(\eta,\xi\right)=\sum_{k=0}^{\infty}\mathcal{I}_{B_{1}\left[0\right]_{k}^{\left(0,k\right)}}\left(\eta,\xi\right)
\end{equation}
where 
\begin{equation}
\mathcal{I}_{B_{1}\left[0\right]_{k}^{\left(0,k\right)}}\left(\eta,\xi\right)=\xi^{k}\left[\chi_{k}\left(\eta\right)-\xi\chi_{k-1}\left(\eta\right)\right]\,,\quad\chi_{k}\left(\eta\right)=\eta^{k}+\eta^{k-2}+\cdots\eta^{-k}\,.
\end{equation}
Meanwhile, at $\lambda=0$, additional semi-short multiplets are present but cancel in the index, and their recombination to long multiplets at generic $\lambda$ is strongly expected (though not proved by the index alone) \cite{Ferrero:2024wld1,Julius:2026rosetta}. 

\section{Discrete symmetries of the 1d defect SCFT}
\label{app:discrete-symmetries}
In this appendix, we describe the discrete symmetries of the Wilson line, namely the so-called transverse and longitudinal parities. In particular, their action on the full vector multiplet of $\mathcal{N}=4$ SYM, tailored to the Neumann boundary conditions at $x^3=0$, are presented, as well as their dual transformations in the background of D3-branes ending on NS5-branes. As an application, the oddness of the candidate axion at strong coupling (see (\ref{F12_map})) under both symmetries is verified.

Transverse parity, denoted $\mathbf{P}$, is a symmetry of 4d $\mathcal{N}=4$ SYM on $\mathbb{R}^{1,3}$ as a consequence of its dimensional reduction from the ten-dimensional $\mathcal{N}=1$ theory. Explicitly, reversing an even number of coordinates in $\mathbb{R}^{1,9}$ preserves the chirality of the 10d Majorana-Weyl fermion $\lambda$ and therefore corresponds to a symmetry. Based on this, an odd number of reflections in both $\mathbb{R}^{1,3}$ and $\mathbb{R}^6$ implements parity in the 4d field theory, at the expense of breaking the R-symmetry. This argument is elaborated in \cite{Gomis:2026ads-cpt}, where a study of $\mathbf{C}$-$\mathbf{P}$-$\mathbf{T}$ in $\mathcal{N}=4$ SYM and their dual bulk transformations in AdS$_5\times S^5$ is carried out. According to our notation, the action of $\mathbf{P}$ on the coordinates and the 4d fields may be given by:\footnote{Following the analysis in \cite{Gaiotto:2009sbc}, the Dirichlet and Neumann boundary conditions on $\vec{X}$ and $\vec{Y}$ respectively (see (\ref{eq:pure-neumann-boundary-conditions})) preserve the supersymmetries which are eigenvectors of $\Gamma_{3456}$. Moreover, in \cite{Gomis:2026ads-cpt}, the 10d fermion is shown to transform with the combination of gamma matrices corresponding to reversed coordinates. A choice of parity transformation compatible with the preserved supersymmetries should therefore reverse either $x^7$ or $x^8$ (corresponding to $A_7 = \Phi_4$ or $A_8 = \Phi_5$), such that the scalar $\Phi_6$ later included in the Wilson line remains invariant. The freedom in choosing $\Phi_4$ or $\Phi_5$ will be irrelevant, due to the $U(1)_Y$ symmetry of the $1d$ SCFT.\label{footnote:transverse-parity}}
\begin{equation}\label{eq:transverse-parity}
    \mathbf{P}:\quad 
    \begin{IEEEeqnarraybox}[\IEEEeqnarraystrutmode][c]{rcl}
        x^\mu = (x^0,x^1,x^2,x^3)\ &\longrightarrow &\ x^\mu_P = (x^0,-x^1,x^2,x^3) \\
        A_{\mu\neq 1}(x)\, ,\,A_1(x)\ &\longrightarrow &\ A_{\mu\neq 1}(x_P)\, ,\,-A_1(x_P) \\
        \vec{X}(x)\ &\longrightarrow &\ \vec{X}(x_P) \\
        \Phi_4(x)\, ,\, \Phi_5(x)\, ,\, \Phi_6(x)\ &\longrightarrow &\ -\Phi_4(x_P)\, ,\, \Phi_5(x_P)\, ,\, \Phi_6(x_P) \\
        \lambda(x)\ &\longrightarrow &\ \Gamma^{17}\lambda(x_P)
    \end{IEEEeqnarraybox}
\end{equation}
with $\vec{X} = (\Phi^1,\Phi^2,\Phi^3)$. After restricting (\ref{eq:transverse-parity}) to the planar 3d boundary at $x^3 = 0$, we introduce a Wilson line $W_{\mathbf{N}}$ in the fundamental representation $\mathbf{N}$ of $SU(N)$ and extended along $x^0 = t$, with the incorporation of $\Phi_6$, as in (\ref{W}). Then,
\begin{equation}
    \mathbf{P}\left(W_{\mathbf{N}}\right) = W_{\mathbf{N}}\ ,
\end{equation}
such that transverse parity is a genuine symmetry of the 1d defect SCFT. Moreover, contrary to the 4d theory, the residual $SU(2)_\mathcal{R}$ R-symmetry is preserved under $\mathbf{P}$. Adapting (\ref{eq:transverse-parity}) to insertions on the line, we get:
\begin{equation}
    \mathbf{P}:\quad 
    \begin{IEEEeqnarraybox}[\IEEEeqnarraystrutmode][c]{rcl}
        \Tilde{\mathbb{F}}_\pm(t)\ &\longrightarrow &\ -\Tilde{\mathbb{F}}_\mp(t) \\
        \Phi_6(t)\, ,\, \Phi_\pm(t)\ &\longrightarrow &\ \Phi_6(t)\, ,\, -\Phi_\mp(t) \\
        F_{12}(t)\ &\longrightarrow &\ -F_{12}(t)
    \end{IEEEeqnarraybox}
\end{equation}
confirming that $F_{12}$ is odd, as expected of the weak-coupling analog of the worldsheet axion. Since both $x^1$ and $x^7$ (whose corresponding gauge field components are $A_1$ and $A_7 = \Phi_4$) are reflected in the ten-dimensional field theory, one obtains that
\begin{equation}
    P^\dagger J_Y P = -J_Y\quad ,\quad P^\dagger J_{\text{rot}}P = -J_{\text{rot}}\ ,
\end{equation}
where $J_Y$ and $J_{\text{rot}}$ are the $U(1)_Y$ and $U(1)_{\text{rot}}$ generators respectively, while $P$ is the unitary operator implementing transverse parity. This implies that the two sets of supercharges (\ref{QQd}) are interchanged under $\mathbf{P}$, namely
\begin{equation}\label{eq:transverse-parity-supercharges}
    P^\dagger\mathcal{Q}_a P = U_{ab}\bar{\mathcal{Q}}^b\quad ,\quad P^\dagger\bar{\mathcal{Q}}^a P = \mathcal{Q}_b(U^\dagger)^{ba}\ ,
\end{equation}
with $U$ an anti-symmetric unitary matrix.

Adapting the discussion of \cite{Gomis:2026ads-cpt} to our setup, the holographic bulk transformations dual to (\ref{eq:transverse-parity}) are realized as follows on the target space coordinates:
\begin{equation}
    \mathbf{P}:\quad X^\mu = (t,\sigma,\bar{x}^1,\bar{x}^2,\theta_1,\phi_1,\theta_2,\phi_2,x,y)\ \longrightarrow\ X^\mu_P = (t,\sigma,-\bar{x}^1,\bar{x}^2,\theta_1,\phi_1,\pi-\theta_2,\phi_2,x,y)\ ,
\end{equation}
such that the CFT$_1$ operators $\mathsf{\Theta}_\pm$ and $\mathsf{\bar{X}}_\pm$ dual to $\theta_\pm = \delta\Tilde{\theta}_2\pm i\delta\Tilde{\phi}_2$ and $\bar{x}_\pm = \delta\Tilde{\bar{x}}^1 \pm i\delta\Tilde{{\bar{x}}}^2$ satisfy:
\begin{equation}\label{eq:transverse-parity-strong-coupling}
    \mathbf{P}:\quad 
    \begin{IEEEeqnarraybox}[\IEEEeqnarraystrutmode][c]{rcl}
        \mathsf{\Theta}_\pm(t)\ &\longrightarrow &\ -\mathsf{\Theta}_\mp(t) \\
        \mathsf{\bar{X}}_\pm(t)\ &\longrightarrow &\ -\mathsf{\bar{X}}_\mp(t) \\
        \mathsf{R}^n(t)\ &\longrightarrow &\ \mathsf{R}^n(t)
    \end{IEEEeqnarraybox}
\end{equation}
like their weak-coupling counterparts. Thanks to (\ref{eq:transverse-parity-supercharges}) and (\ref{eq:transverse-parity-strong-coupling}), one immediately checks:
\begin{align}
    P^\dagger\mathsf{A} P &= \frac{1}{2}P^\dagger\left[\mathcal{Q}_a,\bar{\mathcal{Q}}^a\right]P\left(b_1\left(P^\dagger\mathsf{\Theta}_+P\right)\left(P^\dagger\mathsf{\Theta}_-P\right)+b_2\left(P^\dagger\mathsf{R}^nP\right)\left(P^\dagger\mathsf{R}^nP\right)\right) \label{eq:axion-transverse-parity}\\
    &= -\frac{1}{2}\left[\mathcal{Q}_a,\bar{\mathcal{Q}}^a\right]\left(b_1\mathsf{\Theta}_-\mathsf{\Theta}_+ + b_2\mathsf{R}^n\mathsf{R}^n\right) \\
    &= -\mathsf{A}\ ,
\end{align}
showing that the strong-coupling candidate axion $\mathsf{A}$ (see (\ref{F12_map})) is odd under transverse parity.\\

Longitudinal parity combines charge conjugation $\mathbf{C}$ and (linear) time reversal $\mathbf{T}$. For an $SU(N)$ gauge group, with $N>2$, the former acts as follows on the $\mathcal{N}=4$ fields:
\begin{equation}\label{eq:charge-conjugation}
    \mathbf{C}:\quad \Upsilon :=(A_\mu,\lambda,\Phi_I) \longrightarrow -\Upsilon^T\ ,
\end{equation}
such that the Wilson line undergoes the transformation:
\begin{equation}\label{eq:wilson-line-conjugation}
    \mathbf{C}(W_{\mathbf{N}}) = W_{\mathbf{\bar{N}}}\ ,
\end{equation}
with $\mathbf{\bar{N}}$ the anti-fundamental representation of $SU(N)$. On the other hand, the transformation rules of the 4d fields under time reversal are given by:\footnote{Instead of using the conventions of \cite{Gomis:2026ads-cpt} for anti-unitary time reversal in real time, we implement $\mathbf{T}$ in a similar fashion to transverse parity $\mathbf{P}$. This means that $\vec{X}$ should not be reversed, so as to preserve the supersymmetries of the boundary, as in footnote \ref{footnote:transverse-parity}. However, since $A_0$ changes sign, one chooses to reverse all the components of $\vec{Y}$, including $\Phi_6$.}
\begin{equation}\label{eq:time-reversal}
    \mathbf{T}:\quad 
    \begin{IEEEeqnarraybox}[\IEEEeqnarraystrutmode][c]{rcl}
        x^\mu = (x^0,x^1,x^2,x^3)\ &\longrightarrow &\ x^\mu_T = (-x^0,x^1,x^2,x^3) \\
        A_0(x)\, ,\,A_i(x)\ &\longrightarrow &\ -A_0(x_T)\, ,\,A_i(x_T) \\
        \vec{X}(x)\ &\longrightarrow &\ \vec{X}(x_T) \\
        \vec{Y}(x)\ &\longrightarrow &\ -\vec{Y}(x_T) \\
        \lambda(x)\ &\longrightarrow &\ \Gamma^{0789}\lambda(x_T)
    \end{IEEEeqnarraybox}
\end{equation}
For the Wilson line on the boundary, this in turn implies:
\begin{equation}
    \mathbf{T}(W_{\mathbf{N}}) = W_{\mathbf{\bar{N}}}\ .
\end{equation}
Together with (\ref{eq:wilson-line-conjugation}), this means that the combined action of $\mathbf{C}$ and $\mathbf{T}$ is a symmetry of the 1d SCFT, with its R-symmetry preserved. For insertions on the line, (\ref{eq:charge-conjugation}) and (\ref{eq:time-reversal}) yield:
\begin{equation}
     \mathbf{CT}:\quad 
    \begin{IEEEeqnarraybox}[\IEEEeqnarraystrutmode][c]{rcl}
        \Tilde{\mathbb{F}}_\pm(t)\ &\longrightarrow &\ \Tilde{\mathbb{F}}_\pm^T(-t) \\[2pt]
        \Phi_6(t)\, ,\,\Phi_\pm(t)\ &\longrightarrow &\ \Phi_6^T(-t)\, ,\, \Phi_\pm^T(-t) \\[2pt]
        F_{12}(t)\ &\longrightarrow &\ -F_{12}^T(-t) 
    \end{IEEEeqnarraybox}
\end{equation}
In contrast to transverse parity, $\mathbf{CT}$ does not exchange the two sets of supercharges. Instead,
\begin{equation}\label{eq:longitudinal-parity-supercharges}
    (CT)^\dagger\mathcal{Q}_a(CT) = \eta_{CT}\mathcal{Q}_a\quad ,\quad (CT)^\dagger\bar{\mathcal{Q}}^a(CT) = -\eta_{CT}\bar{\mathcal{Q}}^a\ ,
\end{equation}
with $\eta_{CT}=\pm 1$.

The bulk transformation dual to longitudinal parity is given by:
\begin{equation}
    \mathbf{CT}:\
    \begin{IEEEeqnarraybox}[\IEEEeqnarraystrutmode][c]{rcl}
         X^\mu=(t,\sigma,\bar{x}^1,\bar{x}^2,\theta_1,\phi_1,\theta_2,\phi_2,x,y)\ &\longrightarrow &\ X_{CT}^\mu = (-t,\sigma,\bar{x}^1,\bar{x}^2,\pi-\theta_1,\pi+\phi_1,\theta_2,\phi_2,x,y) \\[2pt]
         (\xi^0,\xi^1)\ &\longrightarrow &\ (-\xi^0,\xi^1)
    \end{IEEEeqnarraybox}
\end{equation}
such that
\begin{equation}\label{eq:longitudinal-parity-strong-coupling}
    \mathbf{CT}:\quad
    \begin{IEEEeqnarraybox}[\IEEEeqnarraystrutmode][c]{rcl}
         \mathsf{\Theta}_\pm(t)\ &\longrightarrow &\ \mathsf{\Theta}_\pm(-t) \\
         \mathsf{\bar{X}}_\pm(t)\ &\longrightarrow &\ \mathsf{\bar{X}}_\pm(-t) \\
         \mathsf{R}^n(t)\ &\longrightarrow &\ -\mathsf{R}^n(-t)
    \end{IEEEeqnarraybox}
\end{equation}
Putting (\ref{eq:longitudinal-parity-supercharges}) and (\ref{eq:longitudinal-parity-strong-coupling}) together and proceeding as in (\ref{eq:axion-transverse-parity}) implies:
\begin{equation}
    (CT)^\dagger\mathsf{A}(CT) = -\mathsf{A}\ ,
\end{equation}
and the strong-coupling candidate $\mathsf{A}$ is indeed odd under both transverse and longitudinal parities.

\bibliographystyle{utphys}
\bibliography{bibliography}

\providecommand{\href}[2]{#2}\begingroup\raggedright\begin{thebibliography}{10}

\bibitem{Barak:2026ym1}
B.~Gabai, V.~Gorbenko, and J.~Qiao, ``{Yang-Mills Flux Tube in AdS},'' 2026.
\newblock \url{https://arxiv.org/abs/2508.08250}.

\bibitem{Barak:2026ym2}
B.~Gabai, V.~Gorbenko, and B.~Offertaler, ``{Yang-Mills Flux Tube in AdS II: Effective String Theory},'' 2026.
\newblock \url{https://arxiv.org/abs/2602.16694}.

\bibitem{Maldacena:1998wilson}
J.~M. Maldacena, ``{Wilson loops in large N field theories},'' \href{http://dx.doi.org/10.1103/PhysRevLett.80.4859}{{\em Phys. Rev. Lett.} {\bfseries 80} (1998) 4859--4862}, \href{http://arxiv.org/abs/hep-th/9803002}{{\ttfamily arXiv:hep-th/9803002}}. \url{https://arxiv.org/abs/hep-th/9803002}.

\bibitem{Rey:1998wilson}
S.-J. Rey and J.-T. Yee, ``{Macroscopic strings as heavy quarks: Large-N gauge theory and anti-de Sitter supergravity},'' \href{http://dx.doi.org/10.1007/s100520100799}{{\em Eur. Phys. J. C} {\bfseries 22} (2001) 379--394}, \href{http://arxiv.org/abs/hep-th/9803001}{{\ttfamily arXiv:hep-th/9803001}}. \url{https://arxiv.org/abs/hep-th/9803001}.

\bibitem{Drukker:1999minimal}
N.~Drukker, D.~J. Gross, and H.~Ooguri, ``{Wilson Loops and Minimal Surfaces},'' \href{http://dx.doi.org/10.1103/PhysRevD.60.125006}{{\em Phys. Rev. D} {\bfseries 60} (1999) 125006}, \href{http://arxiv.org/abs/hep-th/9904191}{{\ttfamily arXiv:hep-th/9904191}}. \url{https://arxiv.org/abs/hep-th/9904191}.

\bibitem{Giombi:2017cqn}
S.~Giombi, R.~Roiban, and A.~A. Tseytlin, ``{Half-BPS Wilson loop and AdS$_2$/CFT$_1$},'' \href{http://dx.doi.org/10.1016/j.nuclphysb.2017.07.004}{{\em Nucl. Phys. B} {\bfseries 922} (2017) 499--527}, \href{http://arxiv.org/abs/1706.00756}{{\ttfamily arXiv:1706.00756 [hep-th]}}.

\bibitem{Grabner:2020nis}
D.~Grabner, N.~Gromov, and J.~Julius, ``{Excited States of One-Dimensional Defect CFTs from the Quantum Spectral Curve},'' \href{http://dx.doi.org/10.1007/JHEP07(2020)042}{{\em JHEP} {\bfseries 07} (2020) 042}, \href{http://arxiv.org/abs/2001.11039}{{\ttfamily arXiv:2001.11039 [hep-th]}}.

\bibitem{Cavaglia:2021bnz}
A.~Cavagli{\`a}, N.~Gromov, J.~Julius, and M.~Preti, ``{Integrability and conformal bootstrap: One dimensional defect conformal field theory},'' \href{http://dx.doi.org/10.1103/PhysRevD.105.L021902}{{\em Phys. Rev. D} {\bfseries 105} no.~2, (2022) L021902}, \href{http://arxiv.org/abs/2107.08510}{{\ttfamily arXiv:2107.08510 [hep-th]}}.

\bibitem{Cavaglia:2022qpg}
A.~Cavagli{\`a}, N.~Gromov, J.~Julius, and M.~Preti, ``{Bootstrability in defect CFT: integrated correlators and sharper bounds},'' \href{http://dx.doi.org/10.1007/JHEP05(2022)164}{{\em JHEP} {\bfseries 05} (2022) 164}, \href{http://arxiv.org/abs/2203.09556}{{\ttfamily arXiv:2203.09556 [hep-th]}}.

\bibitem{Julius:2026rosetta}
J.~Julius and N.~S. Sokolova, ``{A Rosetta Stone for Wilson Line Defects},'' 2026.
\newblock \url{https://arxiv.org/abs/2512.23634}.

\bibitem{Aharony:2013confinement}
O.~Aharony, M.~Berkooz, D.~Tong, and S.~Yankielowicz, ``{Confinement in anti-de Sitter space},'' \href{http://dx.doi.org/10.1007/jhep02(2013)076}{{\em Journal of High Energy Physics} {\bfseries 2013} no.~2, (Feb., 2013) }. \url{http://dx.doi.org/10.1007/JHEP02(2013)076}.

\bibitem{DiPietro:2024confinement}
R.~Ciccone, F.~De~Cesare, L.~Di~Pietro, and M.~Serone, ``{Exploring confinement in Anti-de Sitter space},'' \href{http://dx.doi.org/10.1007/jhep12(2024)218}{{\em Journal of High Energy Physics} {\bfseries 2024} no.~12, (Dec., 2024) }. \url{http://dx.doi.org/10.1007/JHEP12(2024)218}.

\bibitem{Gaiotto:2009sbc}
D.~Gaiotto and E.~Witten, ``{Supersymmetric Boundary Conditions in $\mathcal{N}=4$ Super Yang-Mills Theory},'' \href{http://dx.doi.org/10.1007/s10955-009-9687-3}{{\em Journal of Statistical Physics} {\bfseries 135} no.~5-6, (Feb., 2009) 789–855}. \url{http://dx.doi.org/10.1007/s10955-009-9687-3}.

\bibitem{Gaiotto:2009sdua}
D.~Gaiotto and E.~Witten, ``{$S$-duality of boundary conditions in ${\mathcal N}=4$ super Yang-Mills theory},'' \href{http://dx.doi.org/10.4310/atmp.2009.v13.n3.a5}{{\em Advances in Theoretical and Mathematical Physics} {\bfseries 13} no.~3, (2009) 721–896}. \url{http://dx.doi.org/10.4310/ATMP.2009.v13.n3.a5}.

\bibitem{DHoker:2007ehb1}
E.~D’Hoker, J.~Estes, and M.~Gutperle, ``{Exact half-BPS type IIB interface solutions I: local solution and supersymmetric Janus},'' \href{http://dx.doi.org/10.1088/1126-6708/2007/06/021}{{\em Journal of High Energy Physics} {\bfseries 2007} no.~06, (2007) 021–021}. \url{http://dx.doi.org/10.1088/1126-6708/2007/06/021}.

\bibitem{DHoker:2007ehb2}
E.~D’Hoker, J.~Estes, and M.~Gutperle, ``{Exact half-BPS type IIB interface solutions II: flux solutions and multi-Janus},'' \href{http://dx.doi.org/10.1088/1126-6708/2007/06/022}{{\em Journal of High Energy Physics} {\bfseries 2007} no.~06, (2007) 022–022}. \url{http://dx.doi.org/10.1088/1126-6708/2007/06/022}.

\bibitem{Aharony:2011nhs}
O.~Aharony, L.~Berdichevsky, M.~Berkooz, and I.~Shamir, ``{Near-horizon solutions for D3-branes ending on 5-branes},'' \href{http://dx.doi.org/10.1103/physrevd.84.126003}{{\em Physical Review D} {\bfseries 84} no.~12, (Dec., 2011) }. \url{http://dx.doi.org/10.1103/PhysRevD.84.126003}.

\bibitem{Coccia:2022map}
L.~Coccia and C.~F. Uhlemann, ``{Mapping out the internal space in AdS/BCFT with Wilson loops},'' \href{http://dx.doi.org/10.1007/jhep03(2022)127}{{\em Journal of High Energy Physics} {\bfseries 2022} no.~3, (Mar., 2022) }. \url{http://dx.doi.org/10.1007/JHEP03(2022)127}.

\bibitem{He:2024djr}
D.~He and C.~F. Uhlemann, ``{Solving $ \mathcal{N} $ = 4 SYM BCFT matrix models at large N},'' \href{http://dx.doi.org/10.1007/JHEP12(2024)164}{{\em JHEP} {\bfseries 12} (2024) 164}, \href{http://arxiv.org/abs/2409.13016}{{\ttfamily arXiv:2409.13016 [hep-th]}}.

\bibitem{Dubovsky:2013axion}
S.~Dubovsky, R.~Flauger, and V.~Gorbenko, ``{Evidence from Lattice Data for a New Particle on the Worldsheet of the QCD Flux Tube},'' \href{http://dx.doi.org/10.1103/physrevlett.111.062006}{{\em Physical Review Letters} {\bfseries 111} no.~6, (Aug., 2013) }. \url{http://dx.doi.org/10.1103/PhysRevLett.111.062006}.

\bibitem{Dubovsky:2025axion}
A.~Athenodorou, S.~Dubovsky, C.~Luo, and M.~Teper, ``{Confining Strings and the Worldsheet Axion from the Lattice},'' 2025.
\newblock \url{https://arxiv.org/abs/2411.03435}.

\bibitem{Wang:2020td}
Y.~Wang, ``{Taming defects in $\mathcal{N}$ = 4 super-Yang-Mills},'' \href{http://dx.doi.org/10.1007/jhep08(2020)021}{{\em Journal of High Energy Physics} {\bfseries 2020} no.~8, (Aug., 2020) }. \url{http://dx.doi.org/10.1007/JHEP08(2020)021}.

\bibitem{VanRaamsdonk:2021hlc}
M.~Van~Raamsdonk and C.~Waddell, ``{Holographic and localization calculations of boundary F for $\mathcal{N}$ = 4 SUSY Yang-Mills theory},'' \href{http://dx.doi.org/10.1007/jhep02(2021)222}{{\em Journal of High Energy Physics} {\bfseries 2021} no.~2, (Feb., 2021) }. \url{http://dx.doi.org/10.1007/JHEP02(2021)222}.

\bibitem{Ferrero:2024wld1}
P.~Ferrero and C.~Meneghelli, ``{Unmixing the Wilson line defect CFT. Part I: spectrum and kinematics},'' 2024.
\newblock \url{https://arxiv.org/abs/2312.12550}.

\bibitem{Nathan:2020scd}
N.~B. Agmon and Y.~Wang, ``{Classifying Superconformal Defects in Diverse Dimensions Part I: Superconformal Lines},'' 2020.
\newblock \url{https://arxiv.org/abs/2009.06650}.

\bibitem{Cooke:2017wavy}
M.~Cooke, A.~Dekel, and N.~Drukker, ``{The Wilson loop CFT: insertion dimensions and structure constants from wavy lines},'' \href{http://dx.doi.org/10.1088/1751-8121/aa7db4}{{\em Journal of Physics A: Mathematical and Theoretical} {\bfseries 50} no.~33, (2017) 335401}. \url{http://dx.doi.org/10.1088/1751-8121/aa7db4}.

\bibitem{cordova:2016multiplets}
C.~Cordova, T.~T. Dumitrescu, and K.~Intriligator, ``{Multiplets of Superconformal Symmetry in Diverse Dimensions},'' 2016.
\newblock \url{https://arxiv.org/abs/1612.00809}.

\bibitem{Callan:1991dj}
C.~G. Callan, Jr., J.~A. Harvey, and A.~Strominger, ``{World sheet approach to heterotic instantons and solitons},'' \href{http://dx.doi.org/10.1016/0550-3213(91)90074-8}{{\em Nucl. Phys. B} {\bfseries 359} (1991) 611--634}.

\bibitem{Callan:1991at}
C.~G. Callan, Jr., J.~A. Harvey, and A.~Strominger, ``{Supersymmetric string solitons},'' \href{http://arxiv.org/abs/hep-th/9112030}{{\ttfamily arXiv:hep-th/9112030}}.

\bibitem{Aharony:1998ub}
O.~Aharony, M.~Berkooz, D.~Kutasov, and N.~Seiberg, ``{Linear dilatons, NS five-branes and holography},'' \href{http://dx.doi.org/10.1088/1126-6708/1998/10/004}{{\em JHEP} {\bfseries 10} (1998) 004}, \href{http://arxiv.org/abs/hep-th/9808149}{{\ttfamily arXiv:hep-th/9808149}}.

\bibitem{Dei:2025ilx}
A.~Dei and E.~J. Martinec, ``{On the string theory of a single NS5-brane},'' \href{http://dx.doi.org/10.1007/JHEP10(2025)059}{{\em JHEP} {\bfseries 10} (2025) 059}, \href{http://arxiv.org/abs/2506.22300}{{\ttfamily arXiv:2506.22300 [hep-th]}}.

\bibitem{Polchinski:2011im}
J.~Polchinski and J.~Sully, ``{Wilson Loop Renormalization Group Flows},'' \href{http://dx.doi.org/10.1007/JHEP10(2011)059}{{\em JHEP} {\bfseries 10} (2011) 059}, \href{http://arxiv.org/abs/1104.5077}{{\ttfamily arXiv:1104.5077 [hep-th]}}.

\bibitem{Beccaria:2017rbe}
M.~Beccaria, S.~Giombi, and A.~Tseytlin, ``{Non-supersymmetric Wilson loop in $ \mathcal{N} $ = 4 SYM and defect 1d CFT},'' \href{http://dx.doi.org/10.1007/JHEP03(2018)131}{{\em JHEP} {\bfseries 03} (2018) 131}, \href{http://arxiv.org/abs/1712.06874}{{\ttfamily arXiv:1712.06874 [hep-th]}}.

\bibitem{Gadde:2020yah}
A.~Gadde, ``{Lectures on the Superconformal Index},'' \href{http://dx.doi.org/10.1088/1751-8121/ac42ac}{{\em J. Phys. A} {\bfseries 55} no.~6, (2022) 063001}, \href{http://arxiv.org/abs/2006.13630}{{\ttfamily arXiv:2006.13630 [hep-th]}}.

\bibitem{Gang:2012yr}
D.~Gang, E.~Koh, and K.~Lee, ``{Line Operator Index on $S^{1}\times S^{3}$},'' \href{http://dx.doi.org/10.1007/JHEP05(2012)007}{{\em JHEP} {\bfseries 05} (2012) 007}, \href{http://arxiv.org/abs/1201.5539}{{\ttfamily arXiv:1201.5539 [hep-th]}}.

\bibitem{Dorey:2018klg}
N.~Dorey and A.~Singleton, ``{An Index for Superconformal Quantum Mechanics},'' \href{http://arxiv.org/abs/1812.11816}{{\ttfamily arXiv:1812.11816 [hep-th]}}.

\bibitem{Hatsuda:2025yzp}
Y.~Hatsuda and T.~Okazaki, ``{S-duality of boundary lines in $ \mathcal{N} $ = 4 SYM theories and supersymmetric indices},'' \href{http://dx.doi.org/10.1007/JHEP08(2025)127}{{\em JHEP} {\bfseries 08} (2025) 127}, \href{http://arxiv.org/abs/2505.14962}{{\ttfamily arXiv:2505.14962 [hep-th]}}.

\bibitem{GSW:2012superstring}
M.~B. Green, J.~H. Schwarz, and E.~Witten, {\em Superstring Theory: 25th Anniversary Edition}.
\newblock Cambridge Monographs on Mathematical Physics. Cambridge University Press, 2012.

\bibitem{Dubovsky:2015axion}
S.~Dubovsky, R.~Flauger, and V.~Gorbenko, ``{Flux tube spectra from approximate integrability at low energies},'' \href{http://dx.doi.org/10.1134/s1063776115030188}{{\em Journal of Experimental and Theoretical Physics} {\bfseries 120} no.~3, (Mar., 2015) 399–422}. \url{http://dx.doi.org/10.1134/S1063776115030188}.

\bibitem{Witten:1998confinement}
E.~Witten, ``{Anti-de Sitter Space, Thermal Phase Transition, and Confinement in Gauge Theories},'' \href{http://dx.doi.org/10.4310/ATMP.1998.v2.n3.a3}{{\em Adv. Theor. Math. Phys.} {\bfseries 2} (1998) 505--532}, \href{http://arxiv.org/abs/hep-th/9803131}{{\ttfamily arXiv:hep-th/9803131}}.

\bibitem{Aharony:2000stg}
O.~Aharony, S.~S. Gubser, J.~Maldacena, H.~Ooguri, and Y.~Oz, ``{Large N field theories, string theory and gravity},'' \href{http://dx.doi.org/10.1016/s0370-1573(99)00083-6}{{\em Physics Reports} {\bfseries 323} no.~3-4, (Jan., 2000) 183–386}. \url{http://dx.doi.org/10.1016/S0370-1573(99)00083-6}.

\bibitem{Eberhardt:2018worldsheet}
L.~Eberhardt, M.~R. Gaberdiel, and R.~Gopakumar, ``{The Worldsheet Dual of the Symmetric Product CFT},'' \href{http://dx.doi.org/10.1007/JHEP04(2019)103}{{\em JHEP} {\bfseries 04} (2019) 103}, \href{http://arxiv.org/abs/1812.01007}{{\ttfamily arXiv:1812.01007 [hep-th]}}.

\bibitem{Eberhardt:2019deriving}
L.~Eberhardt, M.~R. Gaberdiel, and R.~Gopakumar, ``{Deriving the $\mathrm{AdS}_3/\mathrm{CFT}_2$ Correspondence},'' \href{http://dx.doi.org/10.1007/JHEP02(2020)136}{{\em JHEP} {\bfseries 02} (2020) 136}, \href{http://arxiv.org/abs/1911.00378}{{\ttfamily arXiv:1911.00378 [hep-th]}}.

\bibitem{Eberhardt:2025localising}
L.~Eberhardt and M.~R. Gaberdiel, ``{A localising AdS$_3$ sigma model},'' \href{http://dx.doi.org/10.21468/SciPostPhys.19.2.060}{{\em SciPost Phys.} {\bfseries 19} (2025) 060}, \href{http://arxiv.org/abs/2505.09226}{{\ttfamily arXiv:2505.09226 [hep-th]}}.

\bibitem{Gomis:2026ads-cpt}
J.~Gomis, ``{The AdS/$\mathsf{C}$-$\mathsf{P}$-${\mathsf T}$ Correspondence},'' 2026.
\newblock \url{https://arxiv.org/abs/2507.12467}.

\end{thebibliography}\endgroup

\end{document}